\documentclass{emulateapj}
\usepackage{lscape}

\newcommand{\kms}{${\rm km \, s^{-1}}$}

\shorttitle{\ion{H}{1} Properties of Bulgeless Disk Galaxies}

\begin{document}

\slugcomment{Submitted to ApJS 2010 October 11; accepted 2011
  April 18}

\title{PROPERTIES OF BULGELESS DISK GALAXIES. I.\ Atomic Gas}

\author{Linda~C.~Watson\altaffilmark{1},
  Eva~Schinnerer\altaffilmark{2}, Paul~Martini\altaffilmark{1,3},
  Torsten~B\"oker\altaffilmark{4}, Ute~Lisenfeld\altaffilmark{5}
}

\altaffiltext{1}{Department of Astronomy, The Ohio State University,
  140 West 18th Avenue, Columbus, OH 43210, USA;
  watson@astronomy.ohio-state.edu}
\altaffiltext{2}{Max-Planck-Institut f\"ur Astronomie, K\"onigstuhl 17,
  D-69117 Heidelberg, Germany}
\altaffiltext{3}{Center for Cosmology and AstroParticle Physics, The
  Ohio State University, 191 West Woodruff Avenue, Columbus, OH 43210,
  USA}
\altaffiltext{4}{European Space Agency, Department of RSSD, Keplerlaan
  1, 2200 AG Noordwijk, The Netherlands}
\altaffiltext{5}{Departamento de F\'isica Te\'orica y del Cosmos, Universidad
  de Granada, Spain}
  
\begin{abstract}

We study the neutral hydrogen properties of a sample of 20
bulgeless disk galaxies (Sd - Sdm Hubble types), an interesting class
that can be used to constrain galaxy formation and evolution,
especially the role of mergers versus internal processes.  Our sample
is composed of nearby (within 32~Mpc), moderately inclined galaxies
that bracket the circular velocity of $120 \, {\rm km \, s}^{-1}$, which
has been found to be associated with a transition in dust scale
heights in edge-on, late-type disks.  Here we present \ion{H}{1}
channel maps, line profiles, and integrated intensity maps.  We also
derive kinematic parameters, including the circular velocity, from
rotation curve analyses and calculate the integrated \ion{H}{1} flux
and \ion{H}{1} mass for each galaxy in the sample.  Three of the
20 galaxies in our sample have kinematically distinct outer
components with major axes that differ by $30\degr - 90\degr$ from the
main disk.  These distinct outer components may be due to a recent
interaction, which would be somewhat surprising because the disks do
not contain bulges.  We will use the data products and derived
properties in subsequent investigations into star formation and
secular evolution in bulgeless disks with circular velocities above
and below $120 \, {\rm km \, s}^{-1}$.

\end{abstract}


\section{Introduction}

The formation of disk galaxies has remained an interesting and
unresolved question in galaxy evolution for many decades. The basic
model is fairly clear: galaxies form in cold dark matter (CDM) halos
through the dissipational collapse of gas \citep{white78}. However,
disks formed in hydrodynamic simulations are generally too highly
concentrated and possess both proportionally larger bulges and lower
angular momentum than the disks observed in the local universe
\citep{katz91,navarro94, navarro00}. Recent simulations that include
feedback and/or high gas fractions have produced disks that survive
hierarchical assembly and more closely resemble observed disk galaxies
\citep{abadi03,robertson06,brook11}, but still some properties of the
modeled disks do not match observed values. A coupled problem is that
the merger rate in CDM models may be too high for many disk-dominated
galaxies to survive to the present day \citep{toth92}, and yet
observations show that these galaxies are common
\citep[e.g.,][]{kautsch06,cameron09,kormendy10}.

Observations of bulgeless disks, including galaxies termed ``flat'' or
``thin'' when viewed edge-on, provide a particularly important
benchmark for theories of disk galaxy evolution. Their structural
homogeneity is valuable for the direct study of disk properties
without the complications associated with removal of a bulge
component. A common conception is that because bulges form due to
mergers, bulgeless galaxies have evolved in isolation.  This view may
be changing as simulations are starting to form bulgeless galaxies
with relatively rich merger histories
\citep[e.g.,][]{robertson06,hopkins09,brook11}.  Nonetheless, the
space density of bulgeless galaxies provides a useful constraint on
the role of mergers in galaxy formation.  In addition, bulgeless disk
galaxies are an excellent comparison sample to study secular
evolution, that is the role of internal processes in the buildup of
bulges \citep{kormendy04}. The properties of bulgeless disks, in
comparison to their slightly earlier-type counterparts with
pseudobulges, may shed new light on the key processes that drive
secular evolution.

Bulgeless galaxies had historically been overlooked in many surveys
because they are often relatively low surface brightness objects, but
more recent work has begun to identify them in large numbers and
demonstrate that these often faint objects are a natural continuation
of the sequence of brighter, earlier-type spirals
\citep{matthews97}. The first large sample of these objects was part
of the edge-on Flat Galaxy Catalog compiled by
\citet{karachentsev93}. Demographic studies of edge-on flat disks in
the Sloan Digital Sky Survey by \citet{kautsch06} provided some of the
first measurements of the frequency of these objects and found that on
order 15\% of all edge-on disk galaxies are thin, bulgeless disks. The
demographics of these galaxies may prove to be a useful constraint on
simulations of disk galaxy formation, particularly as the
sophistication of simulations improves to include major and minor
mergers in orbits derived from cosmological simulations
\citep{stewart08,hopkins08}.

A subset of bulgeless, edge-on galaxies in the Flat Galaxy Catalog was
studied in detail by \citet{dalcanton00}. One extremely interesting
result to arise from this study was that less massive, slowly rotating
bulgeless disks with ${\rm v_{circ}} < 120 \, {\rm km \, s}^{-1}$ do
not have the narrow dust lanes characteristic of both earlier-type
spirals and more rapidly rotating bulgeless disks
\citep{dalcanton04}. These authors conclude that disk stability,
parametrized by a generalized Toomre Q parameter \citep{rafikov01}, is
the most important physical property in determining the dust scale
height of a galaxy. The scale height of the cold interstellar medium
(ISM), and therefore the dust, is set by the balance between the mass
surface density of gas and stars in the disk and the velocity
dispersion of those components.  \citet{dalcanton04} postulated that
this transition in the cold ISM may correlate with substantial changes
in star formation efficiency, metallicity, and bulge formation. Other
studies of bulgeless disks have attempted to measure the cold ISM
directly, in particular the molecular component
\citep{boeker03,matthews05}. These studies found that late-type
spirals follow the same relationships between molecular gas content,
galaxy luminosity, and far-infrared luminosity as earlier-type
spirals, which reinforces the conclusion that bulgeless disks fall on
the end of a continuum of disk galaxy properties, rather than comprise
a unique class.

The goal of our study is to investigate the properties of bulgeless
disk galaxies as a function of circular velocity, yet with a
moderately-inclined sample that is well suited to measurements of star
formation rate (SFR) and gas surface densities, the presence of
bulges, nuclear star clusters, and stellar bars, and precise kinematic
measurements to estimate dynamical masses. Our analysis is based on a
sample of 20 nearby Sd galaxies that bracket the ${\rm v_{circ}} = 120
\, {\rm km \, s^{-1}}$ transition velocity observed by
\citet{dalcanton04}. In the present paper we describe \ion{H}{1}
observations of the complete sample and derive the circular velocity
and atomic gas distribution of each galaxy. In L.\ Watson et al.\
(2011, in preparation), we will present an analysis of the star
formation properties of this sample, which will draw on CO(1-0)
observations from the Institut de Radioastronomie Millim\'etrique
(IRAM) $30 \, {\rm m}$ telescope, H$\alpha$ data from the 2.4~m
Hiltner Telescope of the MDM Observatory, and polycyclic aromatic
hydrocarbon emission measured with the {\it Spitzer Space Telescope}
Infrared Array Camera (IRAC).  Subsequent papers will focus on the
widths of the dust lanes seen in absorption in F606W images from
either the Advanced Camera for Surveys (ACS) or Wide-Field Planetary
Camera 2 instruments on the {\it Hubble Space Telescope} ({\it HST})
and the signposts of secular evolution, such as the presence of
pseudobulges, nuclear star clusters, and large-scale stellar bars.

This paper is organized as follows: we present our sample in
Section~\ref{sec:sample} and outline our data reduction method in
Section~\ref{sec:obs}.  We present channel maps, integrated line
profiles, integrated intensity maps, and rotation curve analyses in
Section~\ref{sec:products}.  We discuss objects with noteworthy
morphology or evidence of interaction in Section~\ref{sec:morph} and
conclude in Section~\ref{sec:summary}.

\section{Sample Selection}
\label{sec:sample}

To assemble our sample, we started with all late-type disk galaxies
(types Scd to Sm) within about 30~Mpc.  This is the maximum distance
at which {\it HST} ACS can resolve features of about 20~pc, which is typical
of dust structures in nearby, earlier-type spiral galaxies and
molecular cloud scale heights in our Galaxy \citep{martini03,stark05}.
We also restricted the selection to galaxies with inclinations between
35 and 55~degrees, such that we can accurately measure the gas and SFR
surface densities as well as derive kinematic parameters from rotation
curve fitting.  Objects near the Galactic plane or with evidence for
tidal disruptions and mergers were removed from the sample, as were
several of the closest systems where the size of the galaxy is larger
than the $202\arcsec \times 202\arcsec$ field of view of the wide
field channel of ACS.  To provide bright stellar emission as the
background to study dust lanes in absorption, we selected the 10
highest surface brightness galaxies with ${\rm v_{circ}} > 120 \, {\rm
km \, s^{-1}}$ and the 10 highest surface brightness galaxies with
${\rm v_{circ}} < 120 \, {\rm km \, s^{-1}}$, with circular velocities
primarily estimated from single-dish observations of the \ion{H}{1}
emission line. Our final sample consists of 20 nearby, moderately
inclined Sd to Sdm galaxies that bracket the circular velocity
transition for dust scale heights found in \citet{dalcanton04}.
Table~\ref{tab:sample} provides the basic properties of the galaxies
in our sample.

\section{Observations and Data Reduction}
\label{sec:obs}

Our \ion{H}{1} data were obtained with the Very Large Array (VLA),
operated by the National Radio Astronomy Observatory\footnote{The
National Radio Astronomy Observatory is a facility of the National
Science Foundation operated under cooperative agreement by Associated
Universities, Inc.}.  Most of these galaxies were observed between
2006 October and 2009 August for projects AM0873 and AM0942.  In
addition, we downloaded data from the VLA archive for two objects:
NGC~5964 was observed in 2001 August for project AZ0133 and UGC~6446
was observed in 2002 June and November for project AL0575.  We most
often used the C configuration, where the largest angular structure
that can be imaged is about $15 \arcmin$ and the resolution is about
$13 \arcsec$.  We used the CnB configuration, with its extended north
arm, to observe four of the southern targets. The B configuration was
used for a portion of the archival data for UGC~6446.  The L-band
observations were generally carried out using spectral line mode 4
with online Hanning smoothing, resulting in four intermediate
frequency bands (IFs), each with a bandwidth of $1.5625 \, {\rm MHz}$,
64 channels, and channel widths of $24.414 \, {\rm kHz}$ ($5.2 \, {\rm
km \, s^{-1}}$).  We tuned the AC and BD IFs to overlap in frequency,
with the line emission approximately centered on the overlap region.
The only exception is UGC~6446, which was observed in mode 2AD with
online Hanning smoothing.  The two IFs were tuned to the same
frequency, each with a bandwidth of $1.5625 \, {\rm MHz}$, 128
channels, and channel widths of $12.207 \, {\rm kHz}$ or $2.6 \, {\rm
km \, s^{-1}}$.  We include a detailed summary of the observations for
each object in Table~\ref{tab:obs_summary}.

The objects in our sample were observed at various stages in the
transition from the VLA to the Expanded Very Large Array (EVLA).  A
number of complications are present in data obtained during the
upgrade, two of which are particularly important for our data.  First,
the EVLA L-band receivers have a wider bandpass than those on VLA
antennas, which led to closure phase errors on VLA-EVLA baselines.
Second, digital EVLA signals had to be converted into analog signals
before being fed into the VLA correlator.  A result of this conversion
is that flux from frequencies 0.5~MHz below a given IF is aliased into
the lower 0.5~MHz of that IF.  The transition to the EVLA system was
completed in 2010 January with the commissioning of the EVLA
correlator.  Thus new data from the EVLA are no longer affected by
these issues.

We broadly followed the typical reduction procedure for \ion{H}{1}
spectral line data, except for a few important changes to address the
effects of closure errors on VLA-EVLA baselines and aliasing.  Our
methods for managing the transition problems depended on the number of
EVLA antennas in the array, as described in the following three
sections.

\subsection{Data Processing with Five or Fewer EVLA Antennas}
\label{sec:dp1}

For the ten objects observed in or before 2006 November, there were
five or fewer EVLA antennas in the array.  We simply excluded the EVLA
antennas from the entire dataset and the results for these objects are
therefore unaffected by closure errors on VLA-EVLA baselines and
aliasing.  These data were processed with the 31DEC05 version of the
Astronomical Image Processing System (AIPS).

We first flagged the calibrator data, looking especially for problems
in the first and last scans and for missing or corrupted data in
blocks of time or in a single antenna.  Next, we carried out the gain
amplitude and phase calibration using the ``channel 0'' dataset, which
is a vector average of the inner 75\% of the frequency channels in the
full spectral dataset.  We used 3C48 and/or 3C286 (two of the
recommended standard VLA calibrators) as our flux density calibrators
and list the phase calibrator used for each object in
Table~\ref{tab:obs_summary}.  Our phase calibrators were observed
approximately every 45 to 60 minutes.  For the flux density
calibrator, we solved for the antenna gain amplitude and phase
corrections by comparing the observed and known amplitude and phase,
using only baselines for which the calibrator is known to be a point
source.  We processed the phase calibrator in the same manner, except
in this case we compared the observed amplitude to a constant but
unknown intrinsic amplitude. We then applied the gain solution to the
phase calibrator.  Finally, we interpolated the phase calibrator gain
amplitude and phase solutions in time to obtain the appropriate
atmospheric correction for the source data.

We derived the bandpass solution by running BPASS on the full spectral
dataset of the flux density calibrator.  This gives the gain amplitude
and phase correction as a function of frequency.  We then divided the
data into individual object sets, applied the gain and bandpass
solutions, and stitched the AC and BD IFs together using SPLIT and
UVGL or UJOIN.  We performed the continuum subtraction in the uv plane
using UVLIN.  We determined the continuum level using a zeroth-order
fit of at least ten line-free channels on both sides of the line.
This low-level fit is appropriate as the average offset between the
IFs is less than 2\%.  Finally, we used IMAGR to create
naturally-weighted (using the parameter {\it robust} = 5) and
robustly-weighted ({\it robust} = 0.5) data cubes, cleaning down to
twice the noise.  We also created an image from an average of the
line-free channels and found that three galaxies have detectable
continuum emission.  We computed the noise in each data cube by
averaging the rms surface brightness in a couple regions in
signal-free channels.  These values are given in Column~6 of
Table~\ref{tab:im_summary} and are on average within 15\% of the
theoretical thermal noise limit.  Finally, we converted the frequency
axis to velocity and shifted the velocities to the heliocentric
reference frame.

We summarize the properties of the final images in
Table~\ref{tab:im_summary}.  For all naturally-weighted data cubes, we
kept the original resolution, except in the case of NGC~3906, where we
averaged two channels together.  To achieve sufficient
signal-to-noise, we averaged two channels together for over half of
the robustly-weighted data cubes to give a final resolution of $10.4
\, {\rm km \, s^{-1}}$.  We produced images that are larger than
typical for NGC~3906 and IC~1291 because both fields contain a bright
continuum source outside the nominal field of view that was not
adequately subtracted.  For these objects, we cleaned two boxes: one
centered on the galaxy and approximately $30\arcmin$ across and the
second centered on the bright continuum source.

To obtain moment maps representing the integrated line intensity, the
intensity-weighted velocity, and the velocity dispersion, we followed
the procedure of \citet{walter08}.  We first convolved the original
cube such that the beam major and minor axes are a factor of two to
four larger.  We then used the AIPS task BLANK to exclude emission
fainter than twice the noise in the lower-resolution cube.  We also
manually blanked the lower-resolution cube to remove any noise peaks
remaining after the flux cut.  Finally, we used the lower-resolution
cube as a mask to blank the original cube.  The advantage of this
procedure over simply applying a $2\sigma$ cut is that our blanked
data cube contains real signal below that cut.  We performed a primary
beam correction on the blanked, naturally-weighted data cube using
PBCOR and then created moment maps using XMOM.  Note that we did not
exclude single-channel emission peaks when creating the moment maps
and we did no flux rescaling to correct for the fact that the cleaned
data cube contains residual flux that is convolved with the dirty beam
\citep{jorsater95}.

\subsection{Data Processing with Fifteen EVLA Antennas}
\label{sec:dp2}

For the eight objects observed in 2008 May, there were fifteen EVLA
antennas in the array.  The reductions of these data were carried out
with the 31DEC08 version of AIPS.  To arrive at calibrated visibility
data, we used the valuable reduction recipe developed by the LITTLE
THINGS (Local Irregulars That Trace Luminosity Extremes - The HI
Nearby Galaxy Survey; PI: Deidre Hunter; D.\ Hunter et al.\ 2011, in
preparation) collaboration, with only minor adjustments.  This recipe
includes methods to reduce the effects of closure errors on VLA-EVLA
baselines and aliasing\footnote{Methods for managing VLA-EVLA
transition data are also described on the VLA Website:
http://www.vla.nrao.edu/astro/guides/evlareturn/.}.  In this section,
we highlight the main differences between the reduction procedure
described in Section~\ref{sec:dp1} and the LITTLE THINGS procedure.

To address the aliasing problem, we excluded EVLA-EVLA baselines in
the calibrator and source data.  However, this did not completely
remove the aliased signal, as we discuss further in
Section~\ref{sec:aliasing}.  To address the closure error problem, our
first major reduction step was to determine the bandpass solution from
the flux density calibrator.  Applying this solution removes the slope
in phase as a function of frequency, which is introduced by the
different VLA and EVLA receiver bandpasses.  We used AVSPC to apply
the bandpass solution and average over frequency to create a new
channel 0 dataset.  We then carried out the gain amplitude and phase
calibration using the new channel 0.  The remaining reduction steps
were similar to those described in Section~\ref{sec:dp1}.

\subsection{Data Processing with Twenty-two EVLA Antennas}
\label{sec:dp3}

The observations of NGC~4519 and UGC~6930 in 2009 July and August were
carried out with twenty-two EVLA antennas, leaving only four VLA
antennas remaining in the array.  We included all functioning EVLA
antennas and EVLA-EVLA baselines for calibration as well as in the
object data because excluding them would have resulted in an
unacceptable loss of signal.  We study the effect of this choice in
Section~\ref{sec:aliasing}.  Otherwise, we used the same reduction
method as in Section~\ref{sec:dp2}, that is we employ the LITTLE
THINGS recipe.

\subsection{More on Aliasing}
\label{sec:aliasing}

Aliasing is a result of the digital to analog conversion that was
required for EVLA signals to be processed by the VLA correlator.  Flux
from 0.5~MHz below a given IF is aliased into the lower 0.5~MHz of the
IF.  Aliasing affects both IFs in the calibrator data because there is
continuum emission at lower frequencies relative to each IF.  It also
affects the object data because we employed band-stitching, where the
AC and BD IFs overlap in frequency, and the line emission is centered
on the overlap region.  The line emission in the lower-frequency IF
should be unaffected by aliasing because there is little source
continuum emission that could be aliased into the IF and the line
emission is almost exclusively in the upper 0.5~MHz of the IF, whereas
any aliased signal would affect the lower 0.5~MHz.  The
higher-frequency IF is affected because line emission from the
lower-frequency IF is aliased into the higher-frequency IF.

For the objects observed in May 2008, we expected that excluding
EVLA-EVLA baselines would remove the aliased signal because it should
not correlate on EVLA-VLA baselines.  Surprisingly, the data are still
affected, in that we see evidence of aliasing in the antenna-averaged
and normalized bandpass correction amplitude determined from the flux
density calibrator (described further in the next paragraph).
Nevertheless, the effect on the integrated flux is the same or smaller
than in the case of the objects observed in 2009, which we describe
further below.

We investigated the effect of the aliased signal on the integrated
flux using UGC~6930 and NGC~4519, the objects observed with only four
VLA antennas remaining in the array.  The aliased signal is strongest
at the low-frequency end of the IF and decreases with increasing
frequency.  As a result, the antenna-averaged and normalized bandpass
correction amplitude determined from the flux density calibrator
spectrum reaches a minimum of about 0.7 to 0.8 in the lowest-frequency
channel and reaches 1.0 at about 0.5~MHz higher frequency.  We applied
this bandpass correction to the object data but it is not necessarily
appropriate for correcting the aliased signal from a line source.  To
illustrate the effect, imagine all the line emission, with a box
profile, is in the upper (lower) frequency IF and the bandpass
correction is 0.8 for all channels.  After bandpass correction, the
integrated flux would be underestimated (overestimated) by 20\%.

The above illustrates the worst case scenario.  The aliased signal
drops off with increasing frequency and we always have line emission
in the upper and lower IFs.  Both these facts decrease the effect of
the aliased signal on the integrated flux.  We found it difficult to
precisely estimate the difference between the final spectrum,
including aliased flux, and the intrinsic spectrum.  Our best estimate
for the contamination of aliased signal to the integrated line flux is
3\% - 11\%.  This estimate is based on the spectrum of NGC~4519, which
has a typical line flux and profile.  For the upper (lower) bound, we
assumed that the peak (minimum) intensity from the upper 0.5~MHz of
the lower-frequency IF was aliased into each channel of the lower
0.5~MHz of the upper-frequency IF according to the amplitude of the
bandpass correction.  Specifically, the estimated aliased intensity for
a channel is $I_{\rm peak} (1/BP_{i} - 1)$, where $I_{\rm peak}$ is
the peak intensity from the lower-frequency IF, $i$ is the channel
number in the upper-frequency IF, and $BP_{i}$ is the bandpass
correction amplitude at channel $i$.  We then compared the integrated
aliased flux to the integrated line flux without bandpass correction.
We expect that the contamination of the aliased signal to the
intrinsic integrated flux is of similar order.

We do not make any correction for aliasing.  Depending on how the line
profile is distributed over the IFs, the integrated flux could be
overestimated or underestimated by 3\% - 11\%.  We take this into
account in our error estimates of the integrated flux.  We expect the
impact of the aliased signal on the rotation curve analysis to be
minimal because any aliased signal should be spatially offset from
real emission and we do not see any evidence of this.

\section{Data Products}
\label{sec:products}

\subsection{Channel Maps}

\begin{figure}
\begin{center}
\plotone{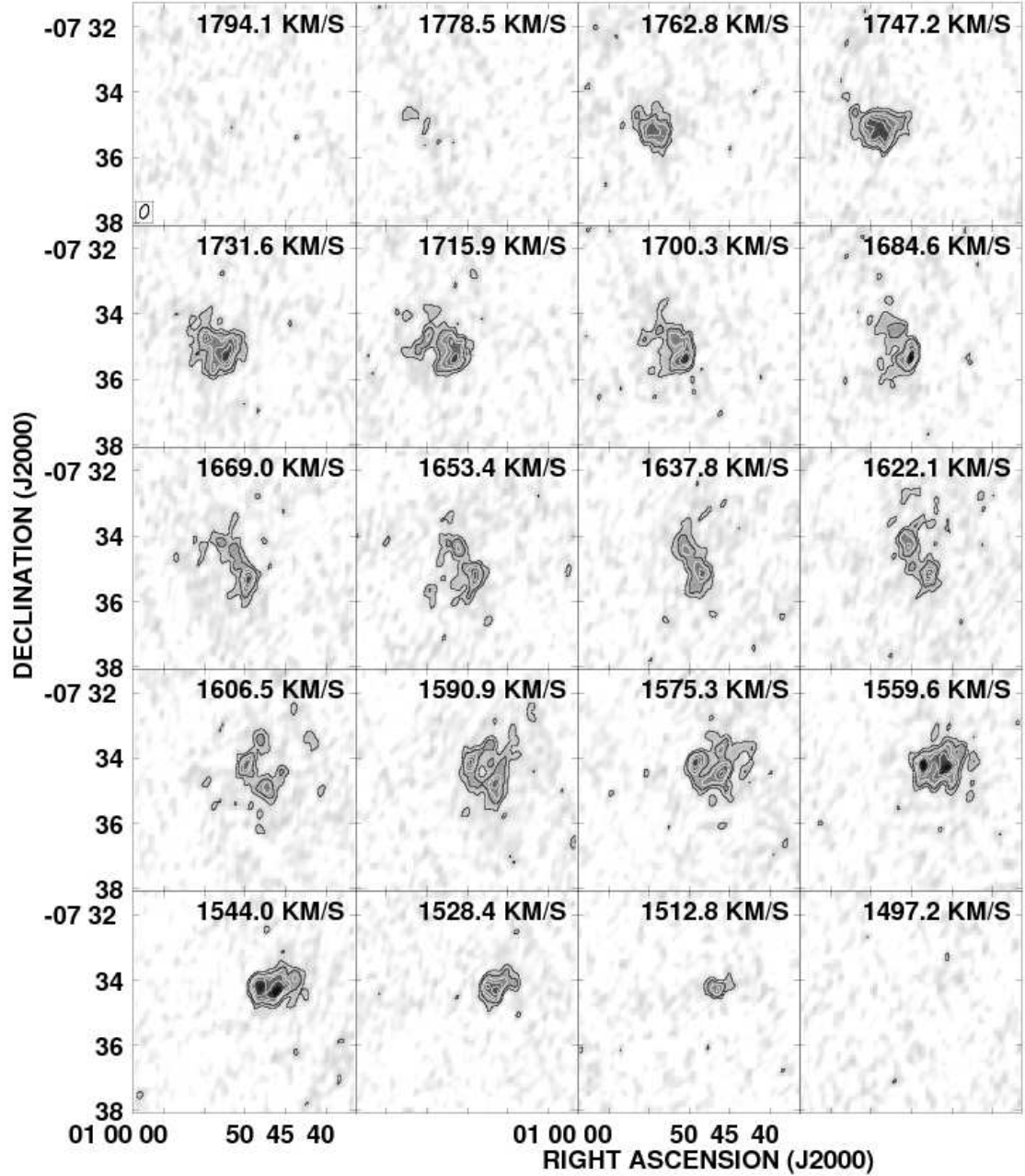}
\caption{NGC~0337 channel maps based on the naturally-weighted cube
  that has not been blanked.  The contours begin at $3\sigma$ and end
  at $11\sigma$, in steps of $2\sigma$, where $\sigma$ is the image
  noise from Column~6 of Table~\ref{tab:im_summary}. The grayscale
  spans -0.02 to 28.33 ${\rm mJy \, beam^{-1}}$.  Every third channel
  is shown and the channel width is 5.2 \kms.  The beam size is shown
  in the lower left corner of the top left panel and the velocity of
  the channel is shown in the upper right of each panel.}
\label{fig:337_chmap}
\end{center}
\end{figure}

\begin{figure}
\begin{center}
\plotone{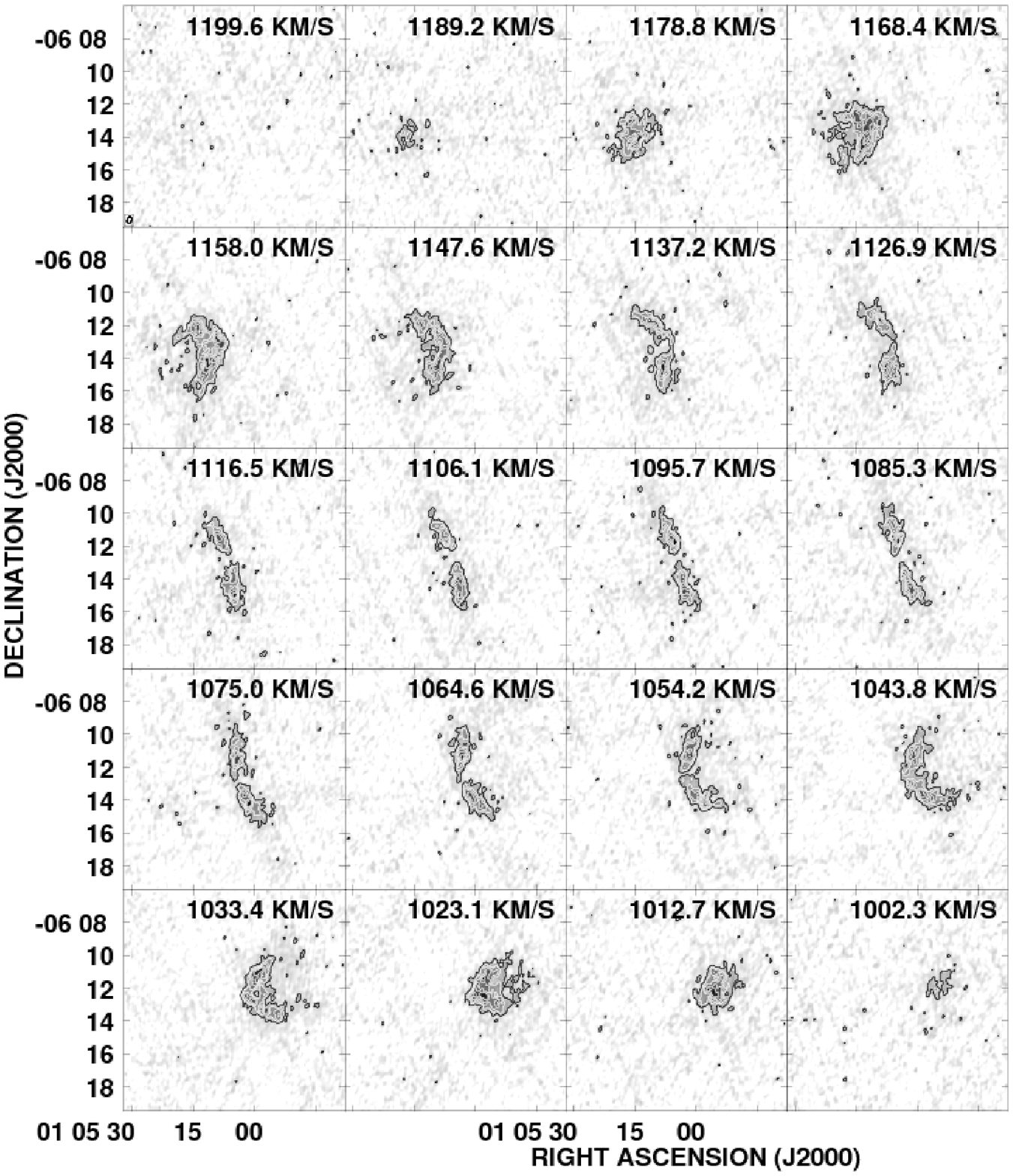}
\caption{As in Figure~\ref{fig:337_chmap}, but for PGC~3853.  The
  grayscale spans -0.02 to 20.30 ${\rm mJy \, beam^{-1}}$.  Every
  second channel is shown and the channel width is 5.2 \kms.}
\label{fig:3853_chmap}
\end{center}
\end{figure}

\begin{figure}
\begin{center}
\plotone{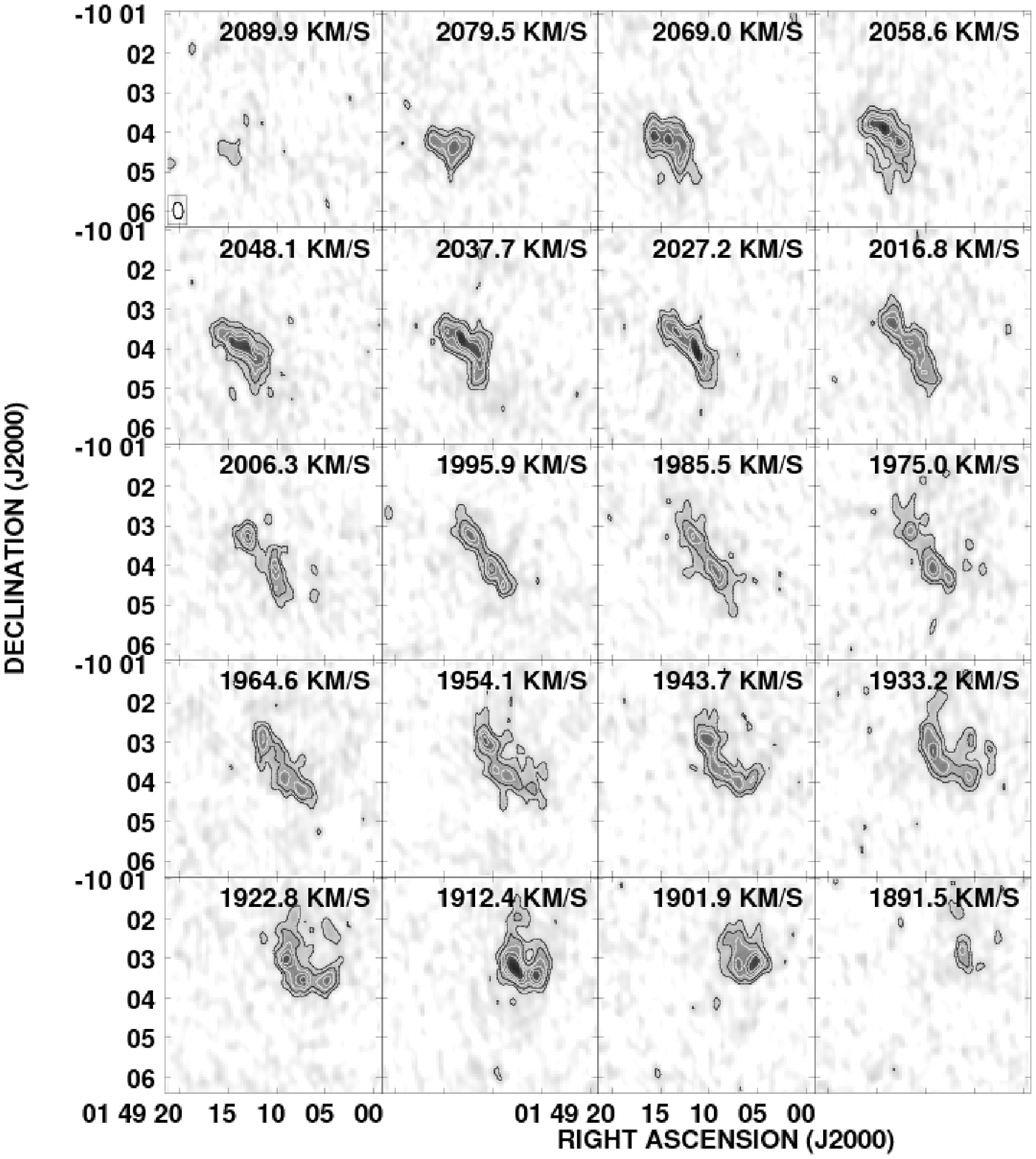}
\caption{As in Figure~\ref{fig:337_chmap}, but for PGC~6667.  The
  grayscale spans -0.02 to 21.13 ${\rm mJy \, beam^{-1}}$.  Every
  second channel is shown and the channel width is 5.2 \kms.}
\label{fig:6667_chmap}
\end{center}
\end{figure}

\begin{figure}
\begin{center}
\plotone{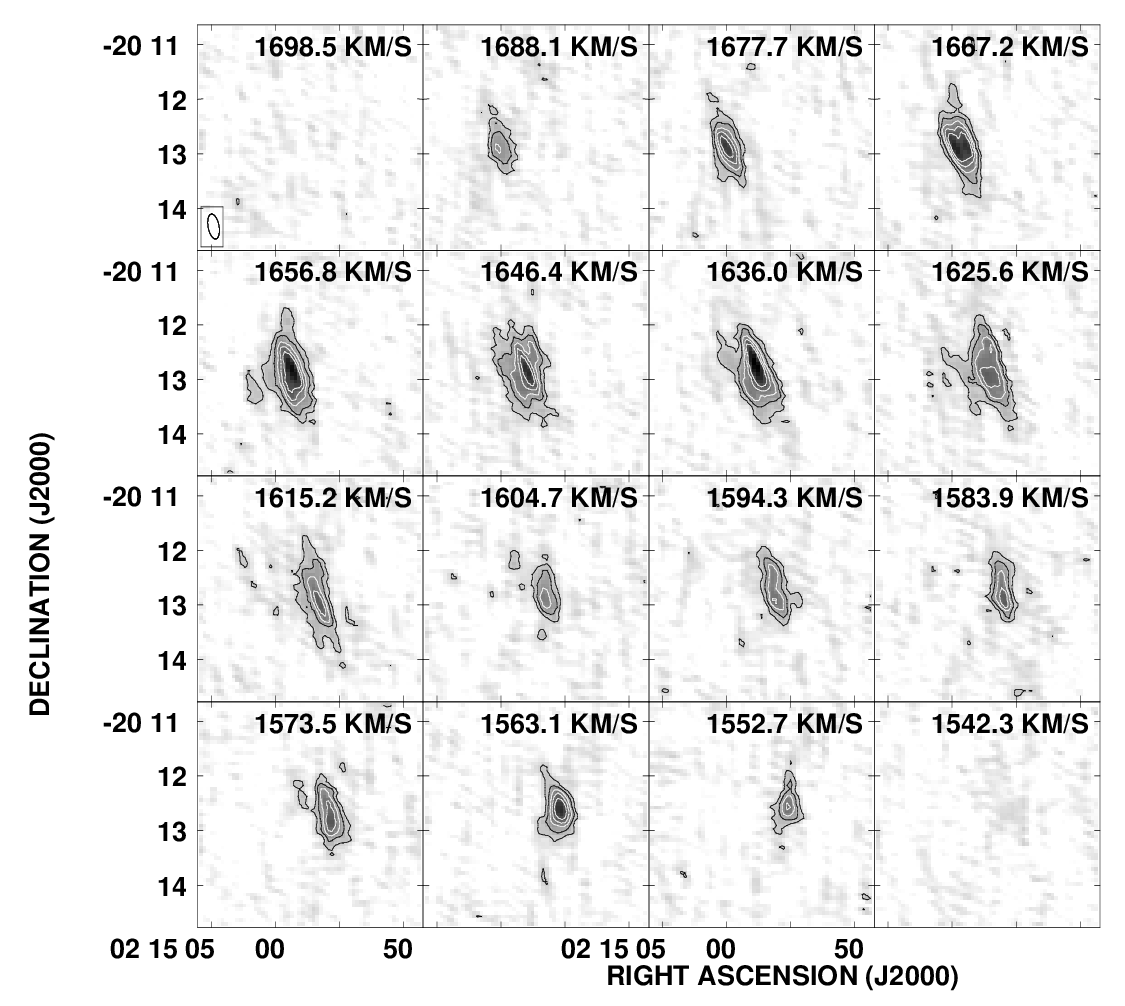}
\caption{As in Figure~\ref{fig:337_chmap}, but for ESO~544-G030.  The
  grayscale spans -0.02 to 16.35 ${\rm mJy \, beam^{-1}}$.  Every
  second channel is shown and the channel width is 5.2 \kms.}
\label{fig:544_chmap}
\end{center}
\end{figure}

\begin{figure}
\begin{center}
\plotone{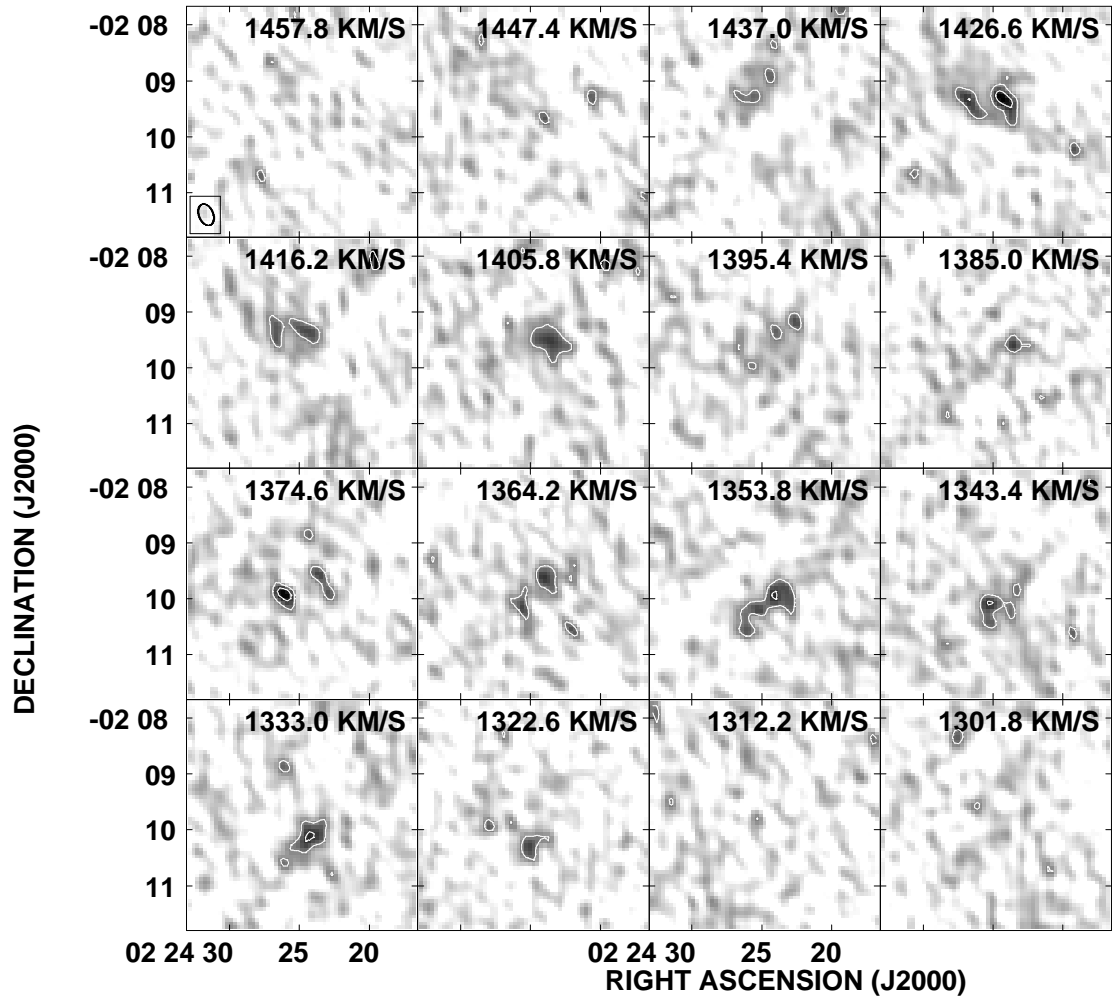}
\caption{As in Figure~\ref{fig:337_chmap}, but for UGC~1862.  The
  grayscale spans -0.02 to 8.43 ${\rm mJy \, beam^{-1}}$.  Every
  second channel is shown and the channel width is 5.2 \kms.}
\label{fig:1862_chmap}
\end{center}
\end{figure}

\begin{figure}
\begin{center}
\plotone{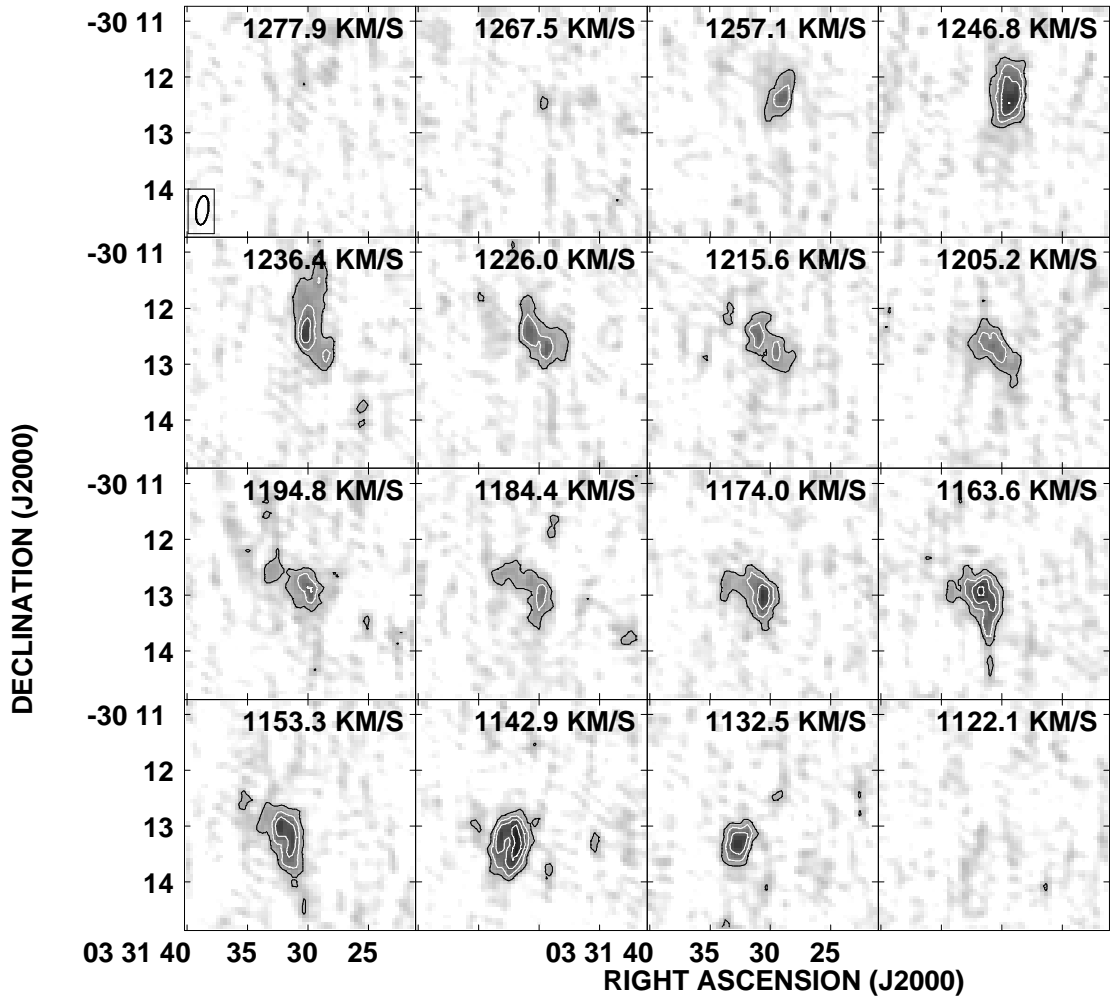}
\caption{As in Figure~\ref{fig:337_chmap}, but for ESO~418-G008.  The
  grayscale spans -0.02 to 14.61 ${\rm mJy \, beam^{-1}}$.  Every
  second channel is shown and the channel width is 5.2 \kms.}
\label{fig:418_chmap}
\end{center}
\end{figure}

\begin{figure}
\begin{center}
\plotone{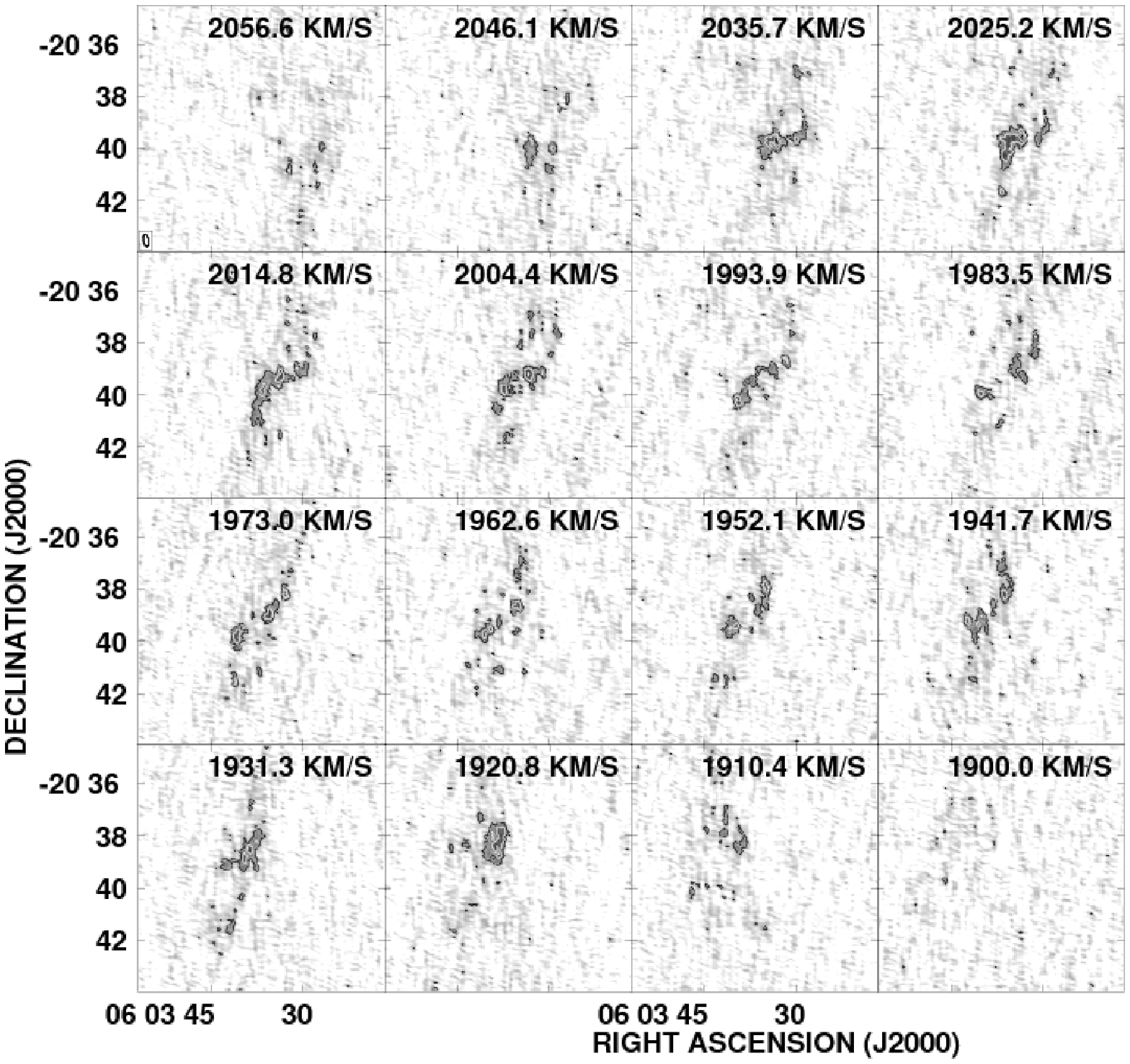}
\caption{As in Figure~\ref{fig:337_chmap}, but for ESO~555-G027.  The
  grayscale spans -0.02 to 12.09 ${\rm mJy \, beam^{-1}}$.  Every
  second channel is shown and the channel width is 5.2 \kms.}
\label{fig:555_chmap}
\end{center}
\end{figure}

\begin{figure}
\begin{center}
\plotone{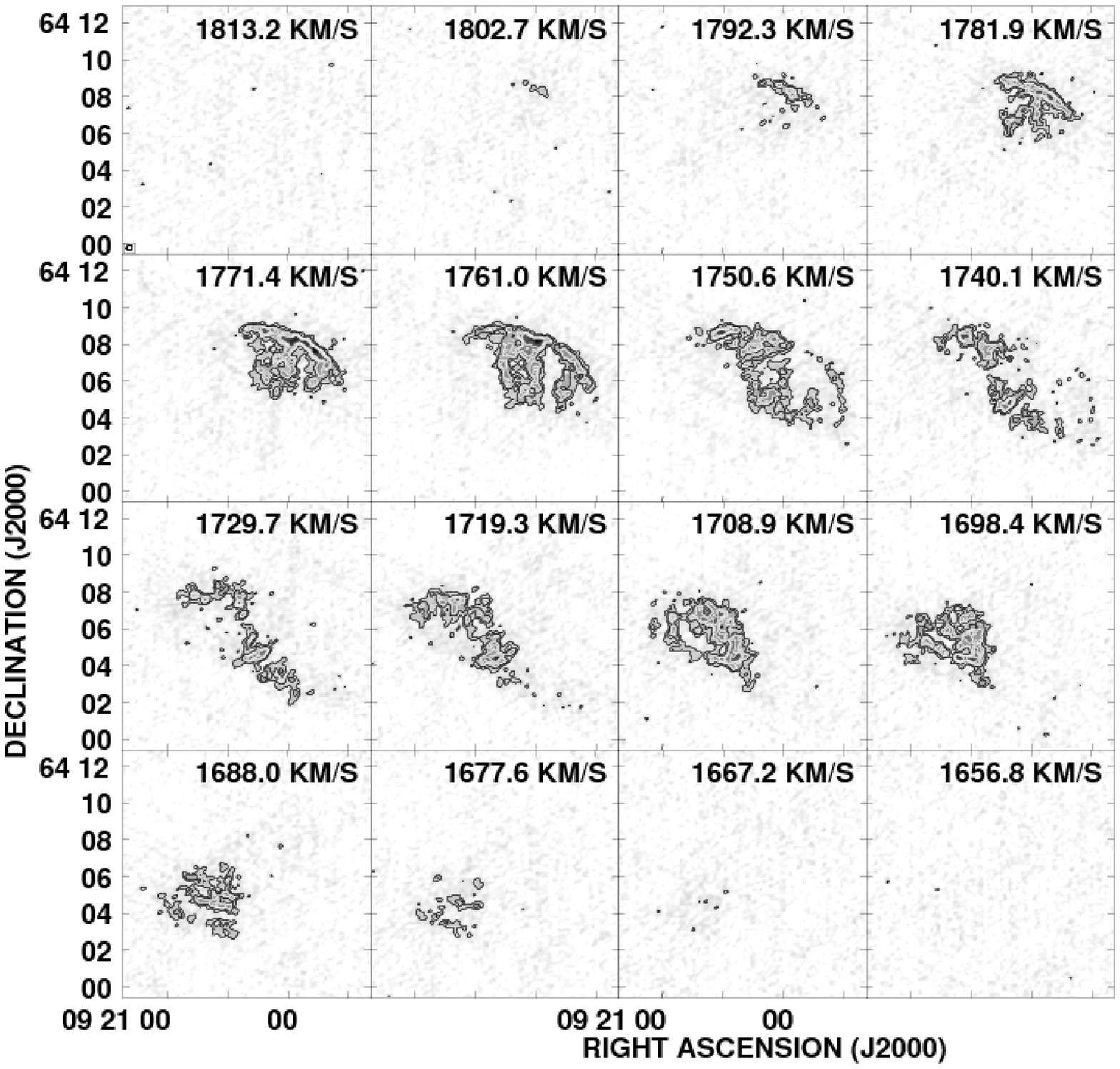}
\caption{As in Figure~\ref{fig:337_chmap}, but for NGC~2805.  The
  grayscale spans -0.02 to 21.31 ${\rm mJy \, beam^{-1}}$.  Every
  second channel is shown and the channel width is 5.2 \kms.}
\label{fig:2805_chmap}
\end{center}
\end{figure}

\begin{figure}
\begin{center}
\plotone{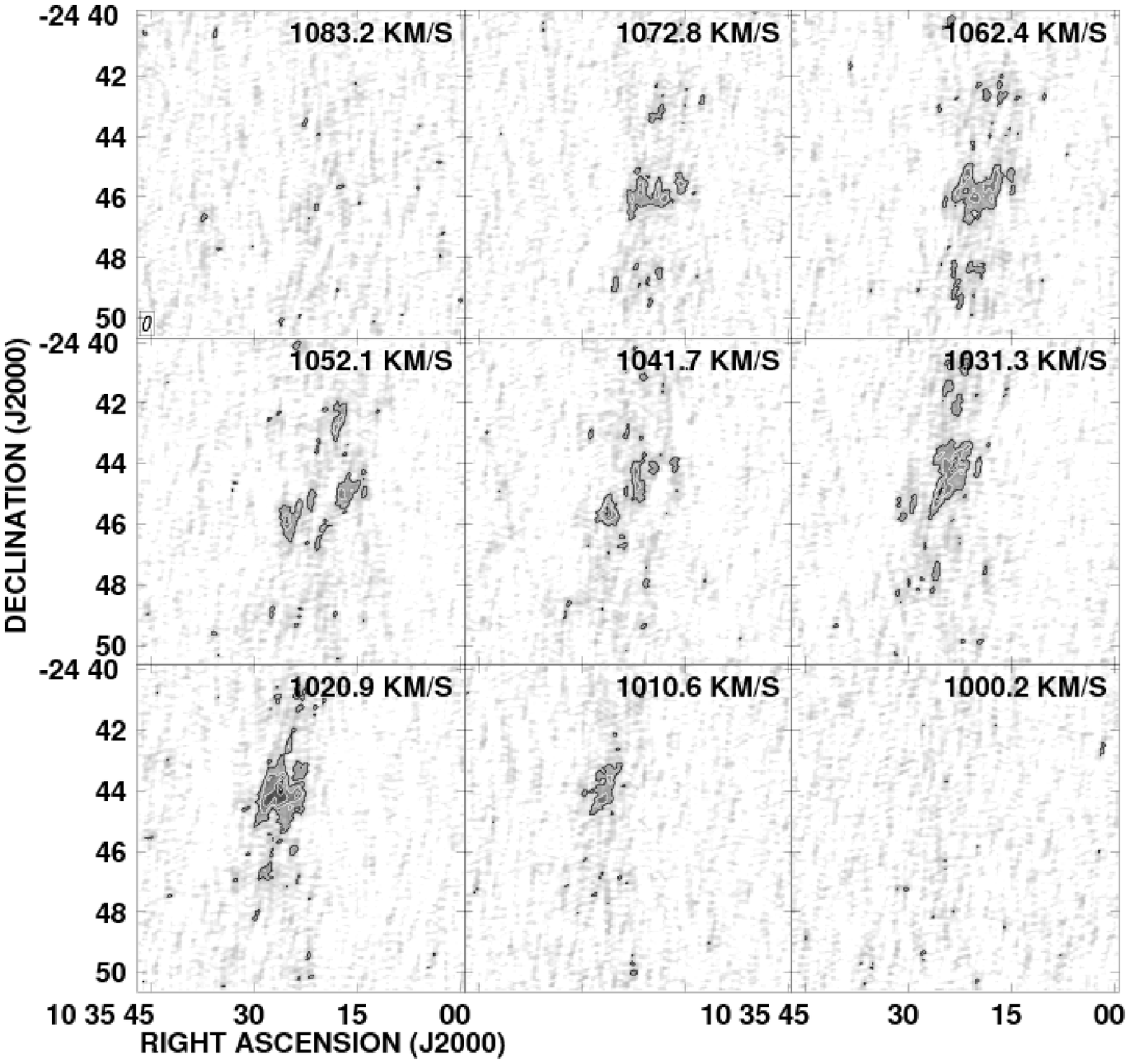}
\caption{As in Figure~\ref{fig:337_chmap}, but for ESO~501-G023.  The
  grayscale spans -0.02 to 16.40 ${\rm mJy \, beam^{-1}}$.  Every
  second channel is shown and the channel width is 5.2 \kms.}
\label{fig:501_chmap}
\end{center}
\end{figure}

\begin{figure}
\begin{center}
\plotone{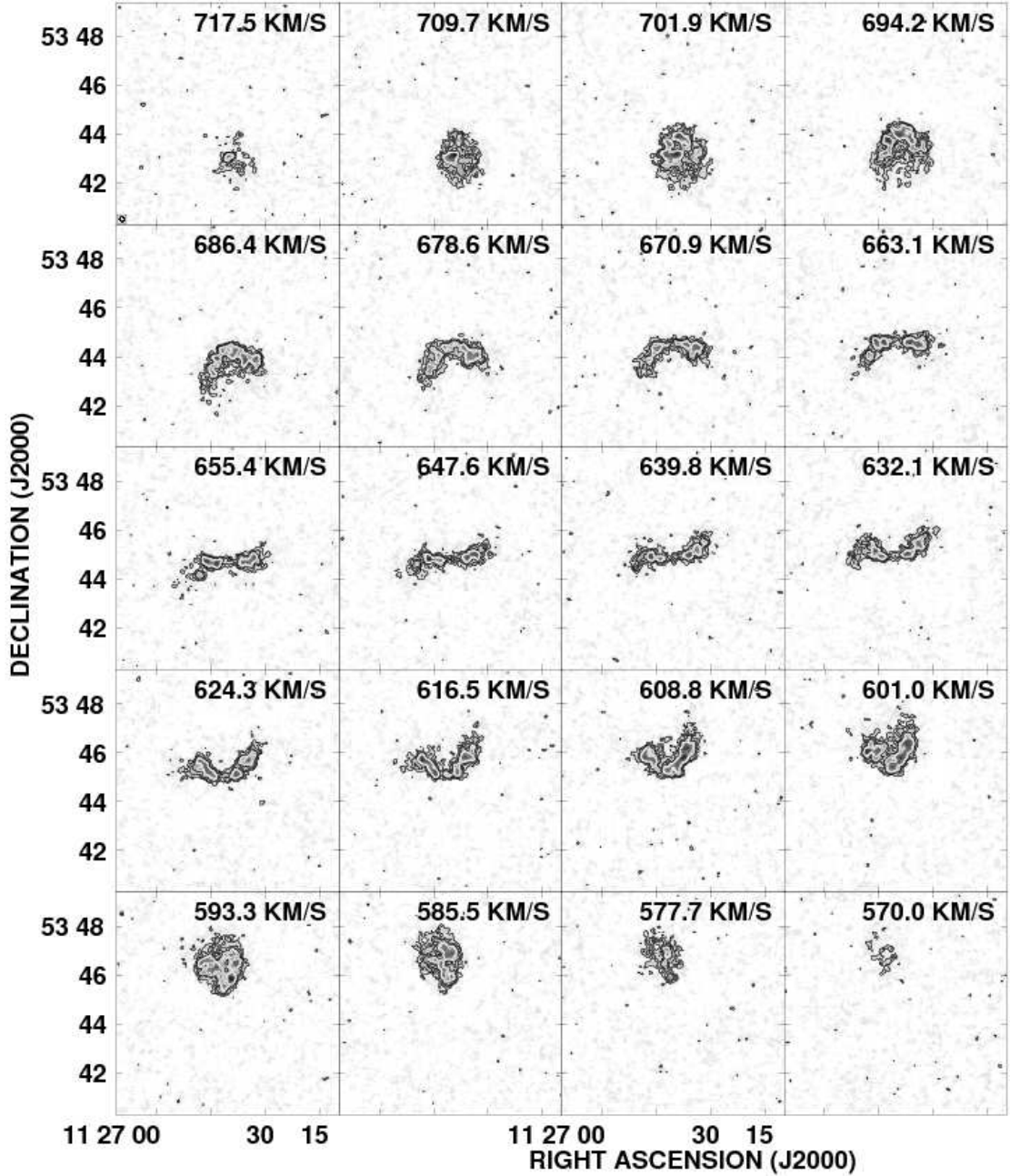}
\caption{As in Figure~\ref{fig:337_chmap}, but for UGC~6446.  The
  grayscale spans -0.02 to 12.52 ${\rm mJy \, beam^{-1}}$.  Every
  third channel is shown and the channel width is 2.6 \kms.}
\label{fig:6446_chmap}
\end{center}
\end{figure}

\begin{figure}
\begin{center}
\plotone{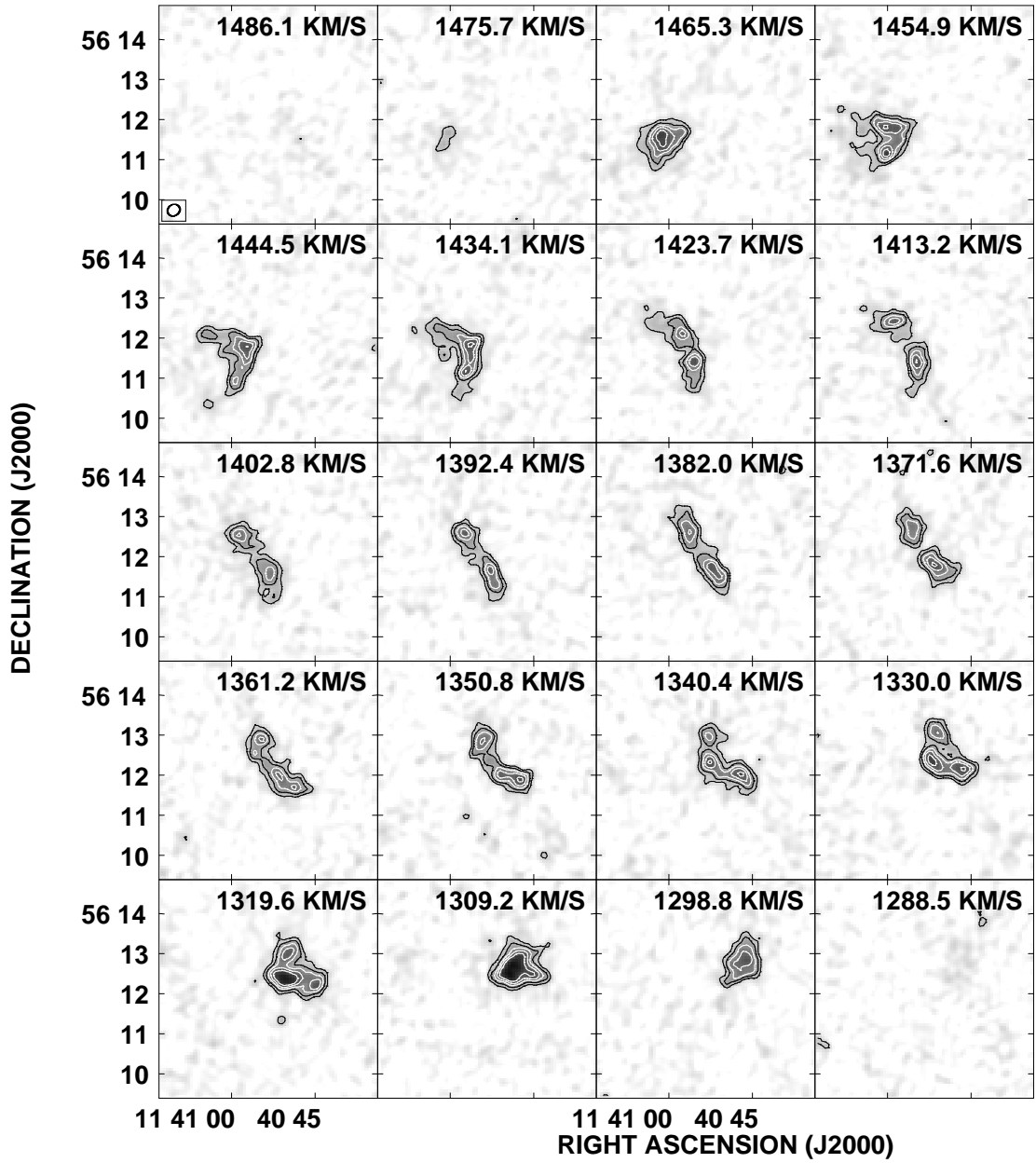}
\caption{As in Figure~\ref{fig:337_chmap}, but for NGC~3794.  The
  grayscale spans -0.02 to 15.67 ${\rm mJy \, beam^{-1}}$.  Every
  second channel is shown and the channel width is 5.2 \kms.}
\label{fig:3794_chmap}
\end{center}
\end{figure}

\begin{figure}
\begin{center}
\plotone{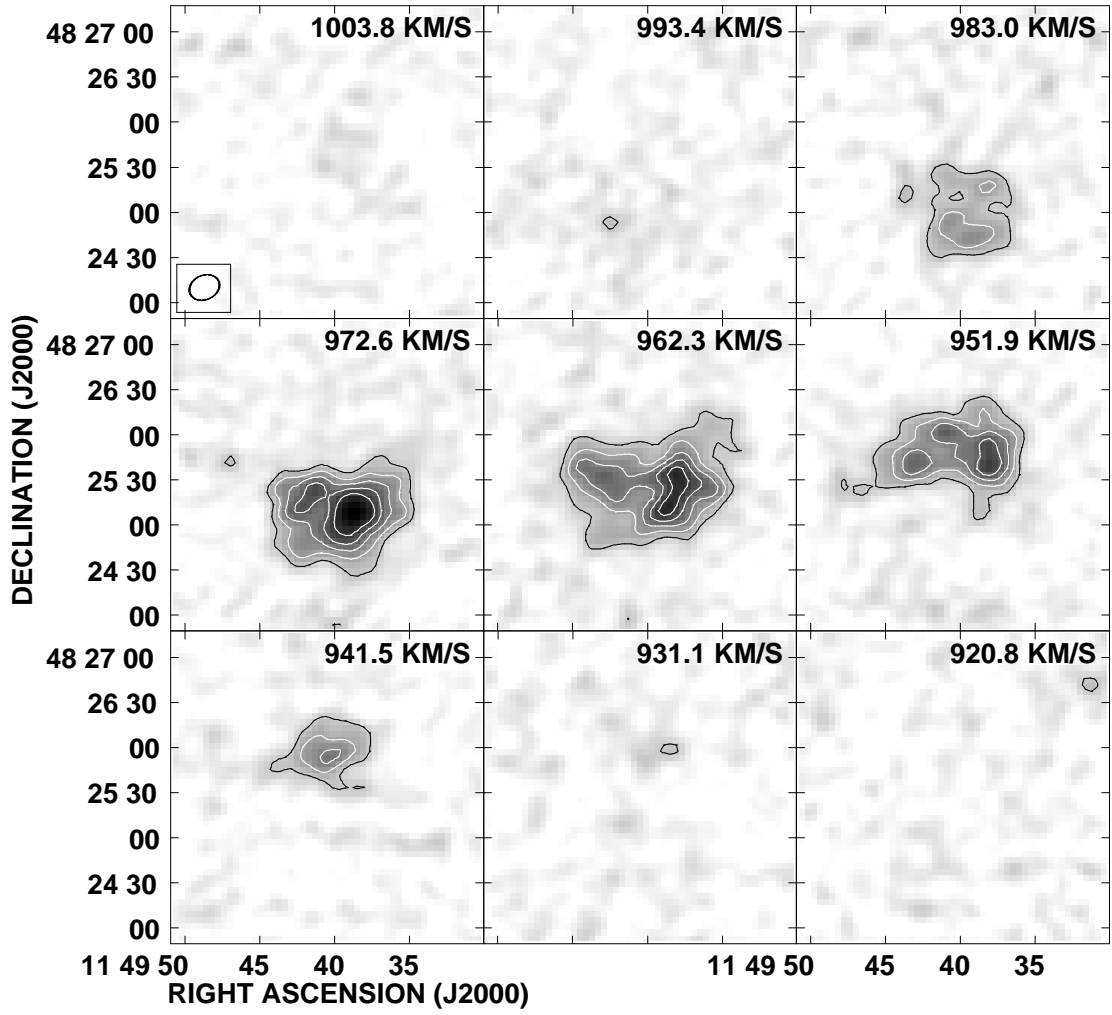}
\caption{As in Figure~\ref{fig:337_chmap}, but for NGC~3906.  The
  grayscale spans -0.02 to 11.48 ${\rm mJy \, beam^{-1}}$.  Every
  channel is shown and the channel width is 10.4 \kms.}
\label{fig:3906_chmap}
\end{center}
\end{figure}

\begin{figure}
\begin{center}
\plotone{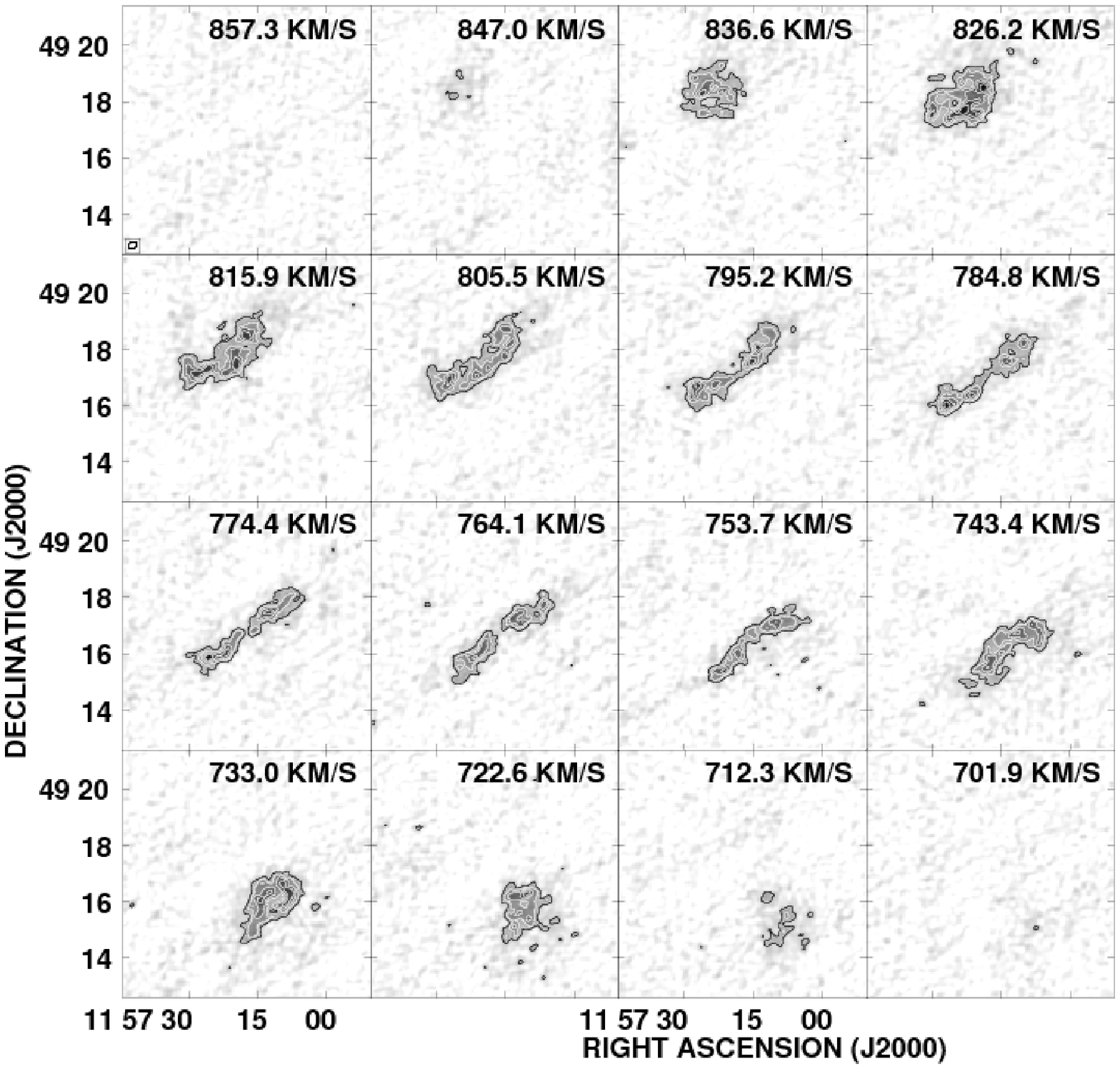}
\caption{As in Figure~\ref{fig:337_chmap}, but for UGC~6930.  The
  grayscale spans -0.02 to 15.06 ${\rm mJy \, beam^{-1}}$.  Every
  second channel is shown and the channel width is 5.2 \kms.}
\label{fig:6930_chmap}
\end{center}
\end{figure}

\begin{figure}
\begin{center}
\plotone{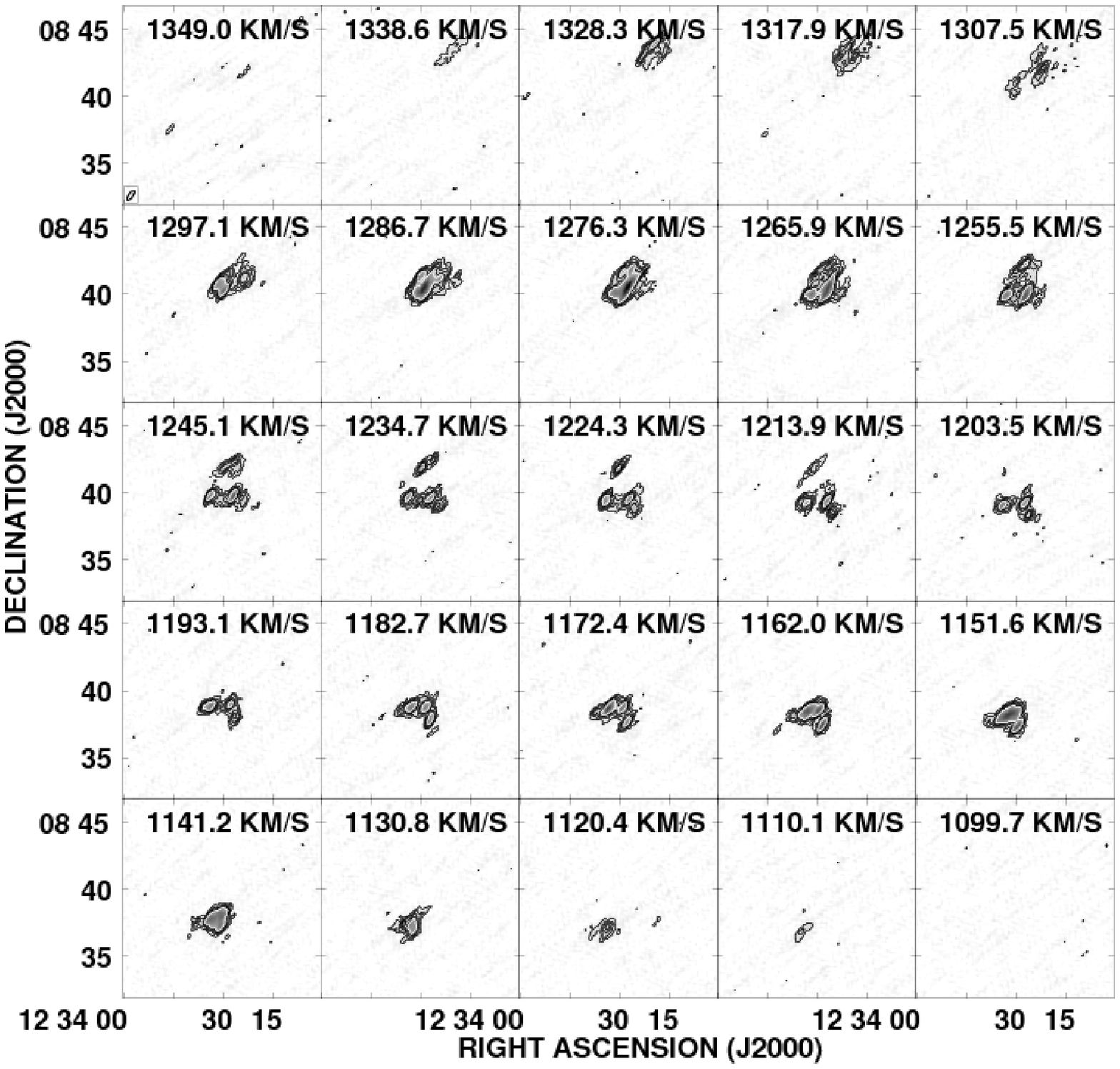}
\caption{As in Figure~\ref{fig:337_chmap}, but for NGC~4519.  The
  grayscale spans -0.02 to 41.37 ${\rm mJy \, beam^{-1}}$.  Every
  second channel is shown and the channel width is 5.2 \kms.}
\label{fig:4519_chmap}
\end{center}
\end{figure}

\begin{figure}
\begin{center}
\plotone{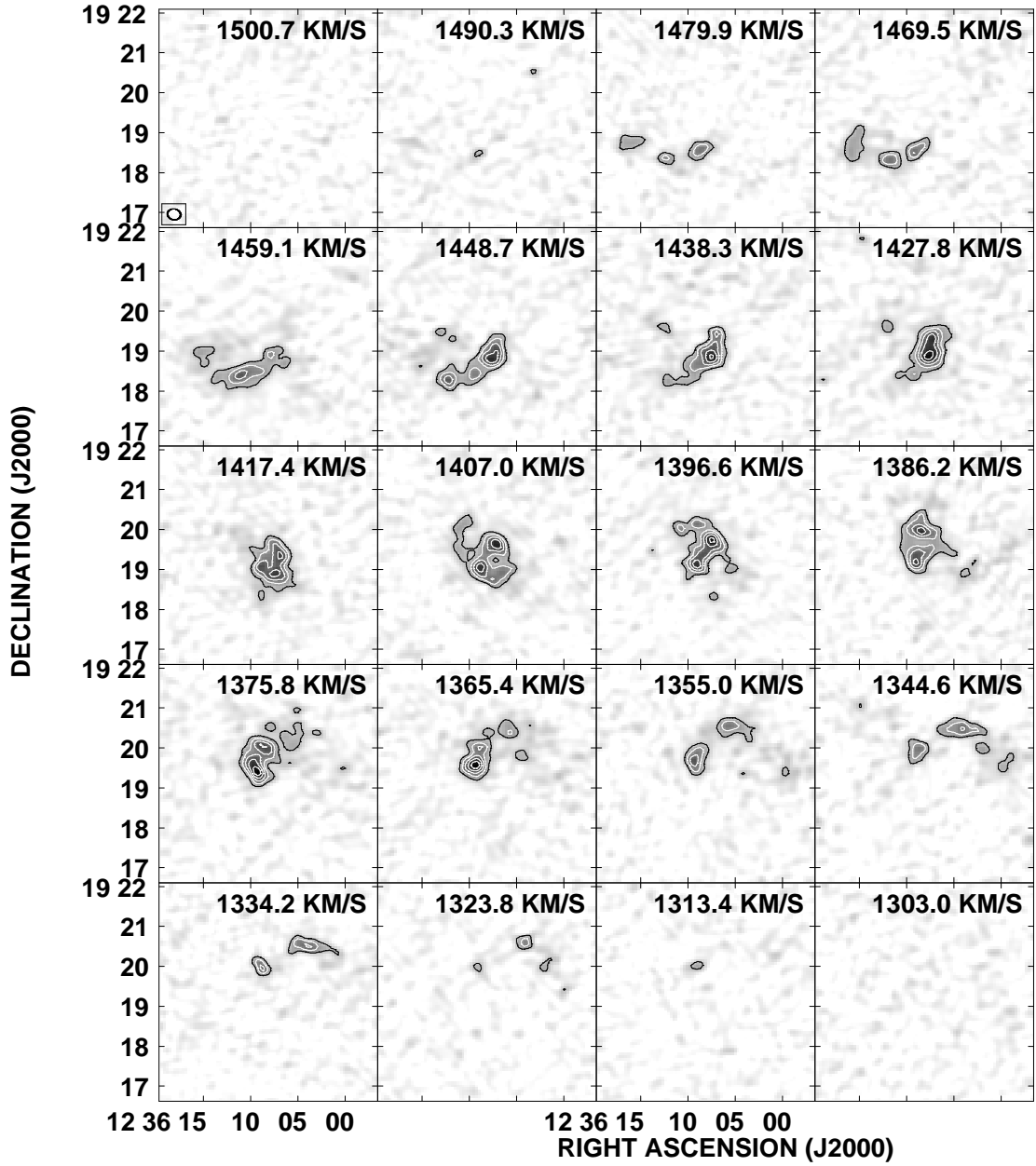}
\caption{As in Figure~\ref{fig:337_chmap}, but for NGC~4561.  The
  grayscale spans -0.02 to 18.81 ${\rm mJy \, beam^{-1}}$.  Every
  second channel is shown and the channel width is 5.2 \kms.}
\label{fig:4561_chmap}
\end{center}
\end{figure}

\begin{figure}
\begin{center}
\plotone{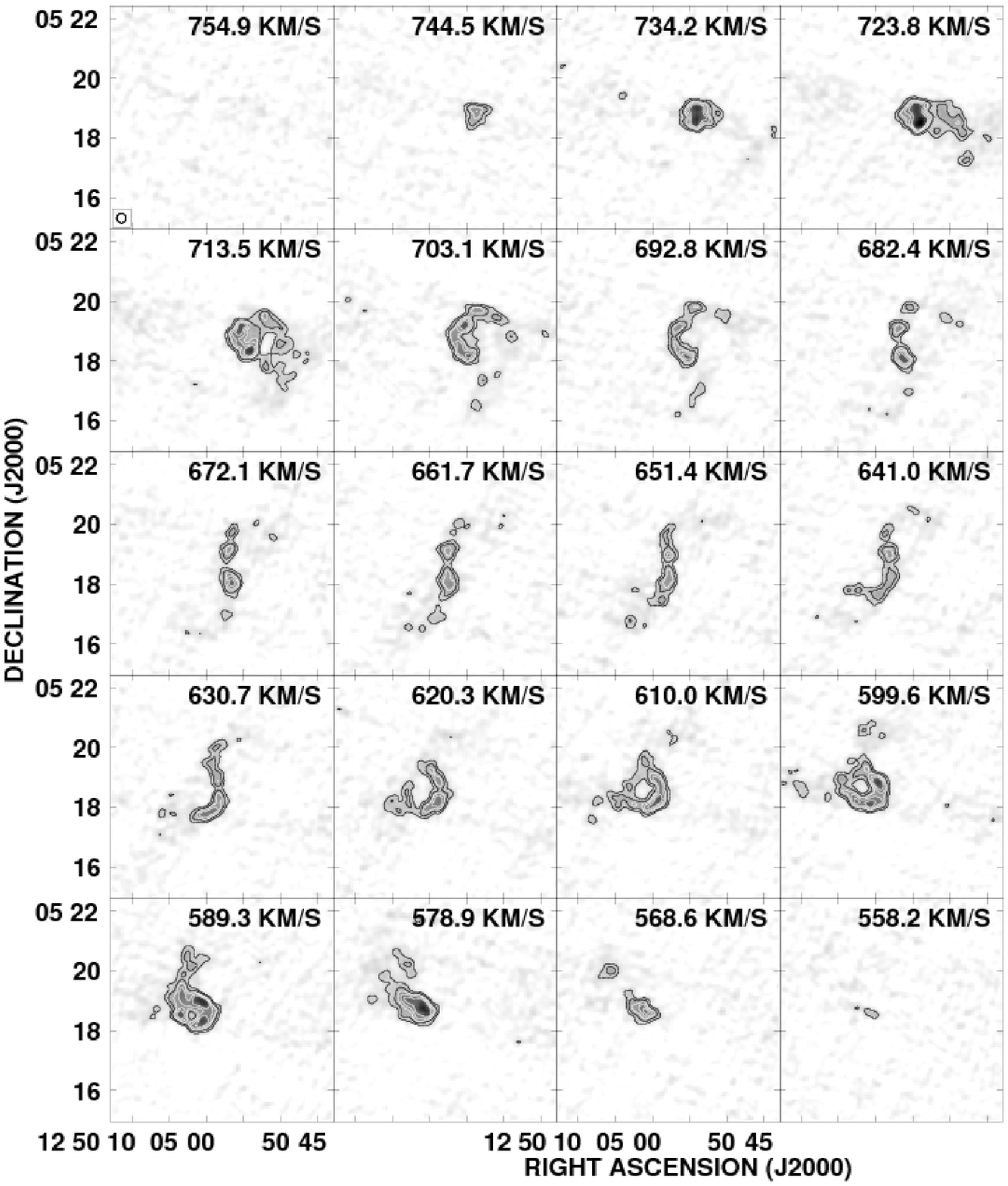}
\caption{As in Figure~\ref{fig:337_chmap}, but for NGC~4713.  The
  grayscale spans -0.02 to 21.68 ${\rm mJy \, beam^{-1}}$.  Every
  second channel is shown and the channel width is 5.2 \kms.}
\label{fig:4713_chmap}
\end{center}
\end{figure}

\begin{figure}
\begin{center}
\plotone{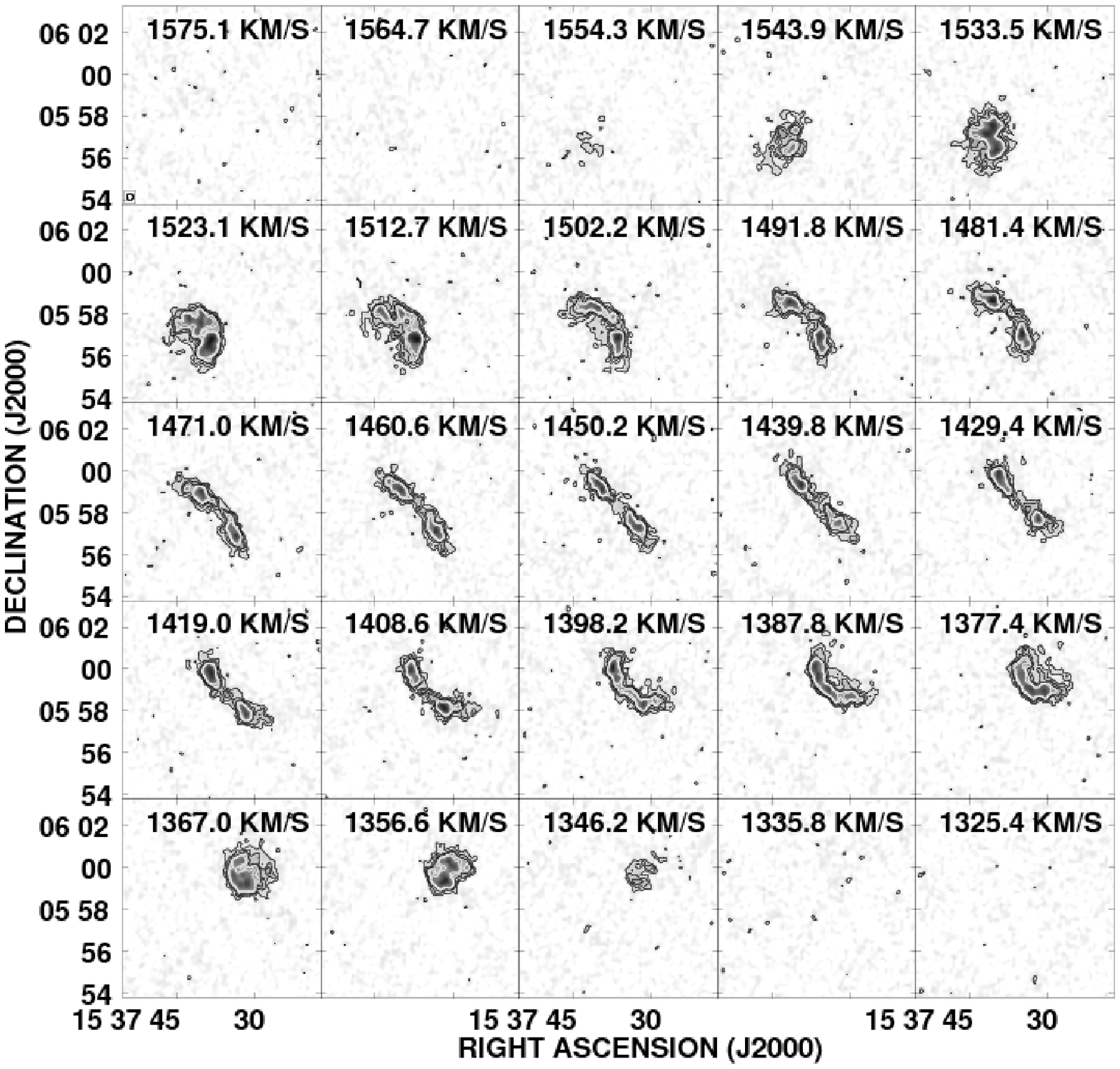}
\caption{As in Figure~\ref{fig:337_chmap}, but for NGC~5964.  The
  grayscale spans -0.02 to 16.25 ${\rm mJy \, beam^{-1}}$.  Every
  second channel is shown and the channel width is 5.2 \kms.}
\label{fig:5964_chmap}
\end{center}
\end{figure}

\begin{figure}
\begin{center}
\plotone{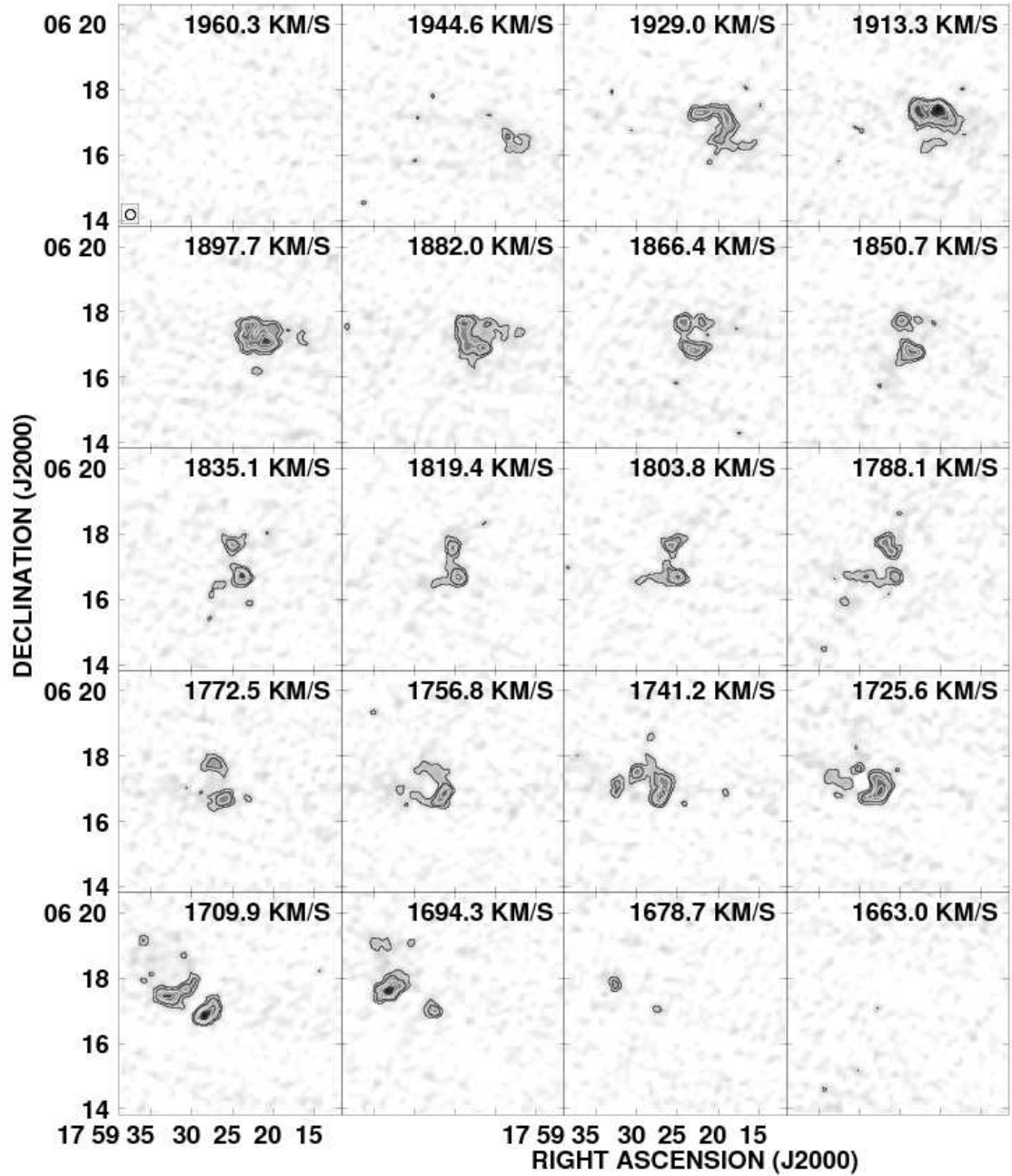}
\caption{As in Figure~\ref{fig:337_chmap}, but for NGC~6509.  The
  grayscale spans -0.02 to 17.43 ${\rm mJy \, beam^{-1}}$.  Every
  third channel is shown and the channel width is 5.2 \kms.  The
  ``hole'' visible especially in the 1725.6 \kms\ channel is due to
  \ion{H}{1} absorption of background continuum emission from a radio
  lobe of 4C~06.63. See Section~\ref{sec:morph} for details.}
\label{fig:6509_chmap}
\end{center}
\end{figure}

\begin{figure}
\begin{center}
\plotone{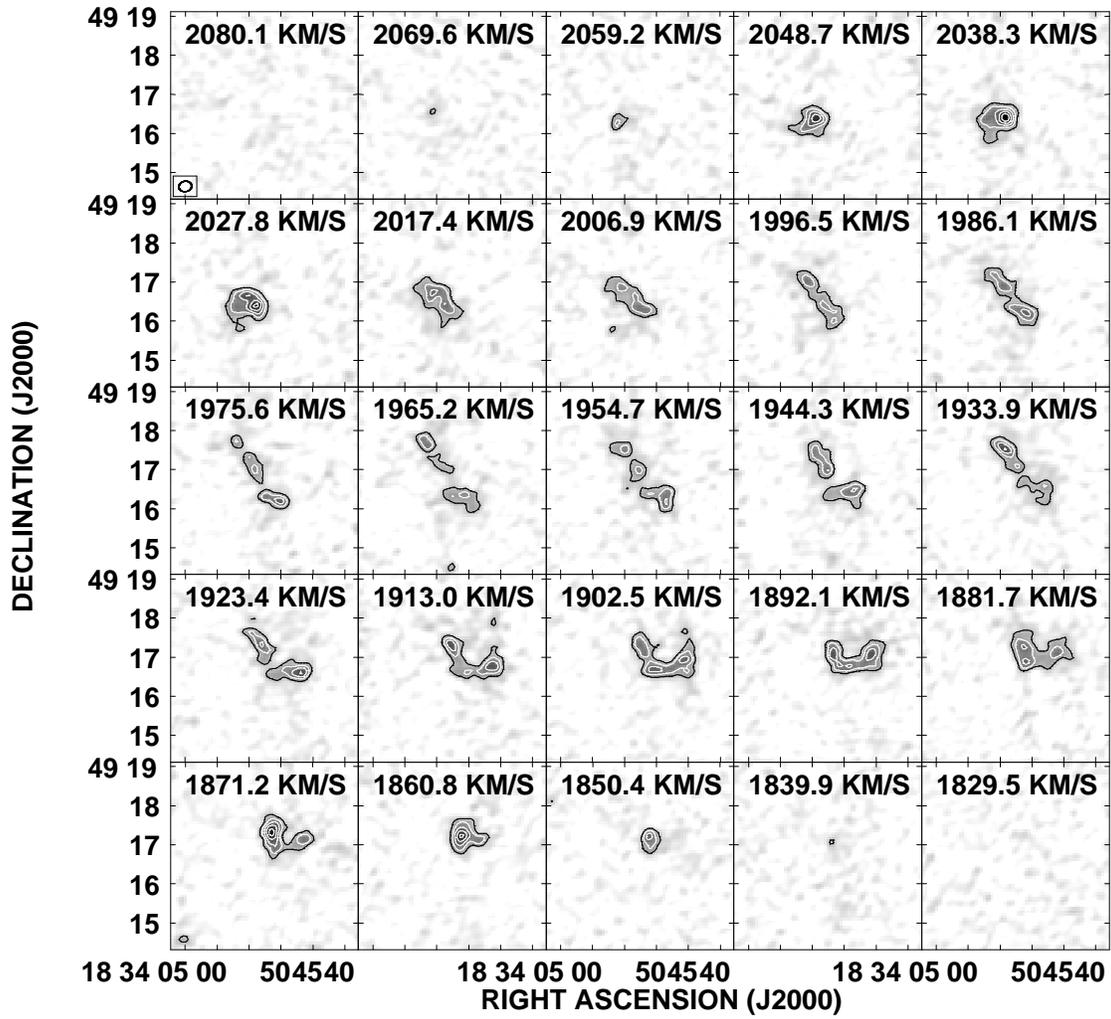}
\caption{As in Figure~\ref{fig:337_chmap}, but for IC~1291.  The
  grayscale spans -0.02 to 14.40 ${\rm mJy \, beam^{-1}}$.  Every
  second channel is shown and the channel width is 5.2 \kms.}
\label{fig:1291_chmap}
\end{center}
\end{figure}

Figures~\ref{fig:337_chmap} - \ref{fig:1291_chmap} show channel maps
for the galaxies, with units of ${\rm mJy \, beam^{-1}}$.  We created
the maps from the naturally-weighted data cubes so low surface
brightness features are more visible.  The cubes have neither been
blanked nor primary beam corrected, so the original noise properties
of the image are apparent.  We use a linear scaling, with the minimum
and maximum surface brightness noted in the caption.  Note that the
ESO~555-G027 and ESO~501-G023 data cubes have residual imaging
artifacts to the north and south of the galaxy that remained after
cleaning.

\subsection{Integrated Line Profiles, Integrated Fluxes, and
  Integrated Intensity Maps}
\label{sec:lineprof}

Figure~\ref{fig:int_line_prof} shows the integrated \ion{H}{1} line
profiles of the galaxies, which were created by summing the emission
in each channel of the naturally-weighted, blanked, and primary beam
corrected data cubes.  The majority of our sample shows the
double-horned profile that indicates a disk with a flat rotation
curve, which generally agrees with our rotation curve analysis in
Section~\ref{sec:rot_curves}.  From these line profiles, we calculated
$W_{20}$, the width of the line at 20\% of the peak flux density, and
corrected each value for spectral resolution.  We make no correction
for turbulent broadening, but note that the typical velocity
dispersion of \ion{H}{1} gas is about $10 \, {\rm km \, s^{-1}}$.  We
assign an uncertainty in $W_{20}$ equal to the channel width, which is
$5\, {\rm km \, s^{-1}}$ for all objects except UGC~6446 and NGC~3906,
where the uncertainty is $3 \, {\rm km \, s^{-1}}$ and $10\, {\rm km
\, s^{-1}}$, respectively.

\begin{figure}
\begin{center}
\plotone{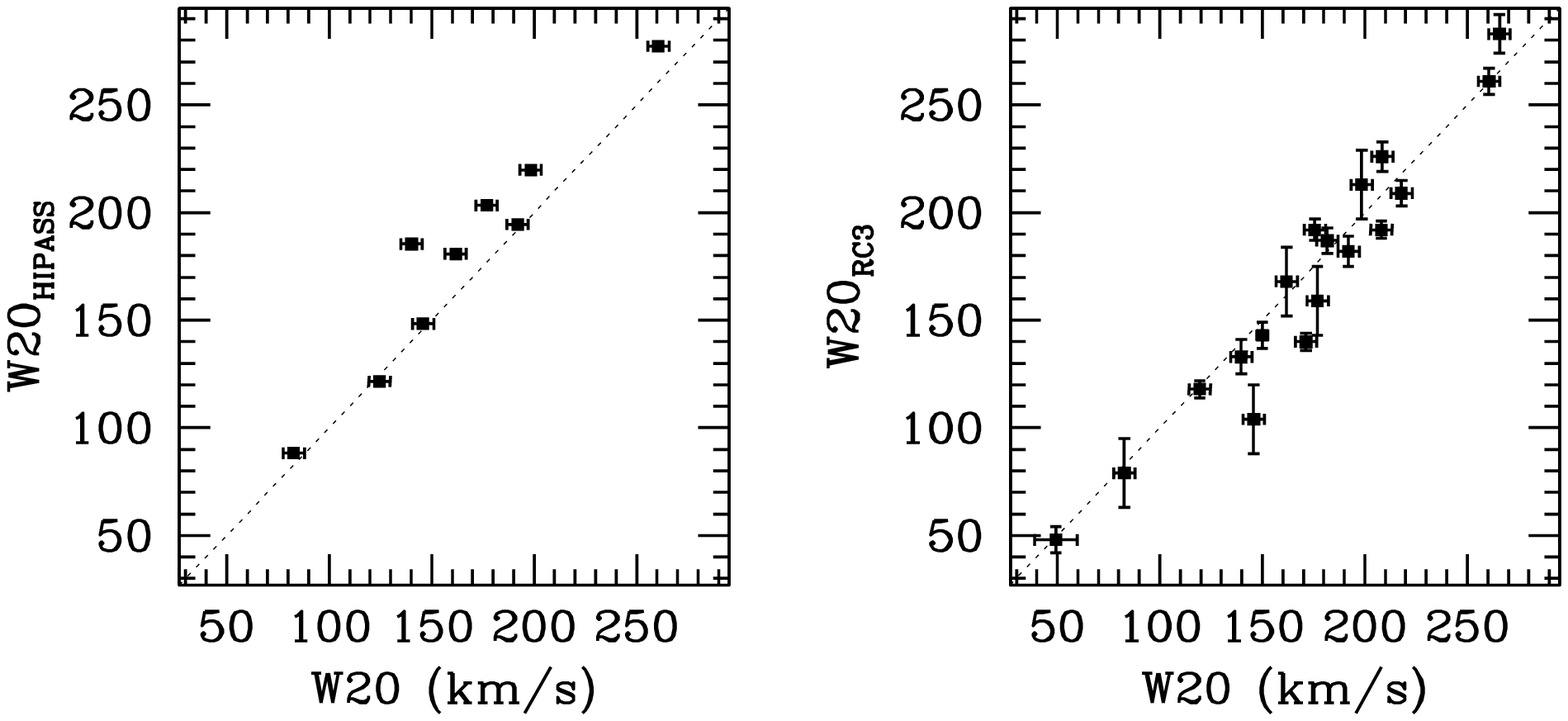}
\end{center}
\caption{Comparison of our $W_{20}$ values to single-dish values from
HIPASS ({\it left}) and RC3 ({\it right}).  The dashed line shows
equality.}
\label{fig:W_20_comp}
\end{figure}

\begin{figure}
\begin{center}
\epsscale{0.65}
\plotone{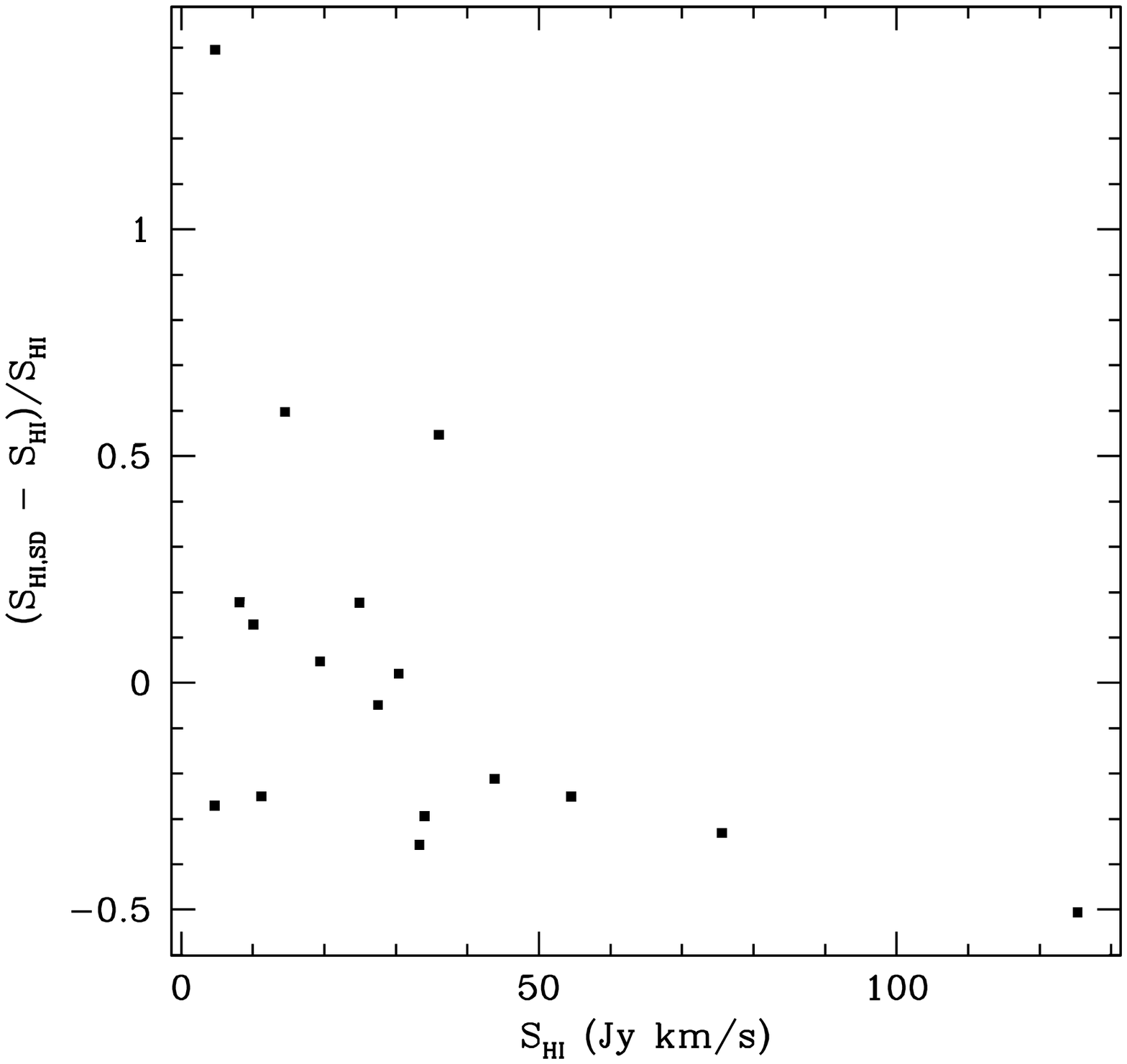}
\end{center}
\caption{Comparison of our total \ion{H}{1} fluxes ($S_{\rm HI}$) to
single-dish measurements ($S_{\rm HI, \, SD}$) from \citet{meyer04} or
\citet{springob05}.  We use the \citet{meyer04} value if both are
available.}
\label{fig:S_HI_comp}
\end{figure}

Figure~\ref{fig:W_20_comp} compares our $W_{20}$ values to single-dish
values from the literature.  The left panel compares to values from
the \ion{H}{1} Parkes All Sky Survey (HIPASS), which used the 64-m
Parkes telescope \citep{meyer04}.  The right panel compares to values
from the Third Reference Catalogue of Bright Galaxies
\citep[RC3;][]{RC3}.  The HIPASS values are generally larger than
ours, whereas the RC3 values agree better, with a scatter of about
8\%.  It is unlikely that we have resolved out flux at large scales,
which might be associated with larger velocities, because we see no
systematic offset between our total \ion{H}{1} fluxes and single-dish
values (see Figure~\ref{fig:S_HI_comp}).  Furthermore, the extent of
each galaxy is less than the largest angular structure for which the C
configuration is sensitive ($\sim 15 \arcmin$; although note that this
angular sensitivity is only strictly appropriate for full-synthesis
data).  Finally, the offset between the HIPASS $W_{20}$ values and our
values is not larger for objects with rotation curves that continue
rising to the edge of the \ion{H}{1} disk (see
Section~\ref{sec:rot_curves}).  We accept our $W_{20}$ values as being
in reasonable agreement with values quoted in the literature, with no
systematic offset relative to values quoted in the RC3.

We measured the total \ion{H}{1} flux by summing the line profiles
over velocity.  The main contributors to the integrated flux
uncertainty are flux calibration (5\% effect) and aliasing (up to an
11\% effect).  For a few objects, we tested whether excluding
single-channel flux peaks affects the total flux.  To do so, we
convolved the mask used in blanking to a larger velocity resolution.
The dominant effect was that the noise was lower, and thus the flux
cut for blanking was decreased and the integrated flux was larger,
10\% larger in the case of a bright galaxy and 50\% larger in the case
of a faint galaxy.  The effect of different flux cuts is not typically
included in error estimates, so we only assign the quadrature sum of
the flux calibration and aliasing errors (12\%) as the generic
uncertainty associated with the integrated fluxes.

Figure~\ref{fig:S_HI_comp} provides a comparison of our integrated
fluxes to values from HIPASS \citep{meyer04} or \citet{springob05},
who compiled measurements from a variety of large single-dish
telescopes.  Our values are in agreement with the literature values,
but with a scatter of about 30\%.  This is typical for comparisons
between data from different instruments \citep[e.g.,][]{walter08}.  We also
compute the total \ion{H}{1} masses, assuming the gas is optically
thin, using:
\begin{equation}
M_{HI} [M_{\odot}] = 2.36 \times 10^{5} D^{2} \times
\sum_{i} S_{i} \, \Delta v,
\end{equation}
where $D$ is the distance in Mpc, $S_{i}$ is total flux density in a
channel in Jy, $\Delta v$ is the velocity width of a channel in ${\rm
km \, s^{-1}}$, and the sum is over all channels that show line
emission.  We made no correction for \ion{H}{1} self absorption.  Note
that the \ion{H}{1} mass in NGC~6509 is a lower limit because a region
is in absorption rather than emission, due to the strong radio lobe of
a background galaxy; see Section~\ref{sec:morph} for more details.

\begin{figure}
\begin{center}
\epsscale{0.65}
\plotone{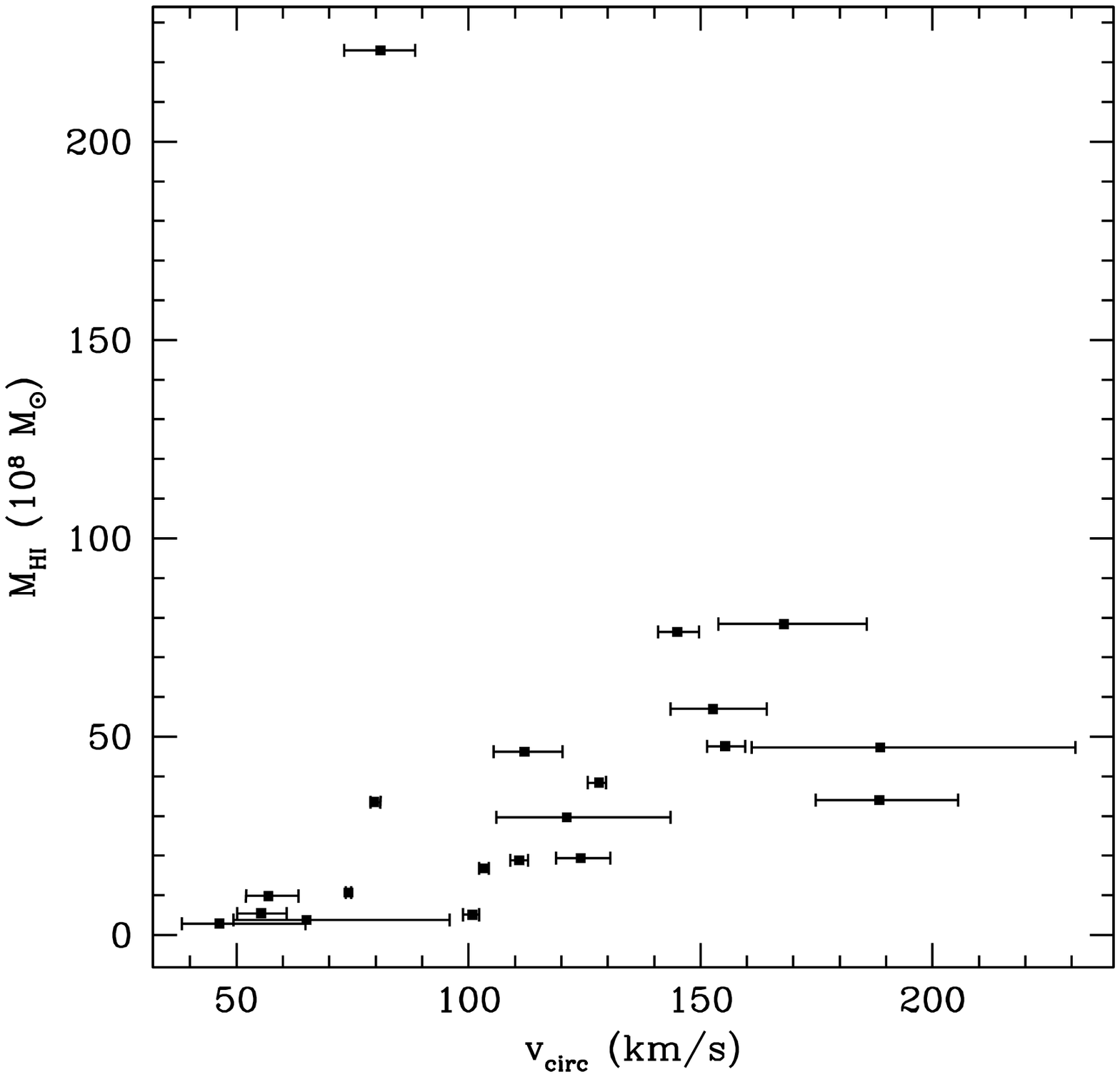}
\end{center}
\caption{\ion{H}{1} mass as a function of circular velocity, derived
  from rotation curve fitting in Section~\ref{sec:rot_curves}.  The
  outlier is NGC~2805, for which the distance may be overestimated.}
\label{fig:M_HI_vs_vcirc}
\end{figure}

Table~\ref{tab:flux} contains our integrated \ion{H}{1} flux,
\ion{H}{1} mass, and $W_{20}$ values, as well as comparison integrated
flux and $W_{20}$ values from single-dish measurements in the
literature.  Figure~\ref{fig:M_HI_vs_vcirc} shows our derived
\ion{H}{1} masses versus circular velocity, which we calculate in
Section~\ref{sec:rot_curves}.  As expected, the \ion{H}{1} mass
generally increases with increasing circular velocity.  The outlier is
NGC~2805, which has a high but reasonable integrated flux value.  For
NGC~2805 to agree with the main distribution in
Figure~\ref{fig:M_HI_vs_vcirc}, the distance would have to be about
$10 \, {\rm Mpc}$ rather than the Table~\ref{tab:sample} value of
$28.0 \, {\rm Mpc}$, which is derived from the Tully-Fisher relation.
However, the recession velocity is consistent with a distance of $28.0
\, {\rm Mpc}$ and the smaller distance would have other implications.
For example, the nuclear star cluster in this galaxy would be somewhat
small and faint compared to the distribution of nuclear star cluster
sizes and magnitudes in \citet{boeker04}.

Figure~\ref{fig:m0} shows the integrated \ion{H}{1} intensity map for
each galaxy, derived from the naturally-weighted, blanked, and primary
beam corrected data cube, with units of ${\rm Jy \, beam^{-1} \, km \,
s^{-1}}$.

\subsection{Rotation Curves}
\label{sec:rot_curves}

In our rotation curve analysis, we aim to derive global kinematic
properties rather than characterize the detailed rotation curve shape.
We therefore simply used the first moment of the data cube as our
velocity field, which provides the intensity-weighted mean velocity at
each spatial pixel.  \citet{deblok08} found that the use of the first
moment can bias the rotation curve fit if the velocity profile is
asymmetric.  They instead favor the use of a third-order Gauss-Hermite
polynomial to fit the line profile and derive the velocity field.
Nonetheless, using the first moment should be sufficient for our
goals.  We primarily used the first-moment map created with the
MOMENTS task within the Groningen Image Processing SYstem
\citep[GIPSY;][]{vanderhulst92}.  We blanked pixels with values less
than three times the image noise and excluded single-channel wide
peaks.  This blanking procedure is different from our primary method,
described in Section~\ref{sec:dp1}.  The method described in
Section~\ref{sec:dp1} retains low-level emission, but may also allow
residual beam pattern from side lobes to contribute to the moments.
Excluding low-level emission should have little effect on the rotation
curve analysis since it has little weight in the first moment.
However, UGC~1862 is so faint that we used the first-moment map
created from the data cube that was blanked with our primary method.

We used first-moment maps created from the robustly-weighted data
cubes, which provide higher spatial resolution, for 12 objects.  The
remaining eight objects (ESO~544-G030, UGC~1862, ESO~418-G008,
ESO~555-G027, ESO~501-G023, UGC~6446, NGC~3906, and IC~1291) are
fainter, so we created moment maps from the naturally-weighted data
cubes, which have better sensitivity.  The first-moment maps were not
corrected for primary beam attenuation.

We carried out velocity field modeling for our galaxies on the
first-moment maps with the ROTCUR task within GIPSY.  ROTCUR performs
a least-squares fit, comparing the data to a model composed of a
series of tilted rings described by
\begin{equation}
\label{eqn:tilted_ring}
V_{\rm los}(x,y) = V_{\rm sys} + V_{\rm rot} \, {\rm sin}(i) \, {\rm cos}(\theta),
\end{equation}
where $V_{\rm los}(x,y)$ is the line of sight velocity at
position (x,y) in the plane of the sky, $V_{\rm sys}$ is the systemic
velocity, $V_{\rm rot}$ is the rotation velocity, $i$ is the
inclination, and $\theta$ is the angle between the major axis and the
position of interest in the plane of the galaxy.

As initial estimates for the center, position angle of the major axis
(PA), and inclination, we used values derived from ellipse fits of
{\it Spitzer Space Telescope} IRAC Channel 1 ($3.6~\mu{\rm m}$) data,
described in L.\ Watson et al.\ (2011, in preparation).  In four cases
(NGC~2805, ESO~501-G023, UGC~6446, and IC~1291), the IRAC fits did not
converge so we used RC3 parameters as initial estimates.  Our initial
estimate for the systemic velocity was obtained from NED\footnote{The
NASA/IPAC Extragalactic Database (NED) is operated by the Jet
Propulsion Laboratory, California Institute of Technology, under
contract with the National Aeronautics and Space Administration.}.

We followed a typical velocity field modeling procedure, outlined
below \citep[see also][]{begeman89,haan08}.  We solved for the
systemic velocity, center, PA, inclination, and rotation velocity by
successively fitting for a single parameter as a function of radius
while most of the other parameters were held fixed.  Except for the
rotation velocity, we calculated the weighted mean of the values at
all radii, excluding the first and last annuli.  The errors used in
the weighted mean are the formal least squares errors provided by
ROTCUR.  In subsequent runs, we held the parameter fixed as a function
of radius at the value of the weighted mean.  In doing so, we ignored
real variations caused by warps and bars, but we accepted this since
we are only interested in global kinematics. In all fits, we forced
the radial velocity to zero.  We excluded points within $20 \degr$ of
the minor axis and used a $|{\rm cos}(\theta)|$ weighting so points
near the major axis were weighted more than points near the minor
axis, where beam smearing becomes severe.  We fit both the approaching
and receding sides of the galaxy simultaneously and sampled the
rotation curve every $(B_{\rm maj} \, B_{\rm min})^{1/2}$, beginning
at $(B_{\rm maj} \, B_{\rm min})^{1/2}/2$, where $B_{\rm maj}$ and
$B_{\rm min}$ are the beam major axis and minor axis FWHM in
arcseconds.

We verified the agreement between each data cube and the derived
velocity field model using the INSPECTOR task within GIPSY.  We also
created two-dimensional model velocity fields described completely by
the derived parameters, using VELFI in GIPSY, and subtracted the model
from the data to form a residual velocity map.  We examined the
residual map for signs of parameter mis-estimation \citep[for
examples, see][]{vanderkruit78} and also noted the rms of the residual
distribution.

In the above procedure, we have allowed the parameters to veer away
from the initial estimates derived from the IRAC ellipse fits.
Therefore, we did one final check: we calculated the rotation
velocities again, but used the IRAC center rather than the fitted
value.  We again created a model two-dimensional velocity field and
residual velocity map.  We declared that the IRAC center fit the data
as well as the fitted value if the rms of the residual map was the
same (within one tenth the channel width) or smaller than the original
rms, and if the residual velocity map had not developed any symptoms
of data-model mismatch.  We carried out this same procedure for the PA
and inclination.  If an IRAC parameter fit the data as well as the
value determined in ROTCUR, we chose the IRAC parameter, as it was
derived from independent data and also fit the kinematics.  We only
chose an RC3 parameter over a ROTCUR parameter if it fit the data
significantly better.  We chose the IRAC center for eleven objects,
the IRAC PA for eight objects (primarily when the fitted value was
very similar to the IRAC value), the IRAC inclination in all sixteen
galaxies where it was available, and the RC3 inclination for one
object.  See Table~\ref{tab:rot_comp} for a comparison of the
parameters derived in the rotation curve analysis, the IRAC values,
and the RC3 values.  The final parameter values are given in
Table~\ref{tab:final_params}, where Column~10 lists which parameters
are derived from the IRAC analysis; all others are from the rotation
curve analysis.

Inclinations are difficult to derive from tilted ring models for
galaxies with $i < 40 \degr$ \citep{begeman89}.  Furthermore, the
rotation velocities derived at these inclinations are less accurate
and precise because the velocity dispersion of the gas becomes a
larger fraction of the line-of-sight velocity and small inclination
errors lead to large rotation velocity errors.  Ten of our objects
have final inclinations below $40 \degr$.  Fortunately, we have
additional inclination information from the IRAC ellipse fits and our
procedure led us to use the IRAC value whenever it was available.
This additional constraint allows us to derive accurate kinematics for
the low-inclination sources.  In addition, the uncertainty in the
inclination is reflected in the circular velocity uncertainty,
described further below.

\begin{figure}
\begin{center}$
\begin{array}{c}
\includegraphics[width=\textwidth]{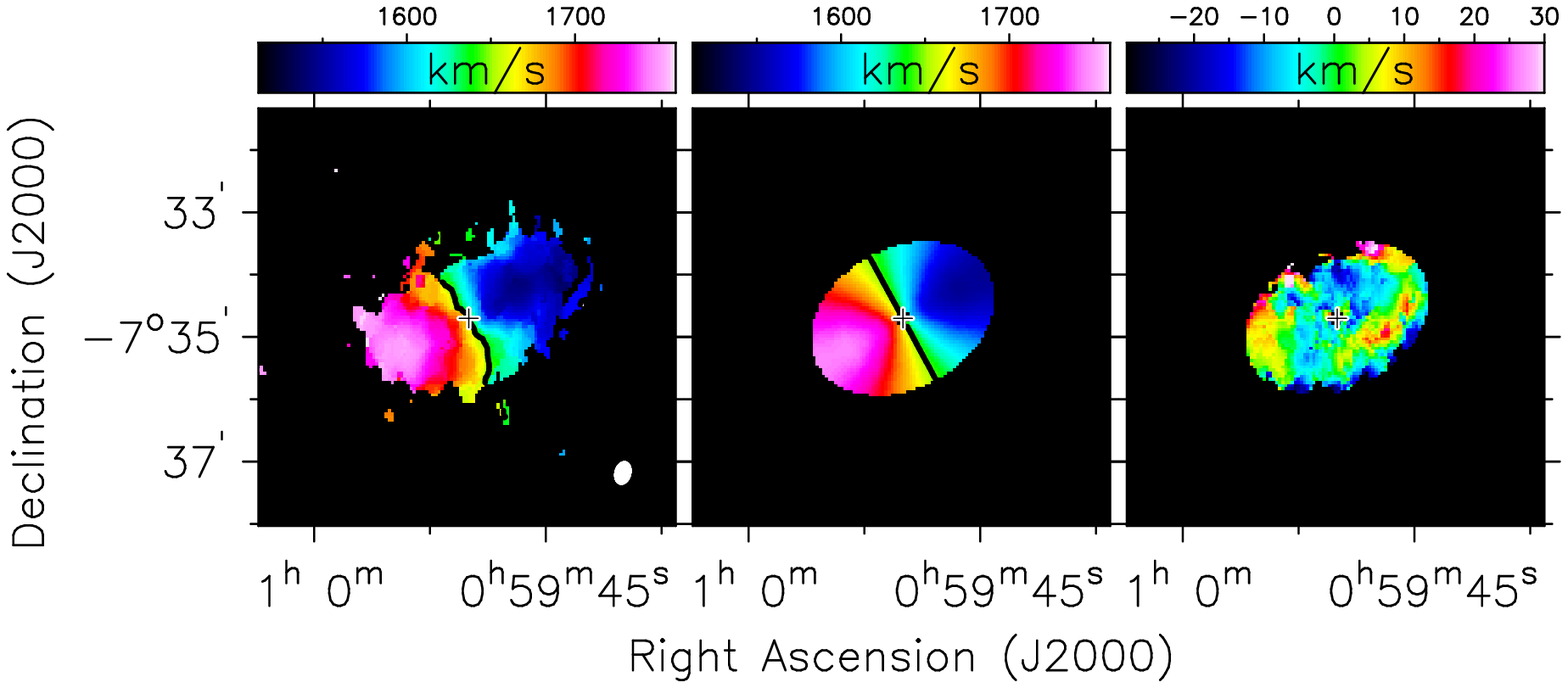} \\
\\
\includegraphics[width=\textwidth]{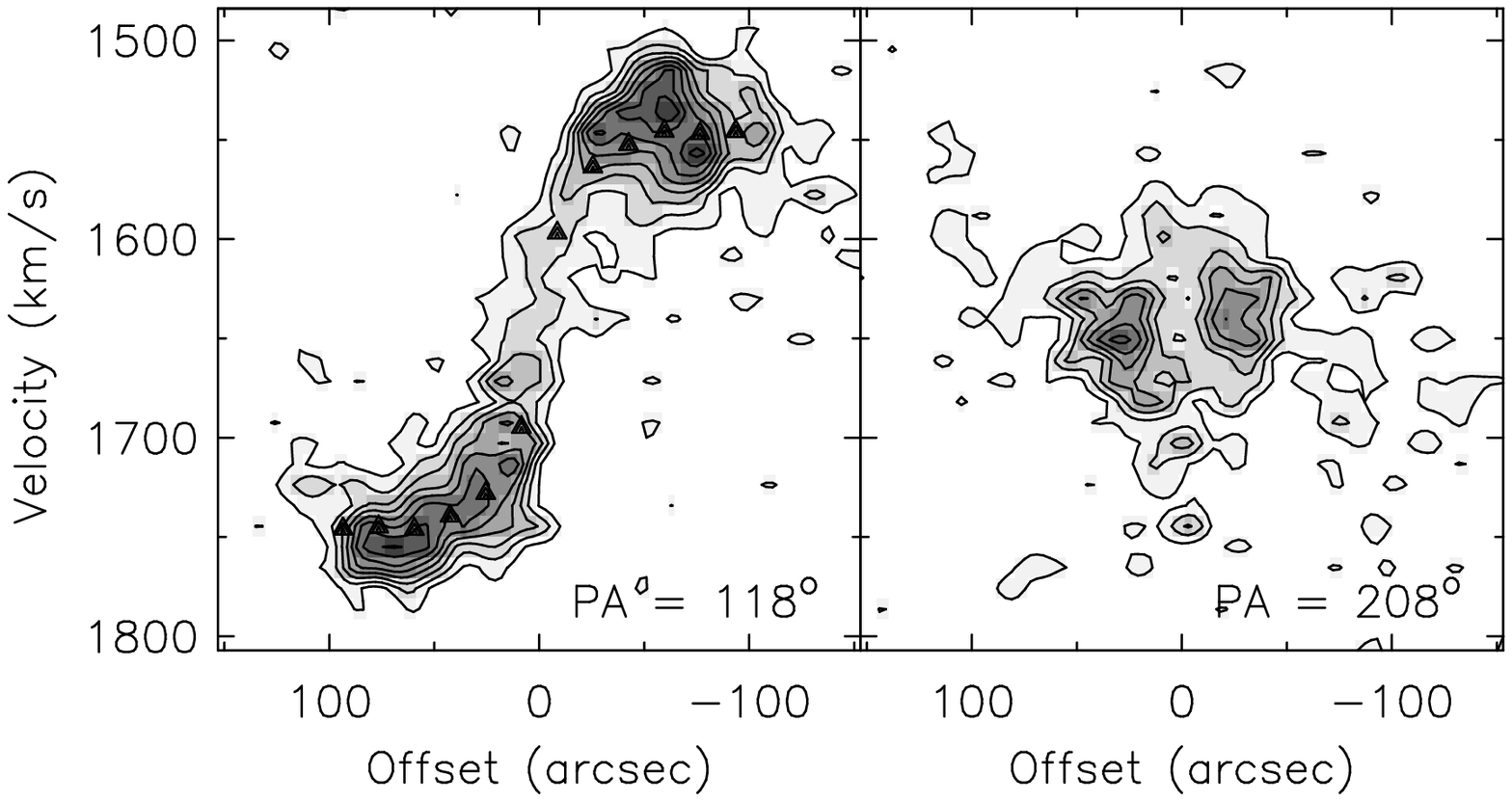} 
\end{array}$
\end{center}
\caption{Velocity field and position-velocity diagrams for NGC~0337.
{\it Top row}: the {\it left panel} shows the first-moment map,
created using the method described in Section~\ref{sec:rot_curves} and
covering the same area as the channel maps.  The beam size is shown in
the lower right corner.  The {\it middle panel} shows the tilted ring
model velocity field.  The {\it right panel} shows the residual
velocity map (the first-moment map minus the velocity field model) in
the region of overlap between the data and model.  The final
dynamical center is shown as a cross in all three panels and the {\it
thick black contour} shows the final systemic velocity in the left and
middle panels.  {\it Bottom row}: position velocity diagrams along the
major axis ({\it left panel}) and minor axis ({\it right panel}).
Contours begin at $2\sigma$ and end at the maximum surface brightness
along the major axis slice plus $2\sigma$, in steps of $2\sigma$,
where $\sigma$ is the image noise from Column~6 of Table~3 (2~${\rm
mJy \, beam^{-1}}$ to 20~${\rm mJy \, beam^{-1}}$ in steps of 2~${\rm
mJy \, beam^{-1}}$ in this case).  The projected tilted ring model fit
is overplotted as {\it solid triangles} in the left panel.}
\label{fig:ngc0337_vel}
\end{figure}

\begin{figure}
\begin{center}$
\begin{array}{c}
\includegraphics[width=\textwidth]{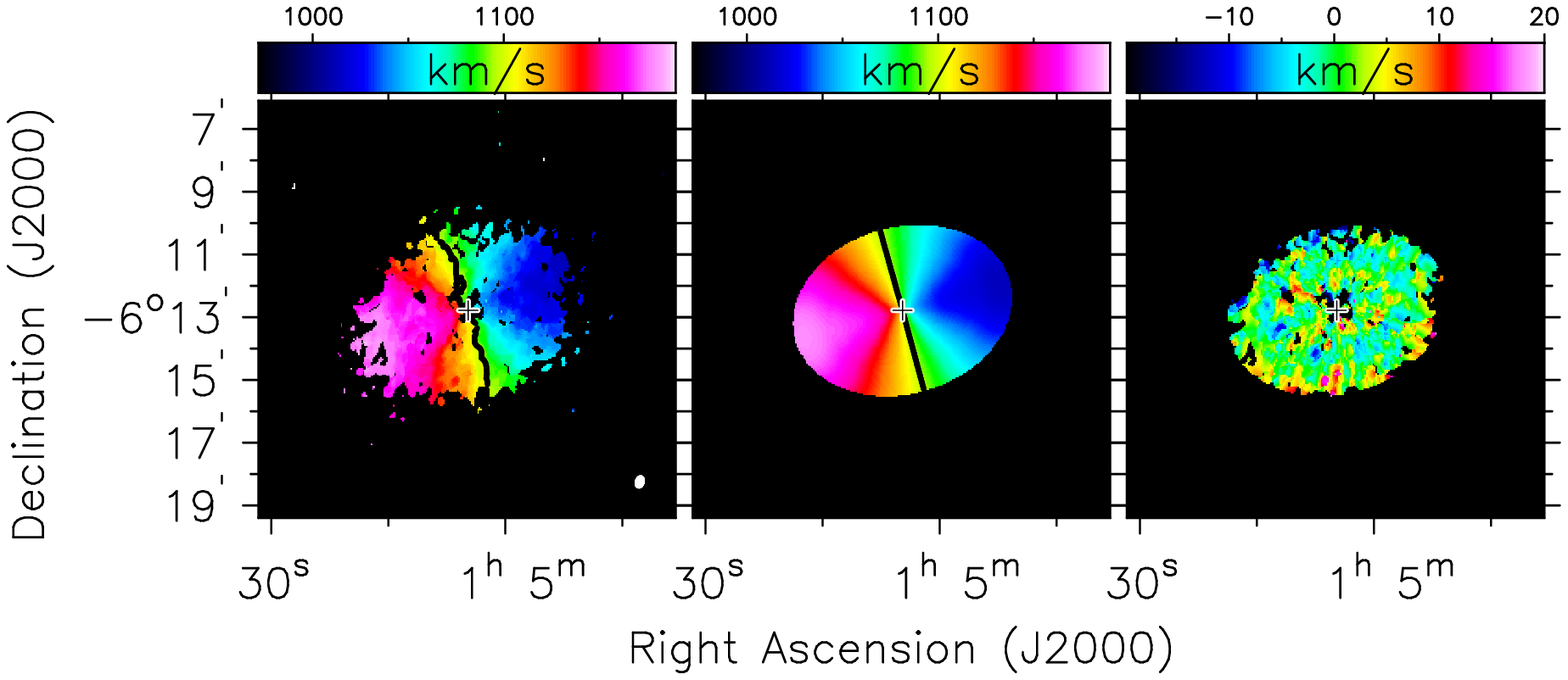} \\
\\
\includegraphics[width=\textwidth]{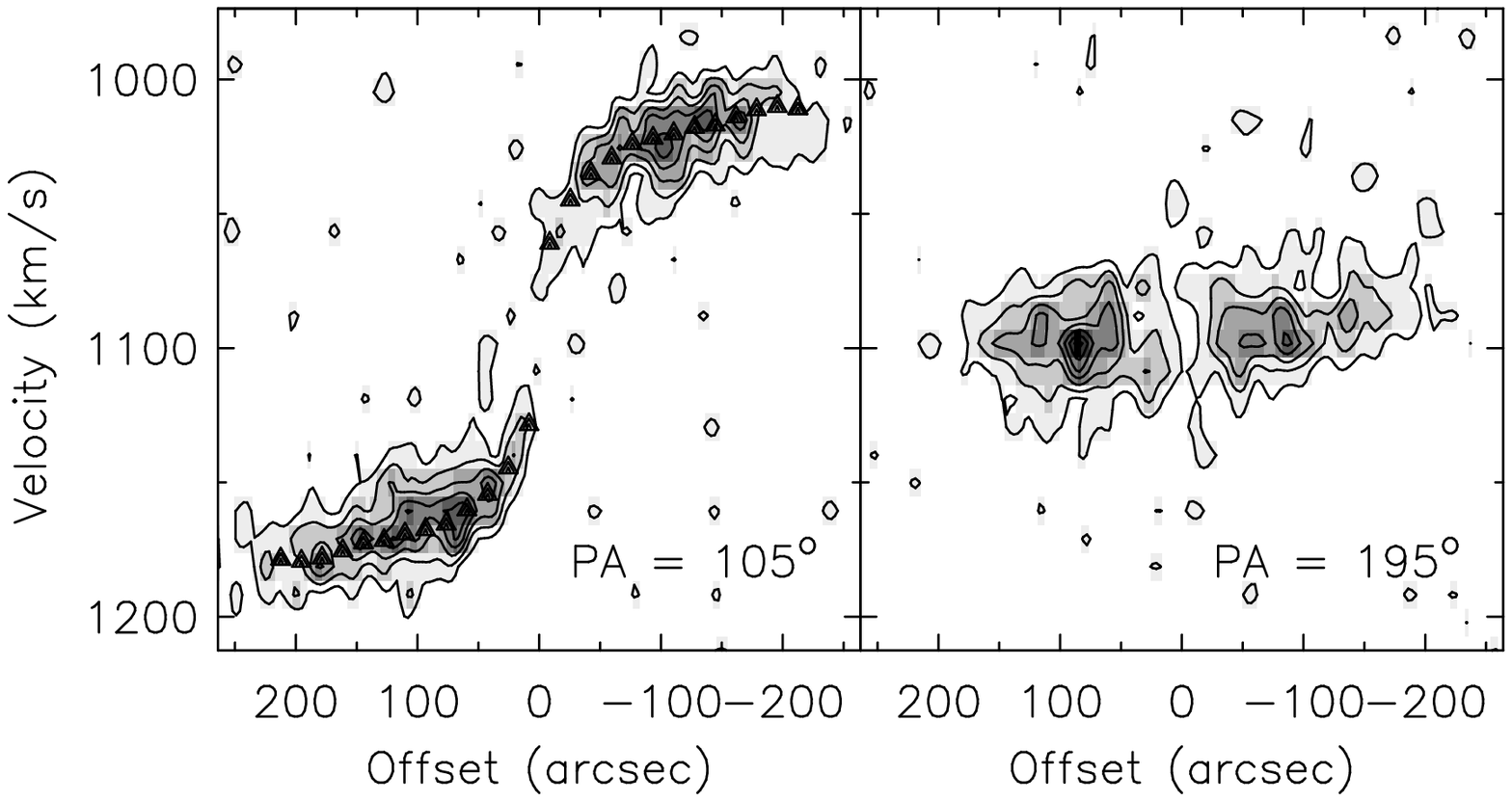} 
\end{array}$
\end{center}
\caption{As in Figure~\ref{fig:ngc0337_vel}, but for PGC~3853.  The
position-velocity diagram contours are from 2~${\rm mJy \, beam^{-1}}$
to 14~${\rm mJy \, beam^{-1}}$ in steps of 2~${\rm mJy \,
beam^{-1}}$.}
\label{fig:pgc3853_vel}
\end{figure}

\begin{figure}
\begin{center}$
\begin{array}{c}
\includegraphics[width=\textwidth]{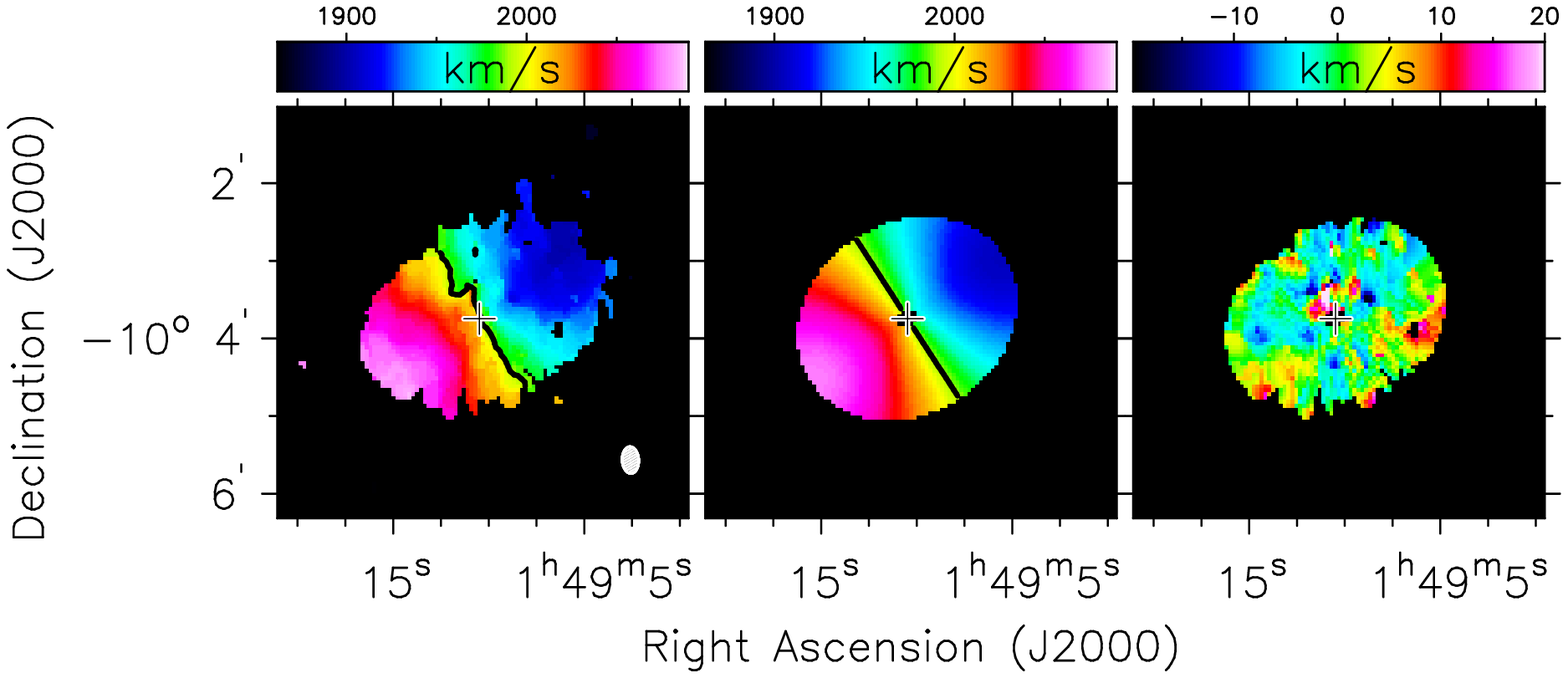} \\
\\
\includegraphics[width=\textwidth]{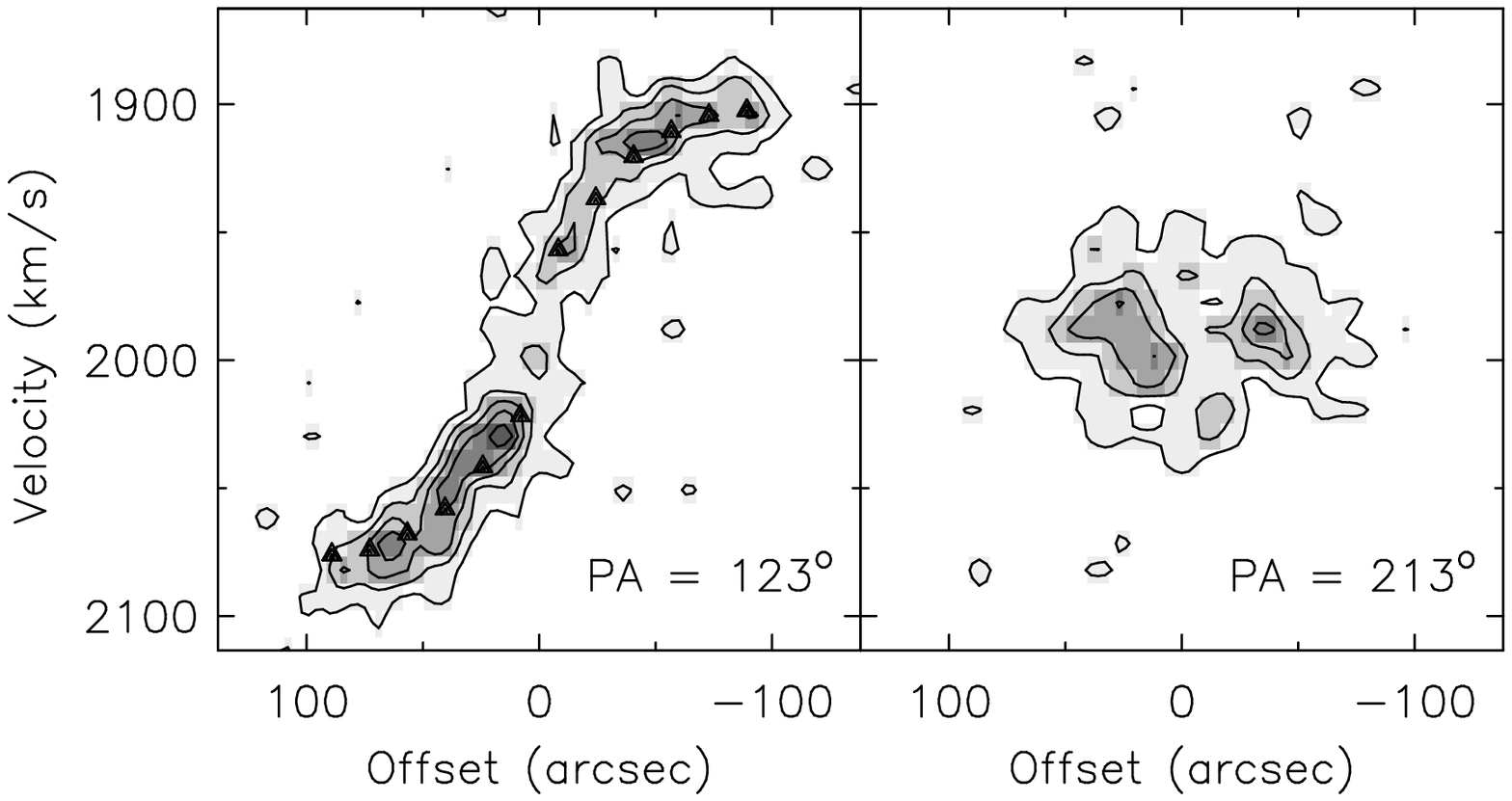} 
\end{array}$
\end{center}
\caption{As in Figure~\ref{fig:ngc0337_vel}, but for PGC~6667.  The
position-velocity diagram contours are from 2~${\rm mJy \, beam^{-1}}$
to 14~${\rm mJy \, beam^{-1}}$ in steps of 2~${\rm mJy \,
beam^{-1}}$.}
\label{fig:pgc6667_vel}
\end{figure}

\begin{figure}
\begin{center}$
\begin{array}{c}
\includegraphics[width=\textwidth]{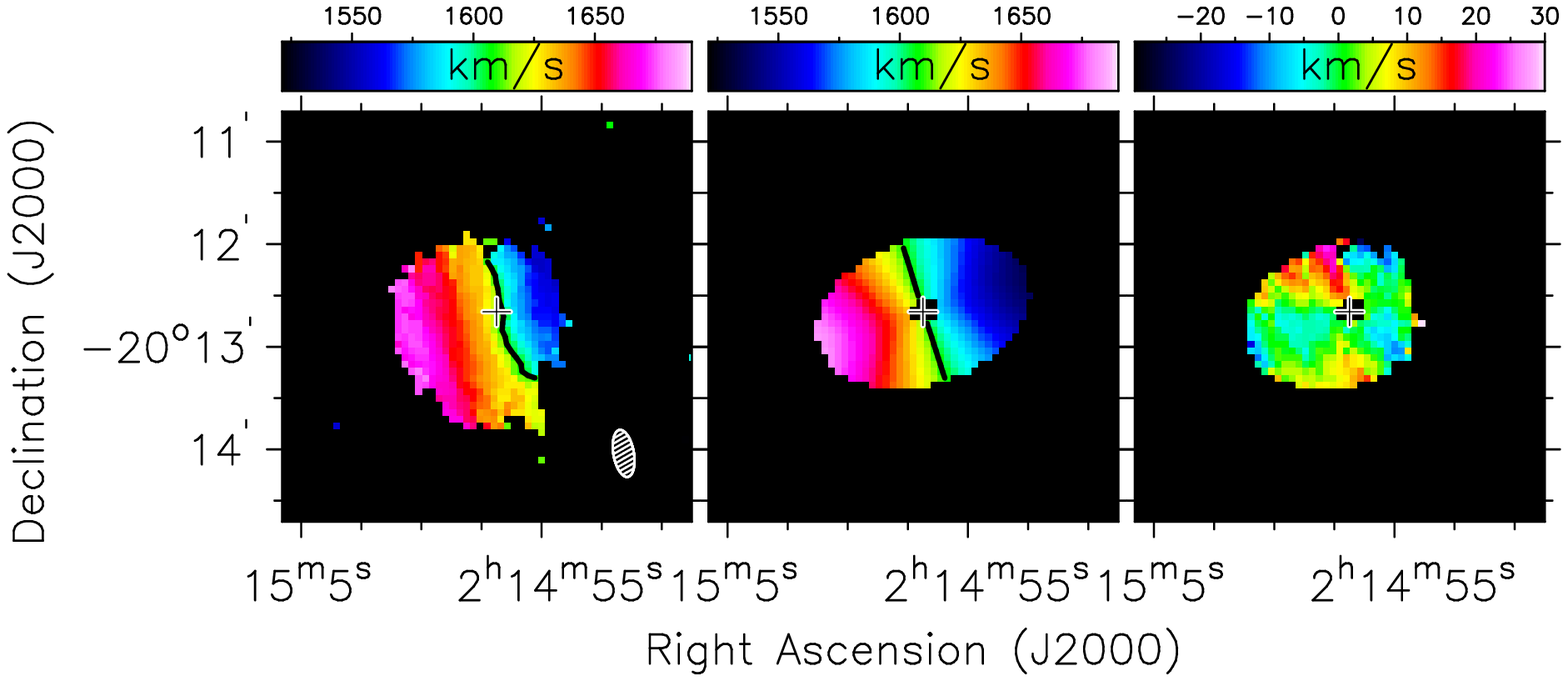} \\
\\
\includegraphics[width=\textwidth]{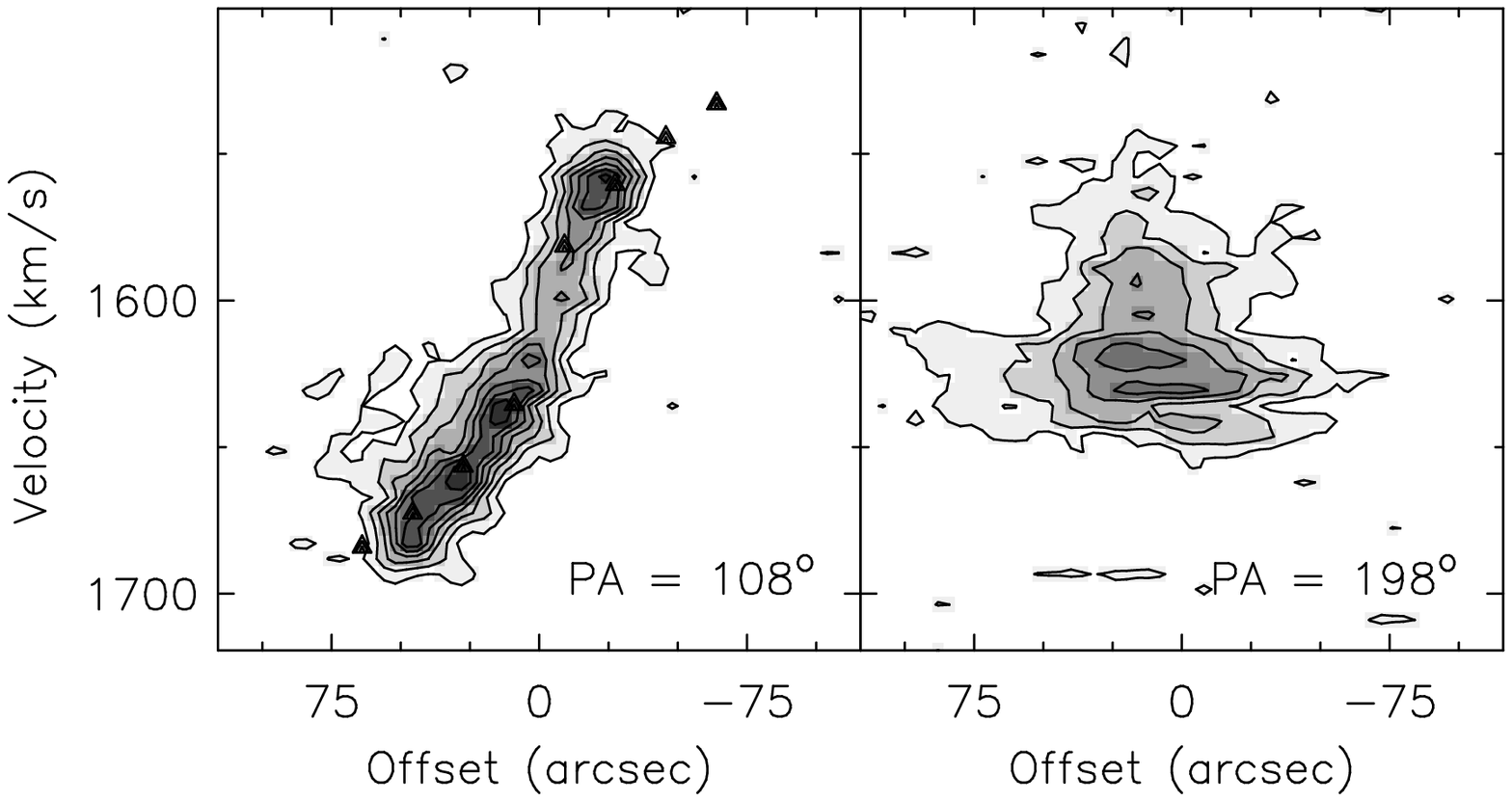} 
\end{array}$
\end{center}
\caption{As in Figure~\ref{fig:ngc0337_vel}, but for ESO~544-G030.
The position-velocity diagram contours are from 2~${\rm mJy \,
beam^{-1}}$ to 18~${\rm mJy \, beam^{-1}}$ in steps of 2~${\rm mJy \,
beam^{-1}}$.}
\label{fig:eso544_vel}
\end{figure}

\begin{figure}
\begin{center}$
\begin{array}{c}
\includegraphics[width=\textwidth]{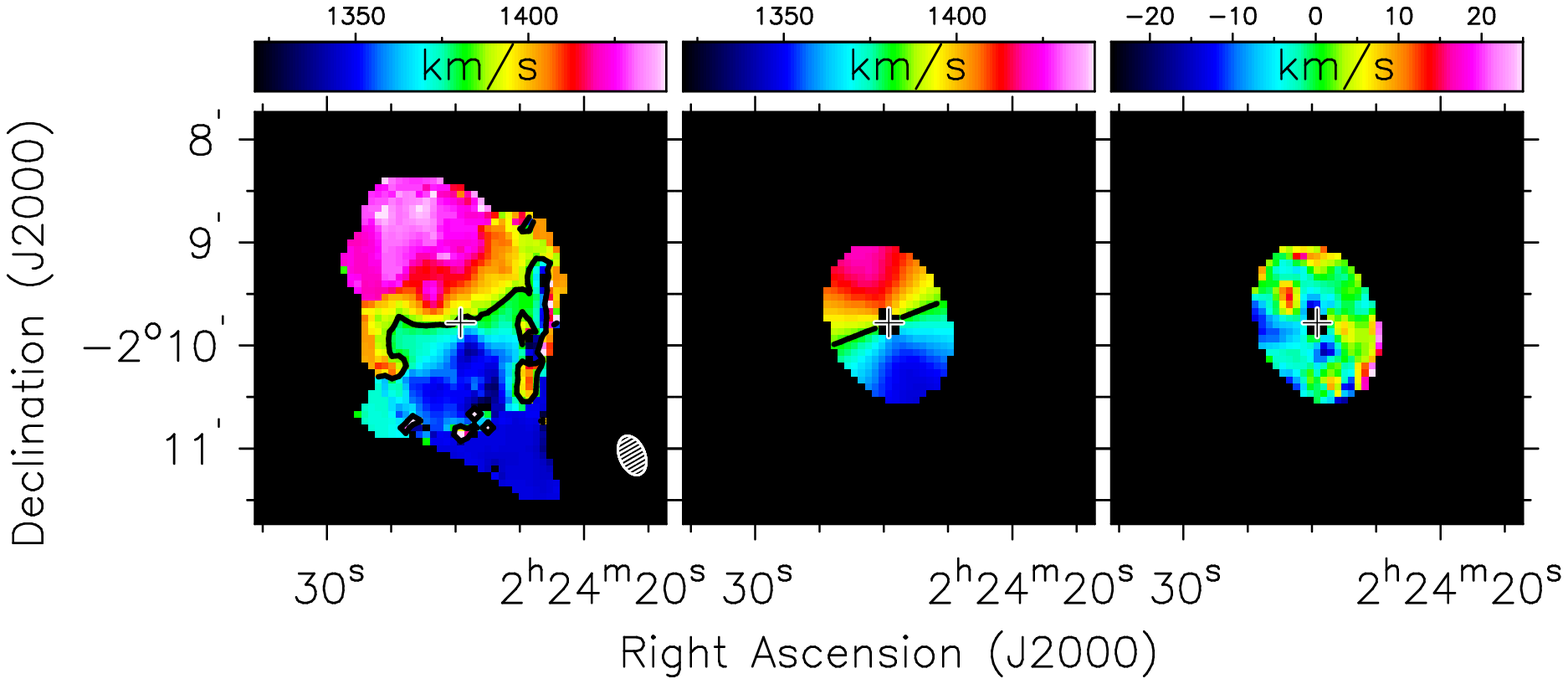} \\
\\
\includegraphics[width=\textwidth]{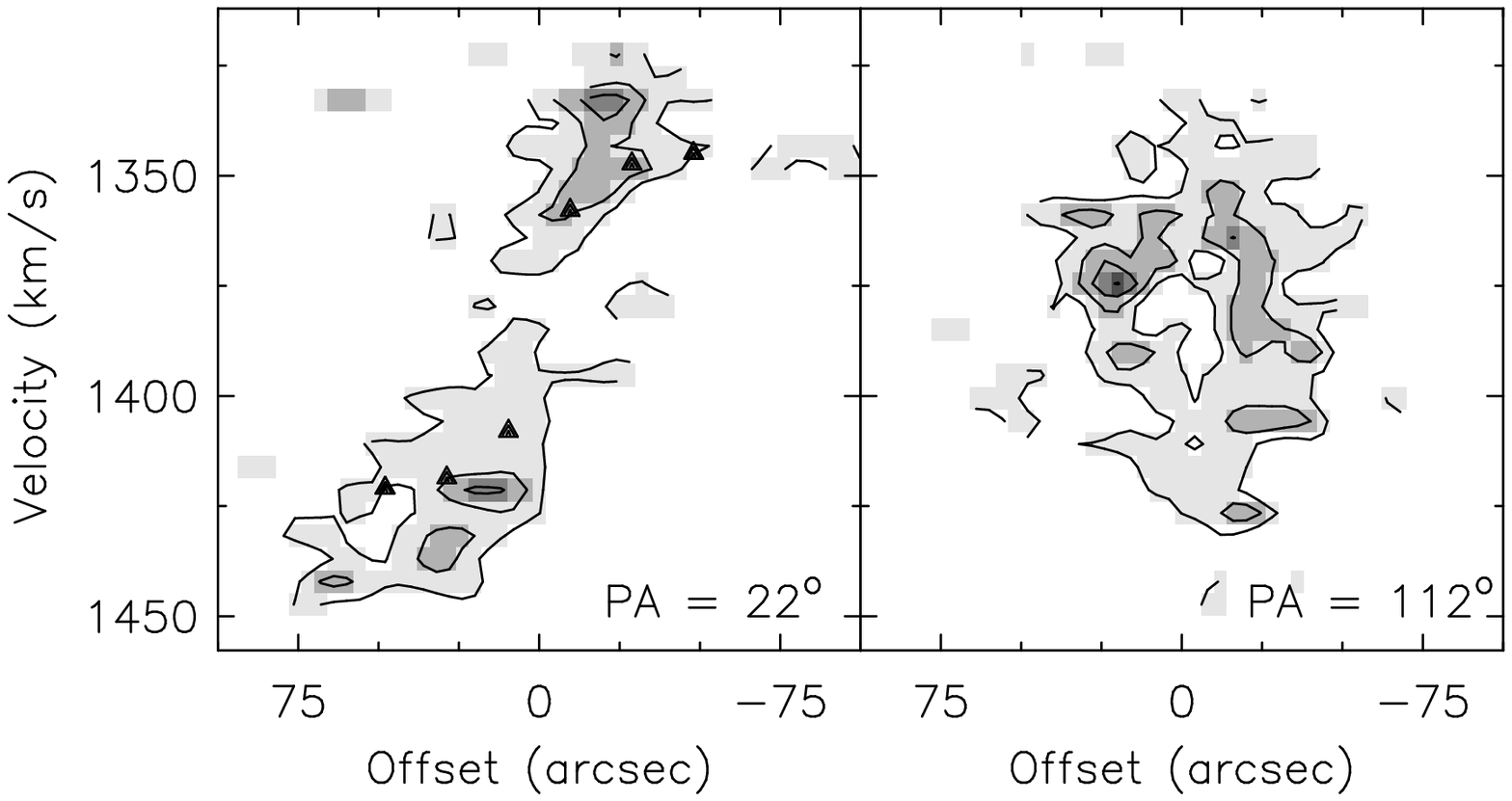} 
\end{array}$
\end{center}
\caption{As in Figure~\ref{fig:ngc0337_vel}, but for UGC~1862.  The
position-velocity diagram contours are from 2~${\rm mJy \, beam^{-1}}$
to 10~${\rm mJy \, beam^{-1}}$ in steps of 2~${\rm mJy \,
beam^{-1}}$.}
\label{fig:ugc1862_vel}
\end{figure}

\begin{figure}
\begin{center}$
\begin{array}{c}
\includegraphics[width=\textwidth]{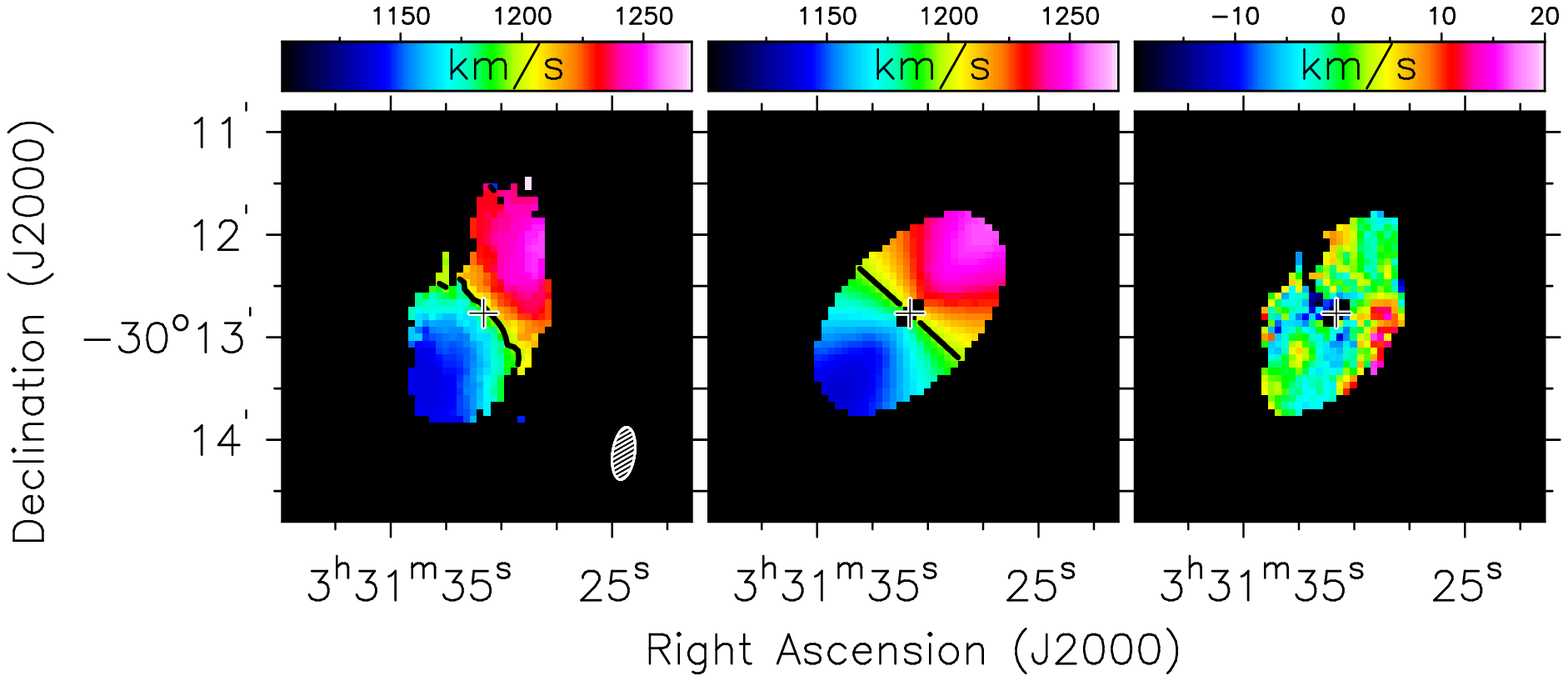} \\
\\
\includegraphics[width=\textwidth]{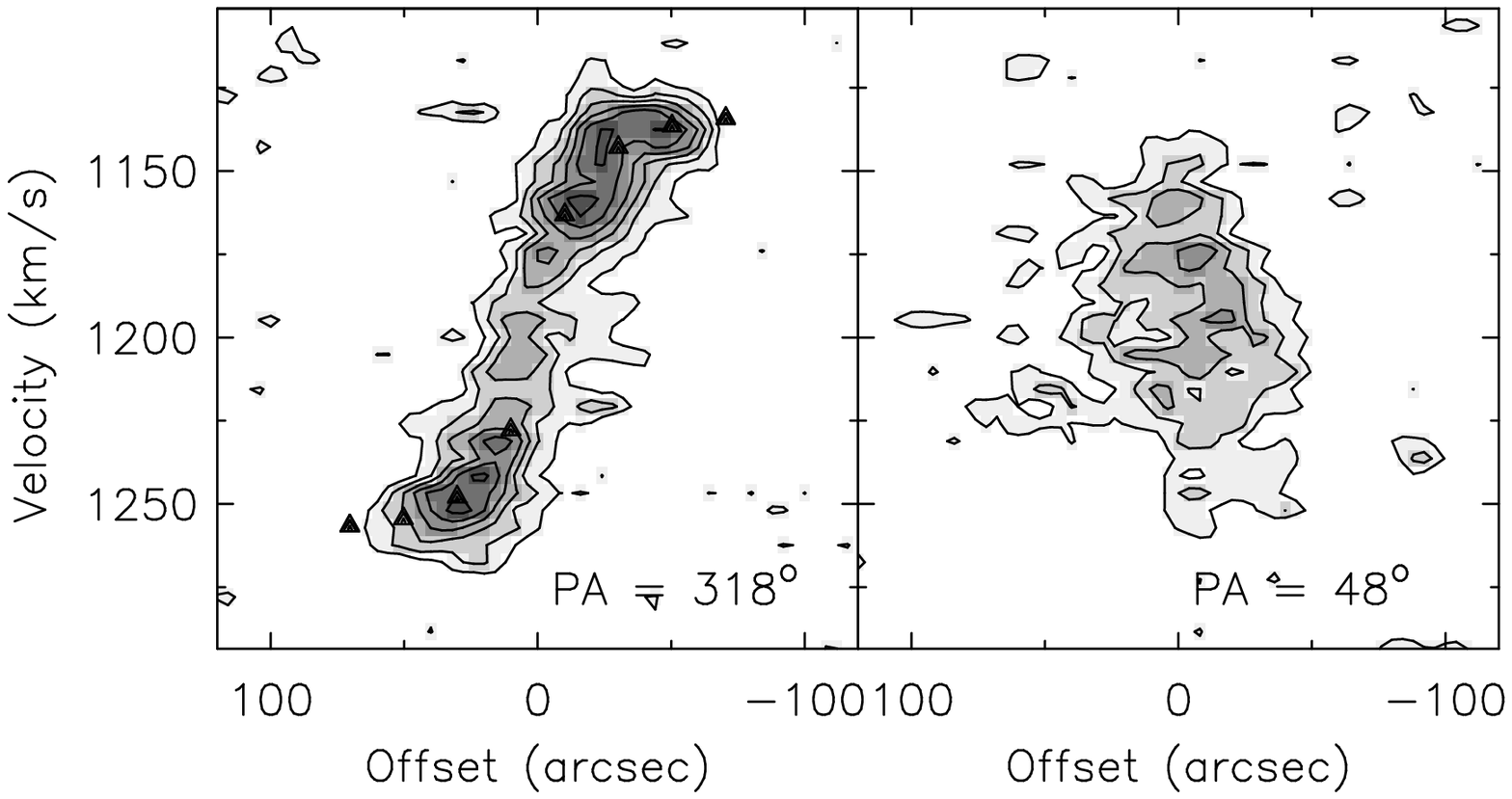} 
\end{array}$
\end{center}
\caption{As in Figure~\ref{fig:ngc0337_vel}, but for ESO~418-G008.
The position-velocity diagram contours are from 2~${\rm mJy \,
beam^{-1}}$ to 16~${\rm mJy \, beam^{-1}}$ in steps of 2~${\rm mJy \,
beam^{-1}}$.}
\label{fig:eso418_vel}
\end{figure}

\begin{figure}
\begin{center}$
\begin{array}{c}
\includegraphics[width=\textwidth]{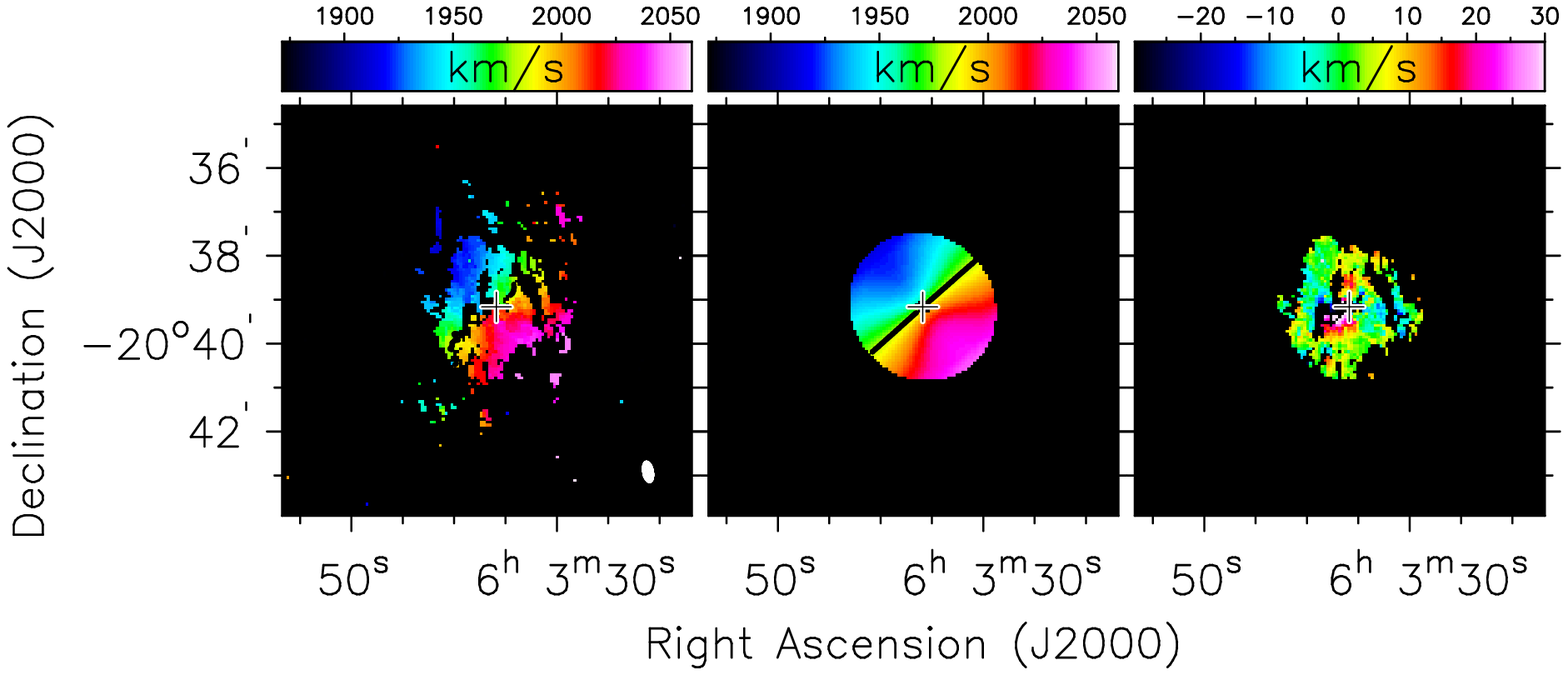} \\
\\
\includegraphics[width=\textwidth]{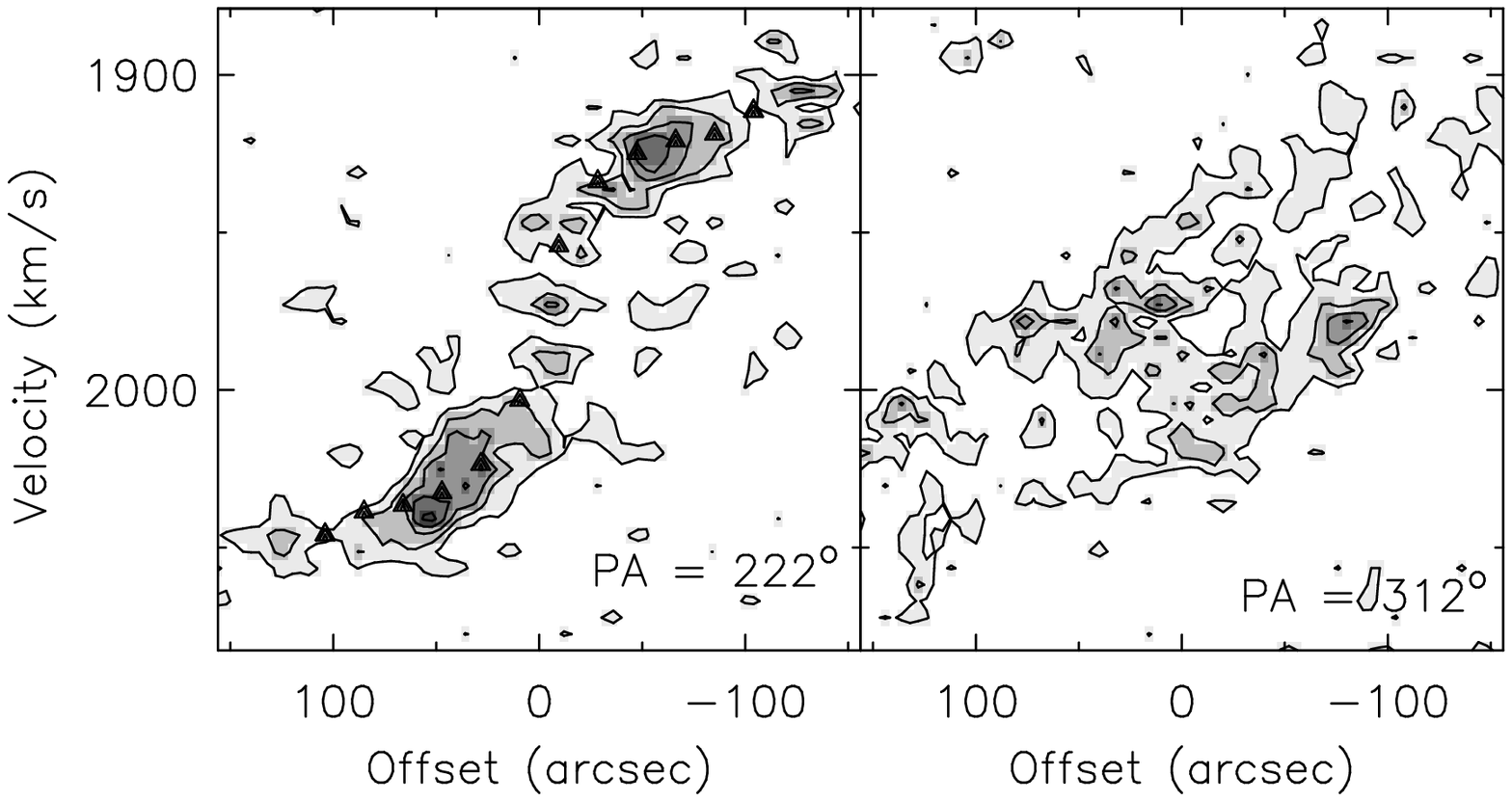} 
\end{array}$
\end{center}
\caption{As in Figure~\ref{fig:ngc0337_vel}, but for ESO~555-G027.
The position-velocity diagram contours are from 2~${\rm mJy \,
beam^{-1}}$ to 13~${\rm mJy \, beam^{-1}}$ in steps of 2~${\rm mJy \,
beam^{-1}}$.}
\label{fig:eso555_vel}
\end{figure}

\begin{figure}
\begin{center}$
\begin{array}{c}
\includegraphics[width=\textwidth]{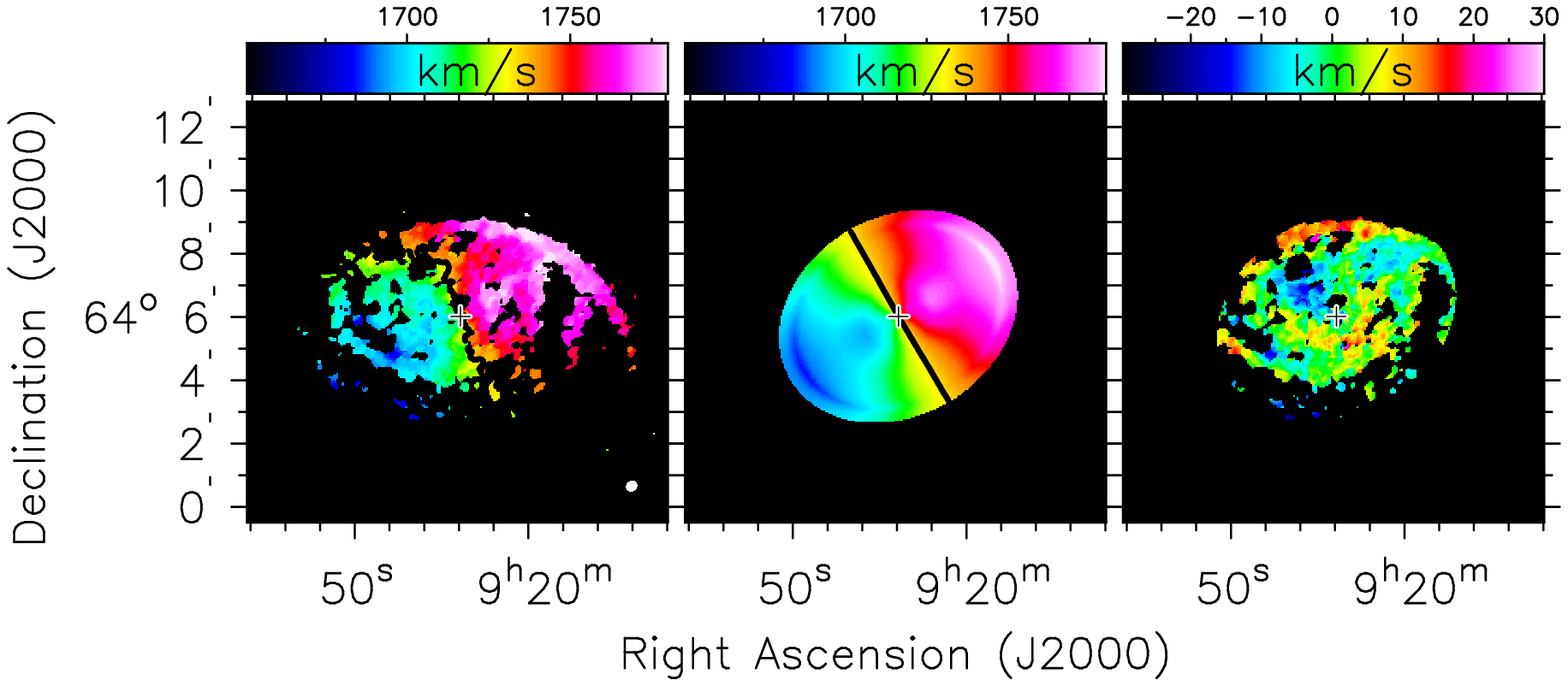} \\
\\
\includegraphics[width=\textwidth]{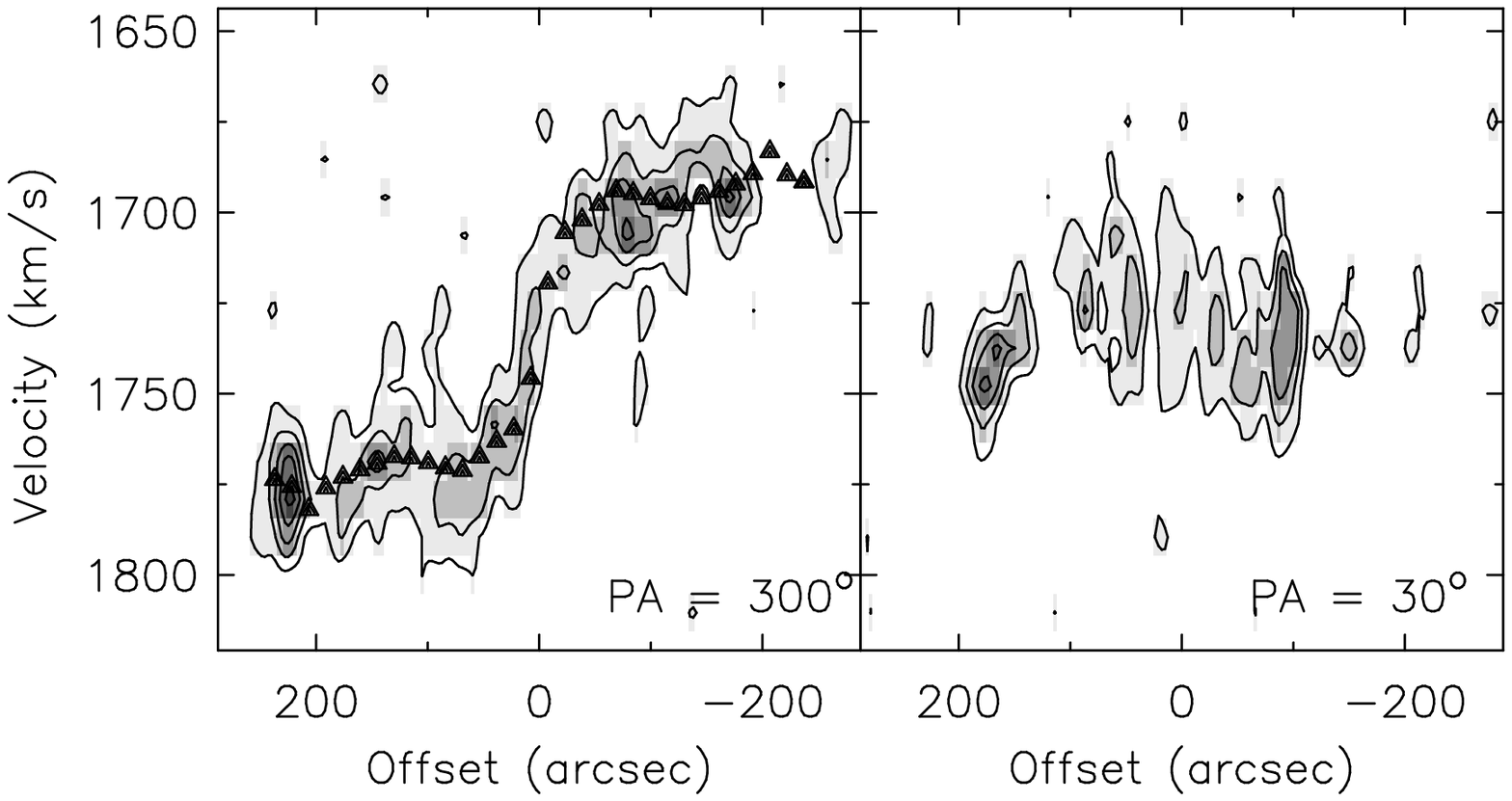} 
\end{array}$
\end{center}
\caption{As in Figure~\ref{fig:ngc0337_vel}, but for NGC~2805.  The
position-velocity diagram contours are from 2~${\rm mJy \, beam^{-1}}$
to 13~${\rm mJy \, beam^{-1}}$ in steps of 2~${\rm mJy \,
beam^{-1}}$.}
\label{fig:ngc2805_vel}
\end{figure}

\begin{figure}
\begin{center}$
\begin{array}{c}
\includegraphics[width=\textwidth]{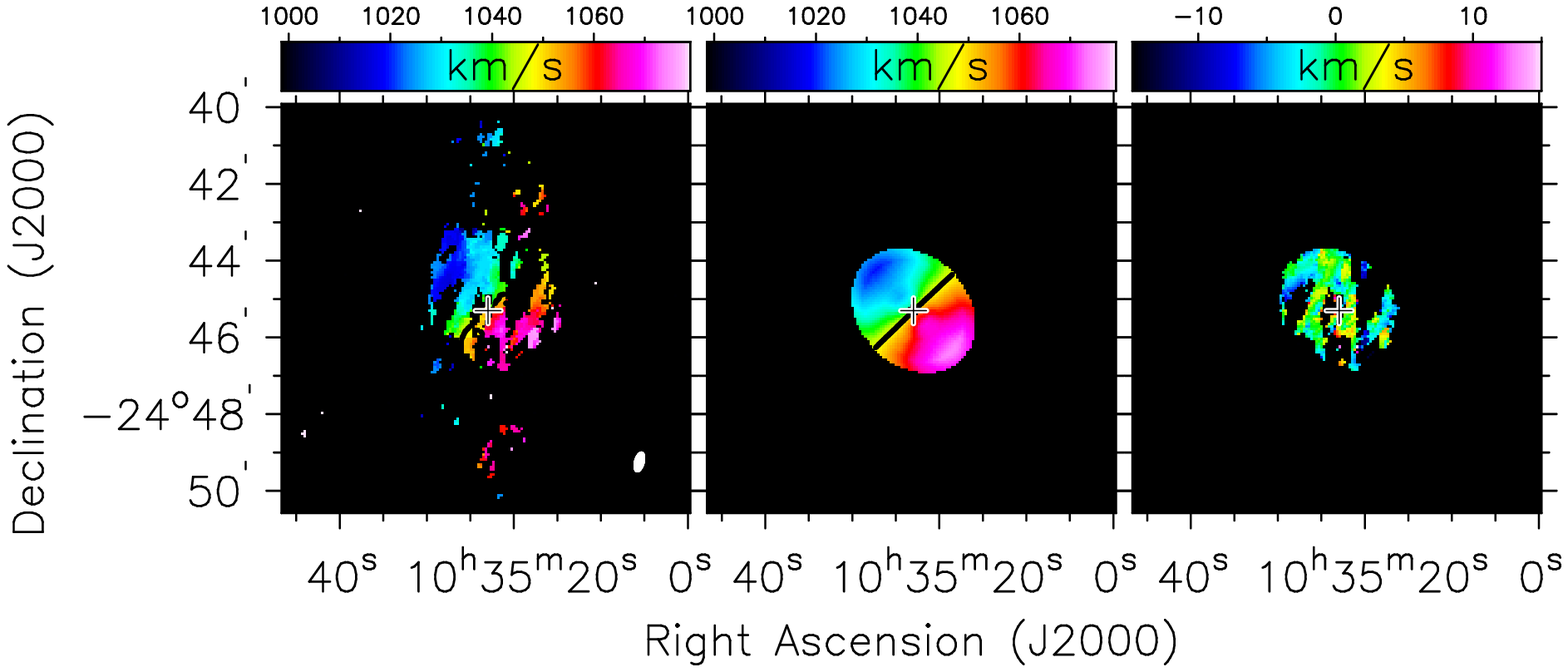} \\
\\
\includegraphics[width=\textwidth]{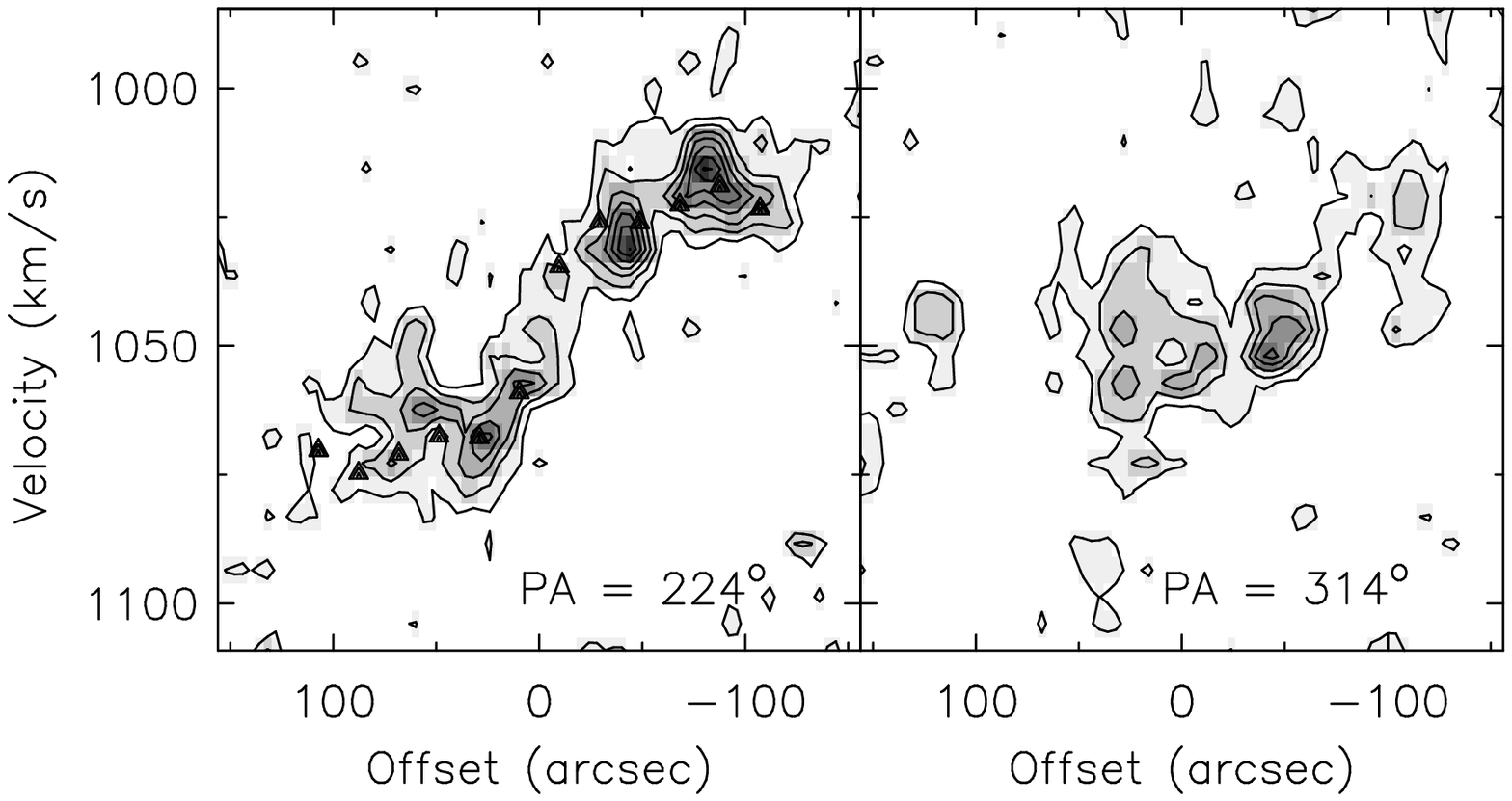} 
\end{array}$
\end{center}
\caption{As in Figure~\ref{fig:ngc0337_vel}, but for ESO~501-G023.
The position-velocity diagram contours are from 2~${\rm mJy \,
beam^{-1}}$ to 16~${\rm mJy \, beam^{-1}}$ in steps of 2~${\rm mJy \,
beam^{-1}}$.}
\label{fig:eso501_vel}
\end{figure}

\begin{figure}
\begin{center}$
\begin{array}{c}
\includegraphics[width=\textwidth]{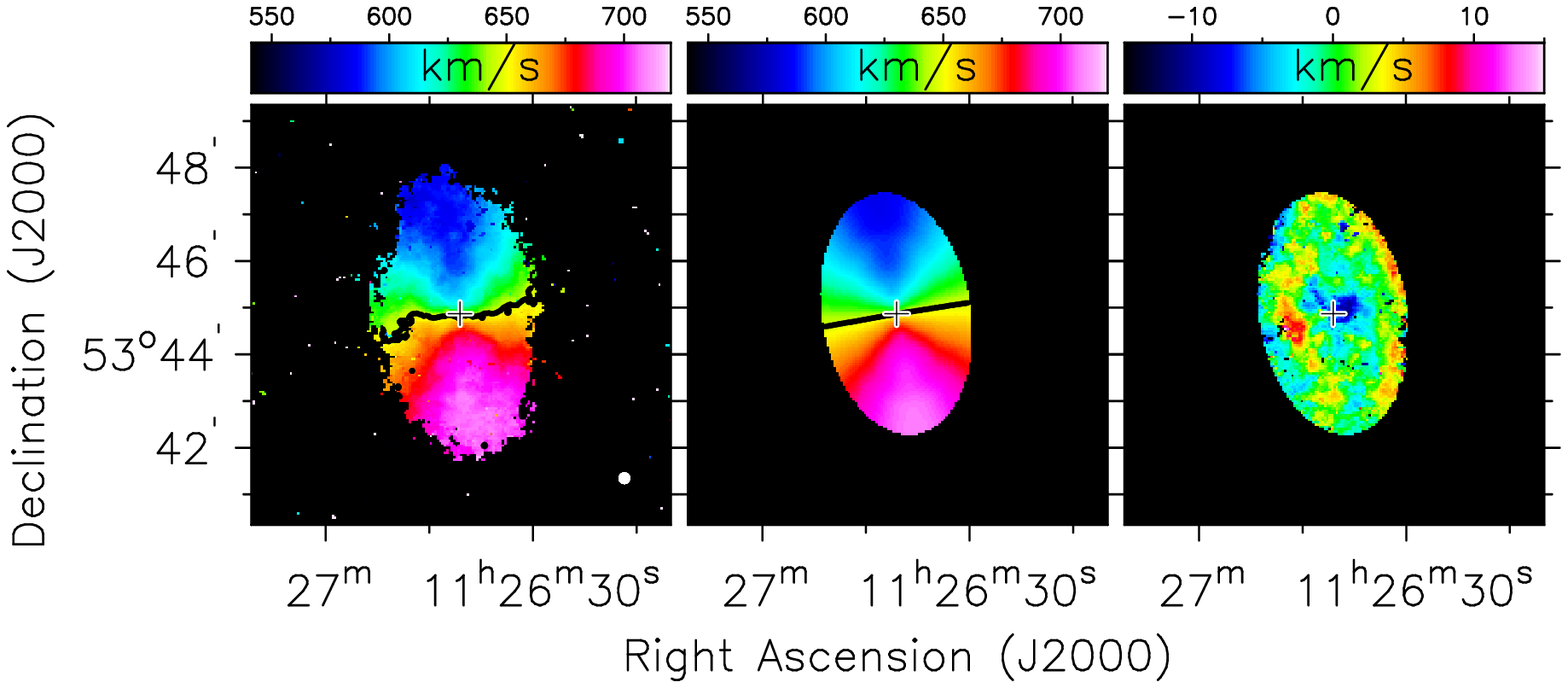} \\
\\
\includegraphics[width=\textwidth]{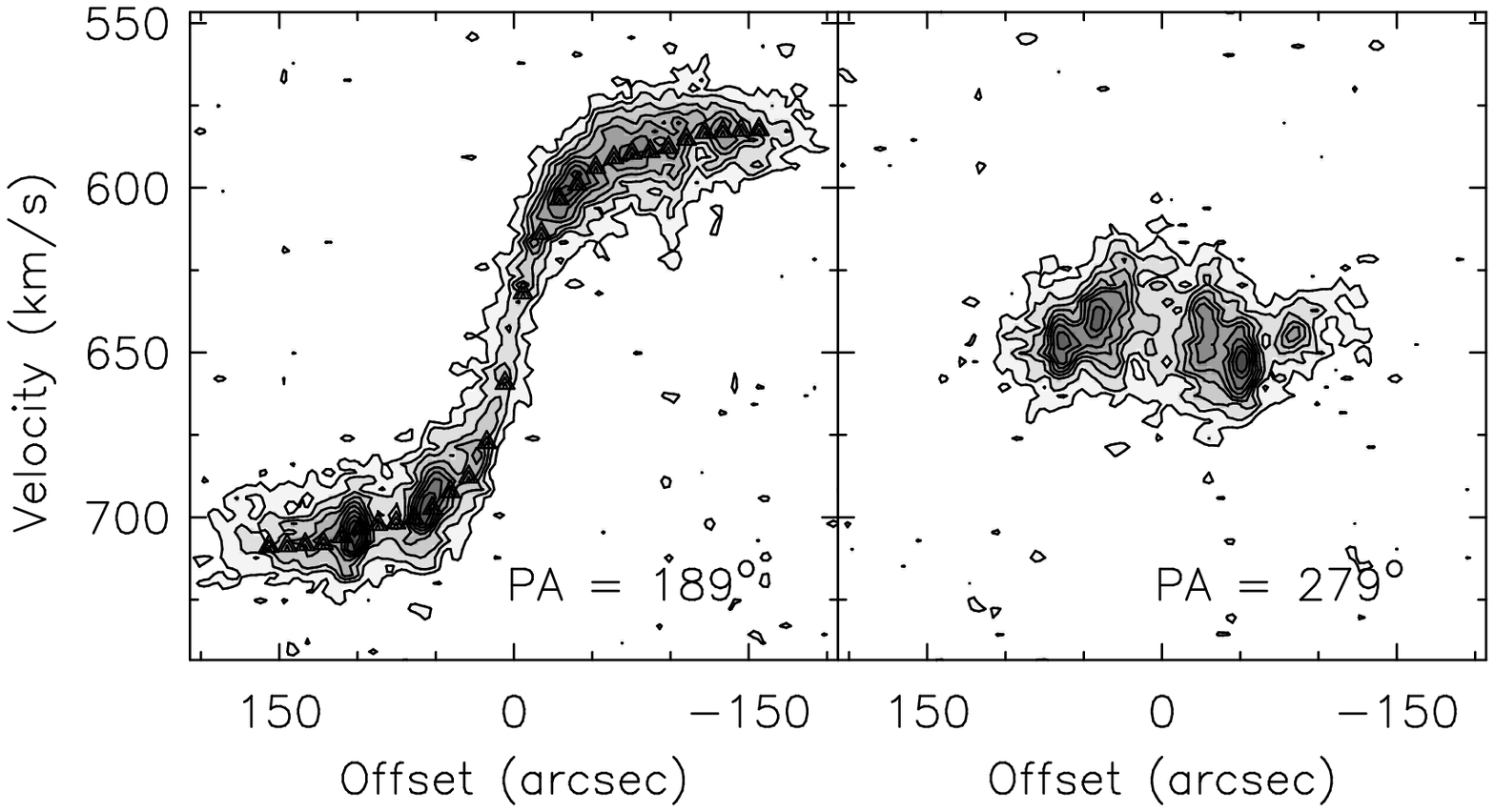} 
\end{array}$
\end{center}
\caption{As in Figure~\ref{fig:ngc0337_vel}, but for UGC~6446.  The
position-velocity diagram contours are from 1~${\rm mJy \, beam^{-1}}$
to 12~${\rm mJy \, beam^{-1}}$ in steps of 1~${\rm mJy \,
beam^{-1}}$.}
\label{fig:ugc6446_vel}
\end{figure}

\begin{figure}
\begin{center}$
\begin{array}{c}
\includegraphics[width=\textwidth]{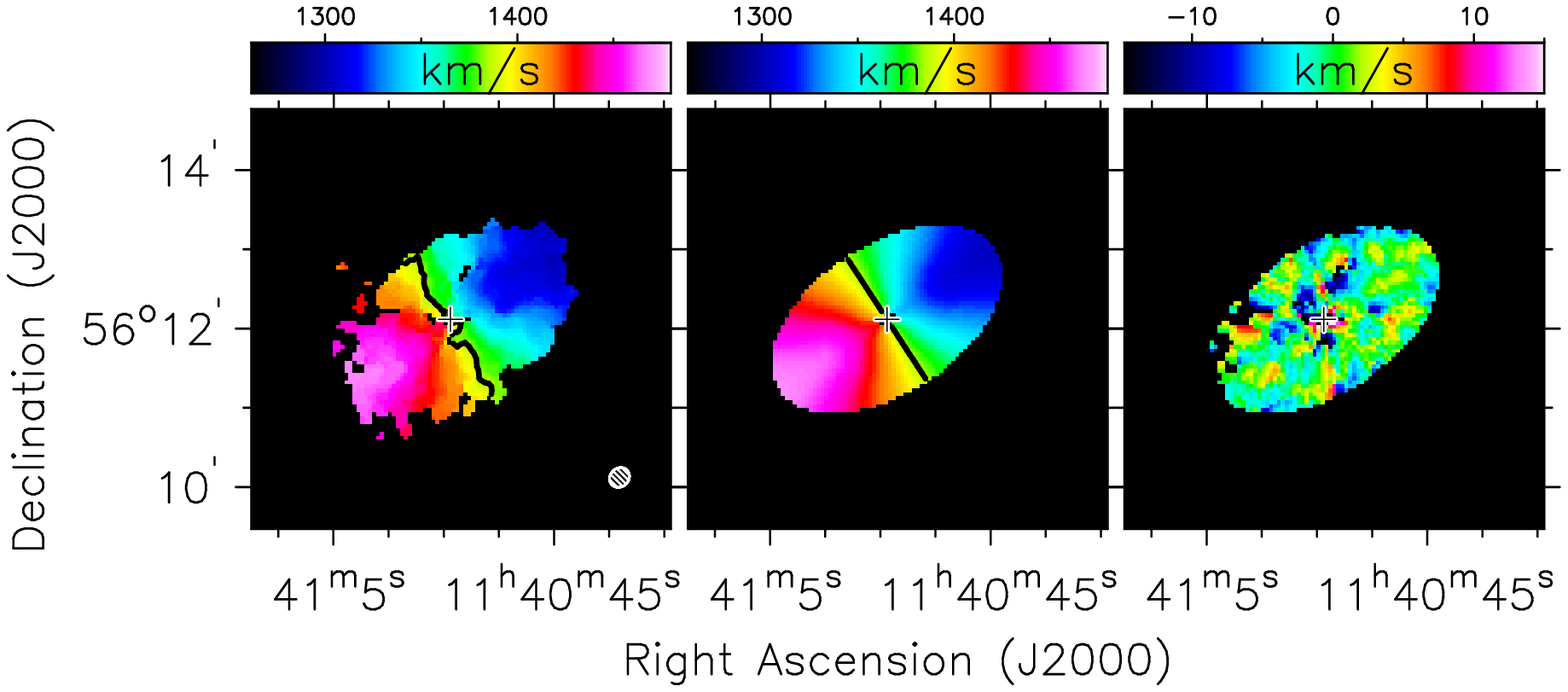} \\
\\
\includegraphics[width=\textwidth]{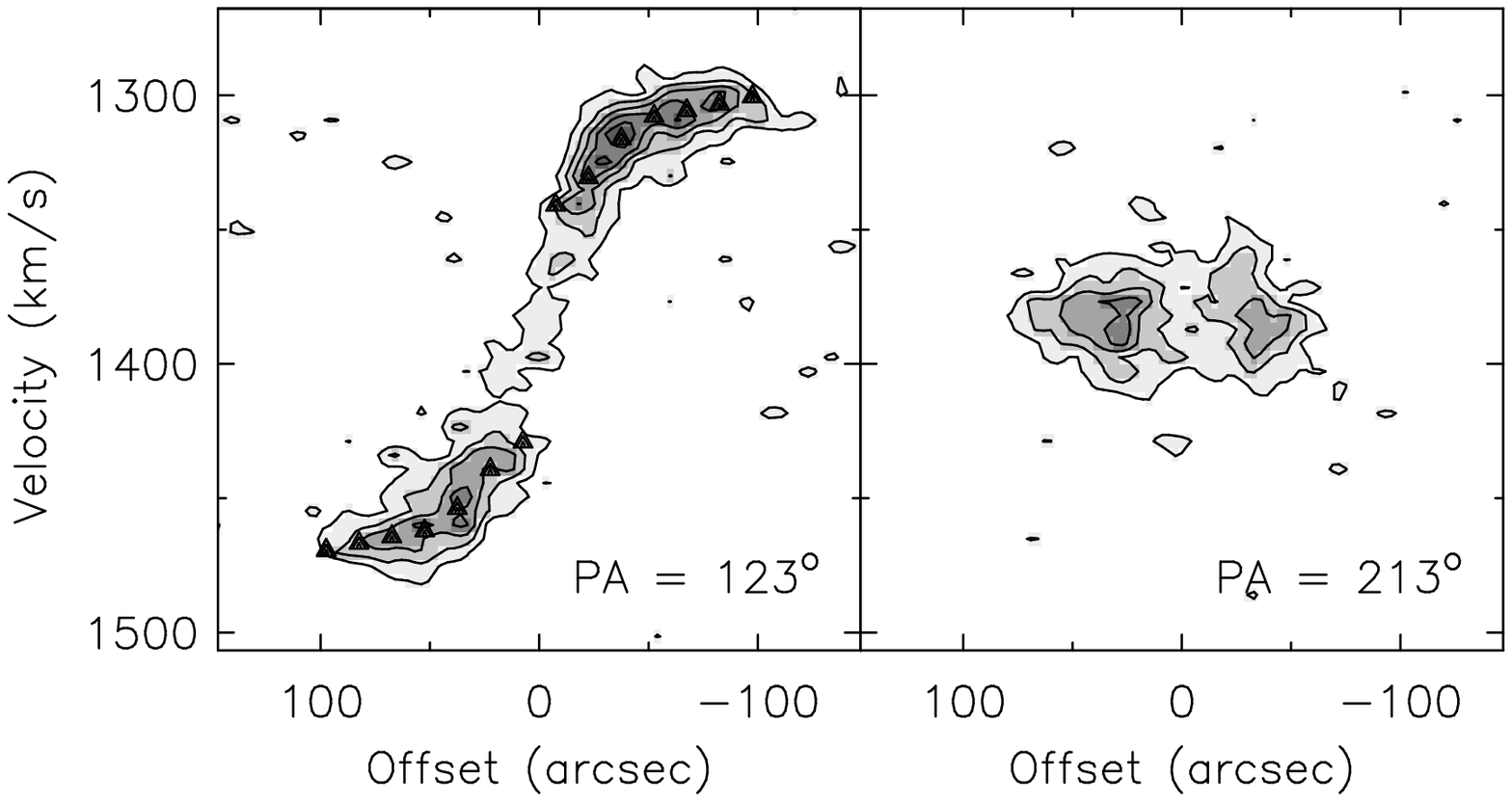} 
\end{array}$
\end{center}
\caption{As in Figure~\ref{fig:ngc0337_vel}, but for NGC~3794.  The
position-velocity diagram contours are from 2~${\rm mJy \, beam^{-1}}$
to 14~${\rm mJy \, beam^{-1}}$ in steps of 2~${\rm mJy \,
beam^{-1}}$.}
\label{fig:ngc3794_vel}
\end{figure}

\begin{figure}
\begin{center}$
\begin{array}{c}
\includegraphics[width=\textwidth]{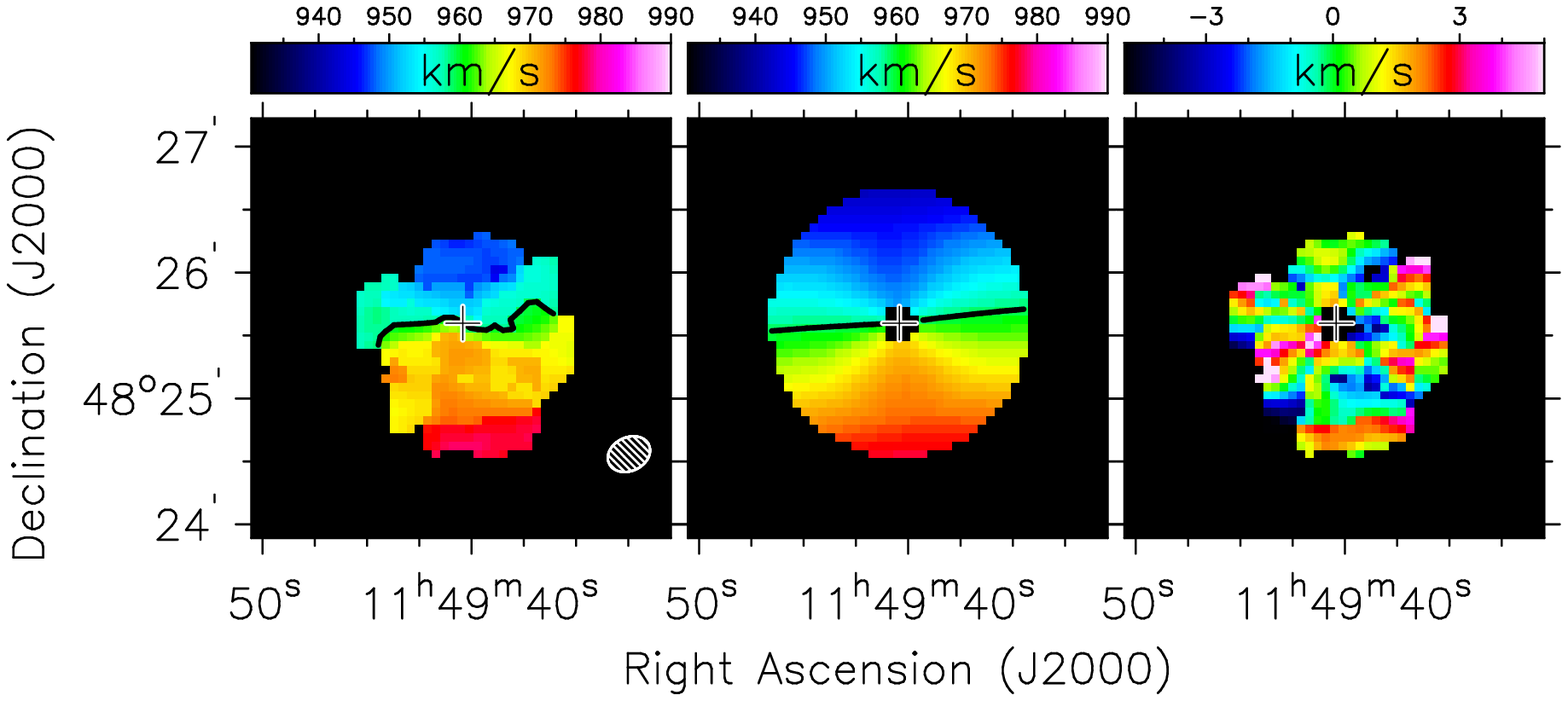} \\
\\
\includegraphics[width=\textwidth]{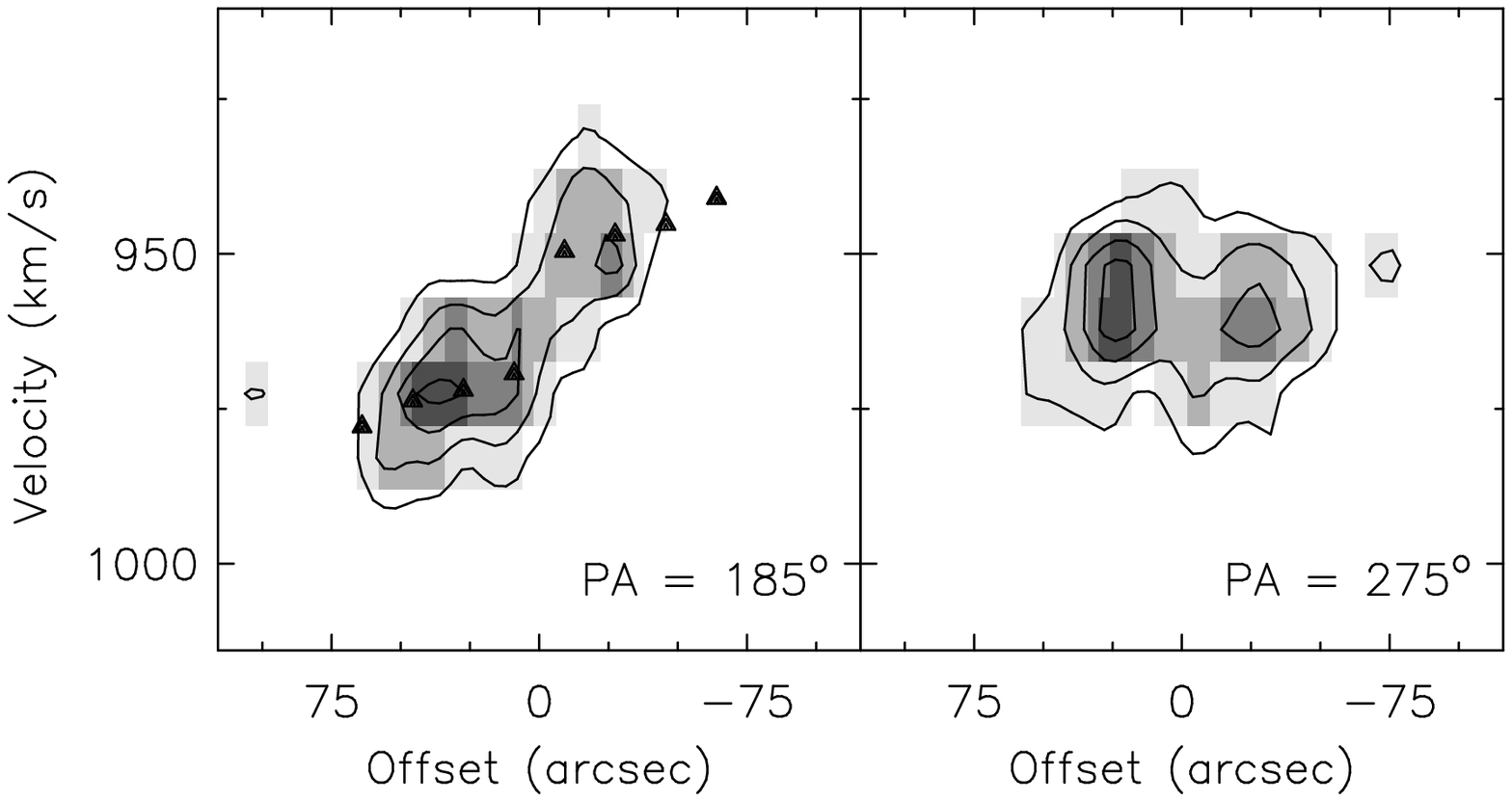} 
\end{array}$
\end{center}
\caption{As in Figure~\ref{fig:ngc0337_vel}, but for NGC~3906.  The
position-velocity diagram contours are from 2~${\rm mJy \, beam^{-1}}$
to 11~${\rm mJy \, beam^{-1}}$ in steps of 2~${\rm mJy \,
beam^{-1}}$.}
\label{fig:ngc3906_vel}
\end{figure}

\begin{figure}
\begin{center}$
\begin{array}{c}
\includegraphics[width=\textwidth]{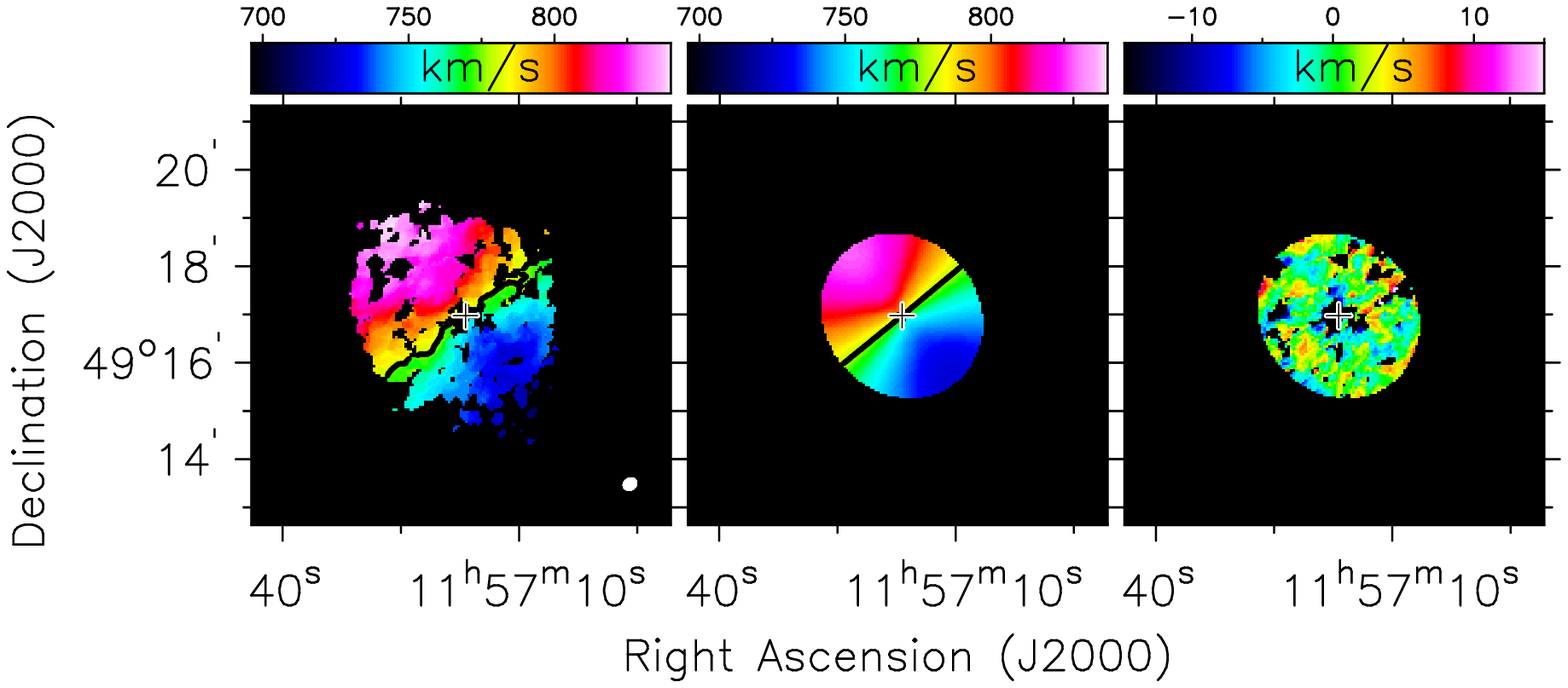} \\
\\
\includegraphics[width=\textwidth]{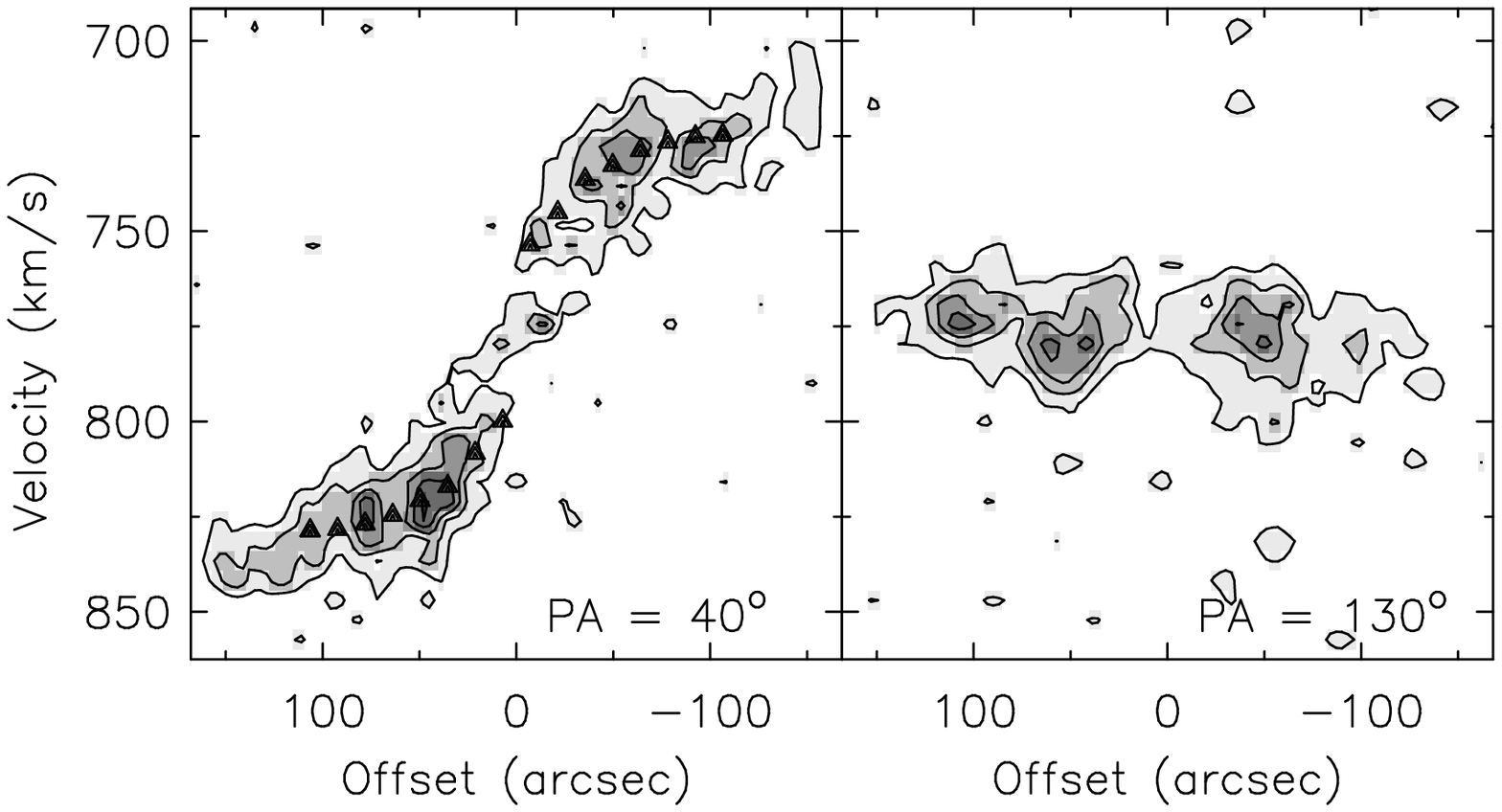} 
\end{array}$
\end{center}
\caption{As in Figure~\ref{fig:ngc0337_vel}, but for UGC~6930. The
position-velocity diagram contours are from 2~${\rm mJy \, beam^{-1}}$
to 12~${\rm mJy \, beam^{-1}}$ in steps of 2~${\rm mJy \,
beam^{-1}}$.}
\label{fig:ugc6930_vel}
\end{figure}

\begin{figure}
\begin{center}$
\begin{array}{c}
\includegraphics[width=\textwidth]{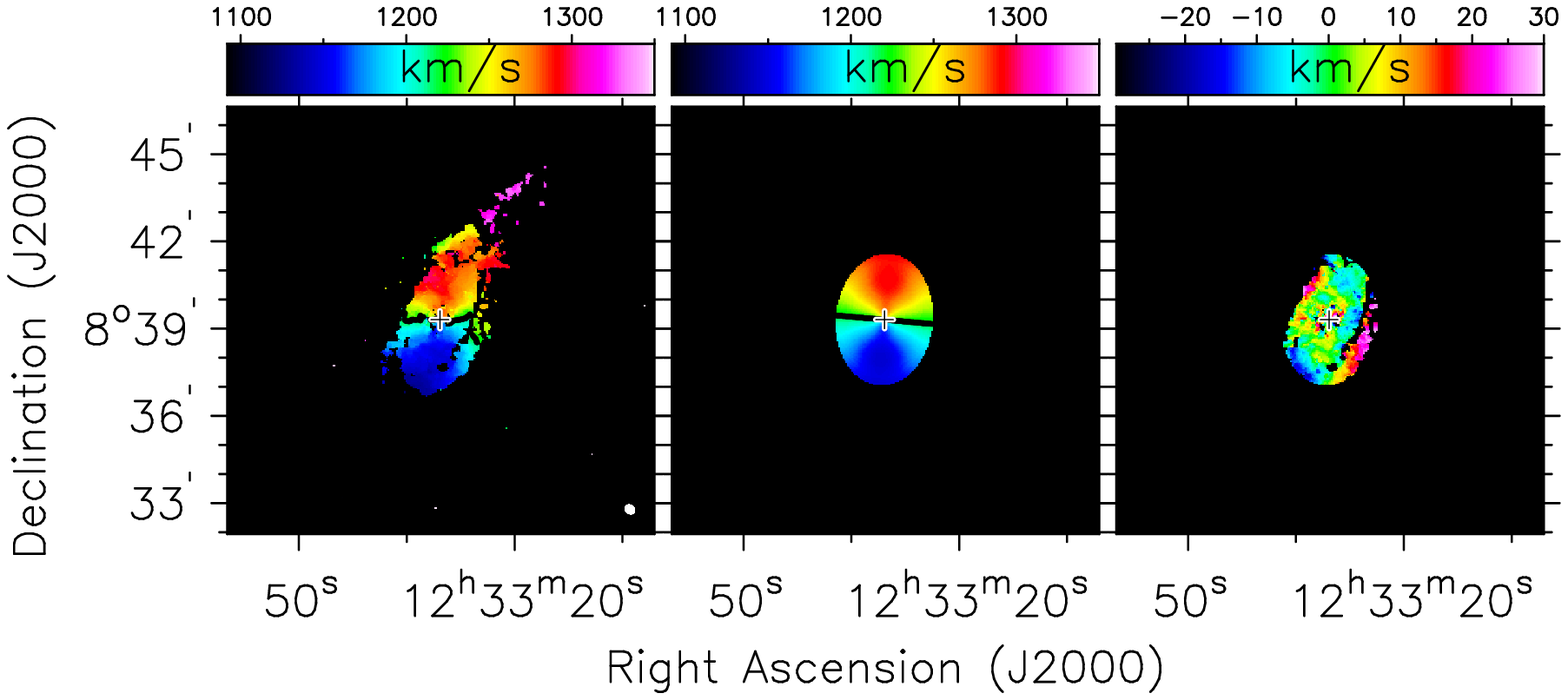} \\
\\
\includegraphics[width=\textwidth]{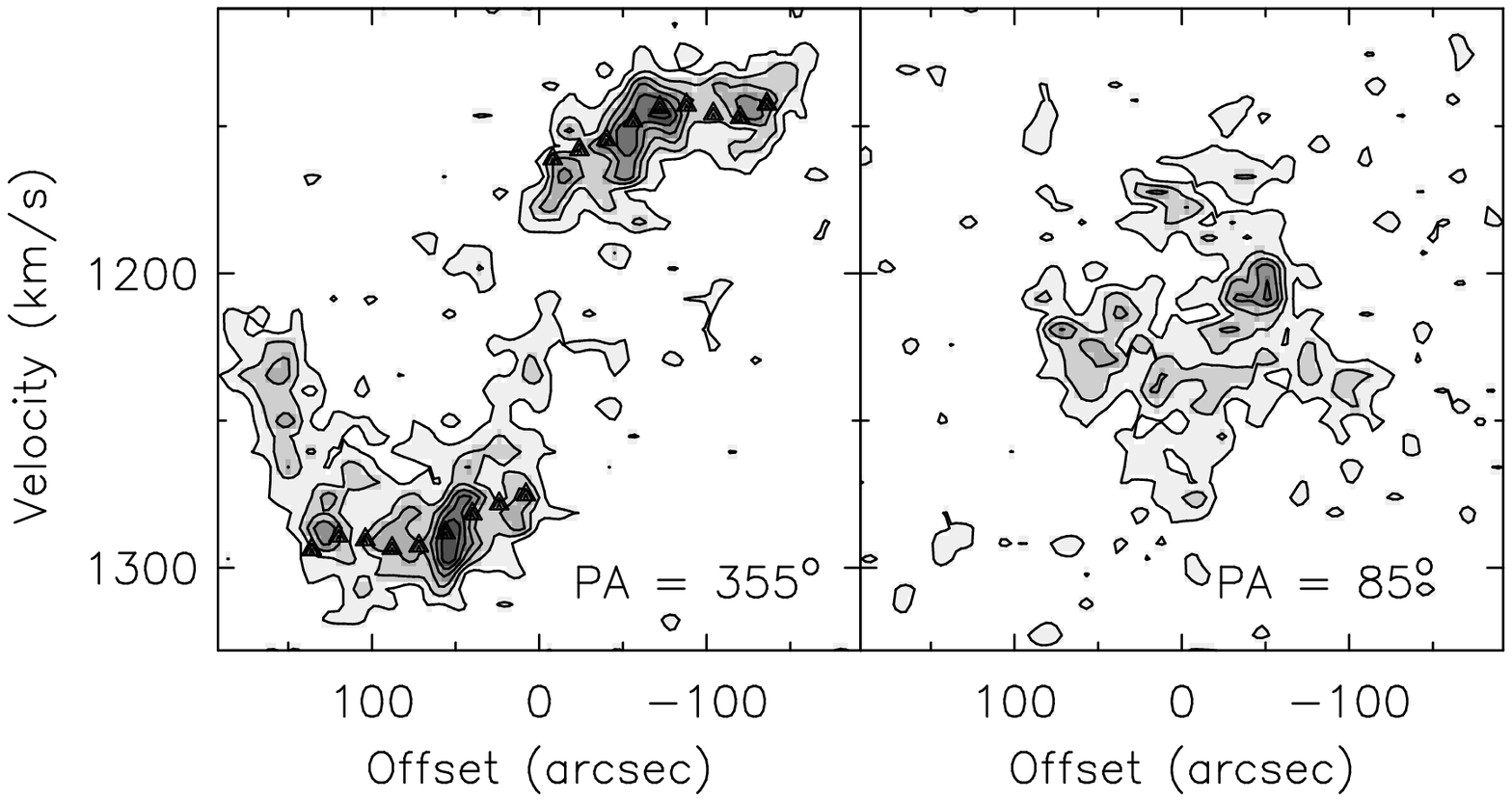} 
\end{array}$
\end{center}
\caption{As in Figure~\ref{fig:ngc0337_vel}, but for NGC~4519. The
position-velocity diagram contours are from 2~${\rm mJy \, beam^{-1}}$
to 16~${\rm mJy \, beam^{-1}}$ in steps of 2~${\rm mJy \,
beam^{-1}}$.}
\label{fig:ngc4519_vel}
\end{figure}

\begin{figure}
\begin{center}$
\begin{array}{c}
\includegraphics[width=\textwidth]{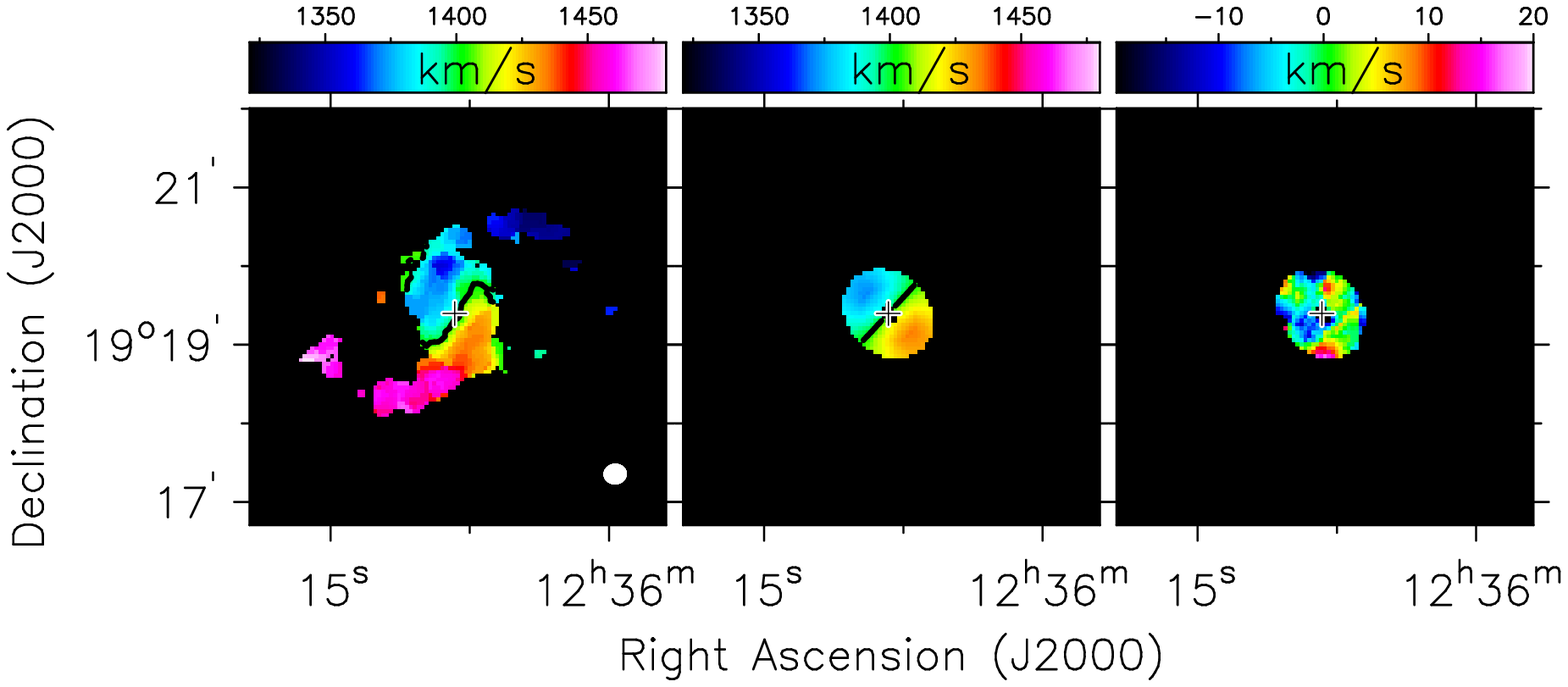} \\
\\
\includegraphics[width=\textwidth]{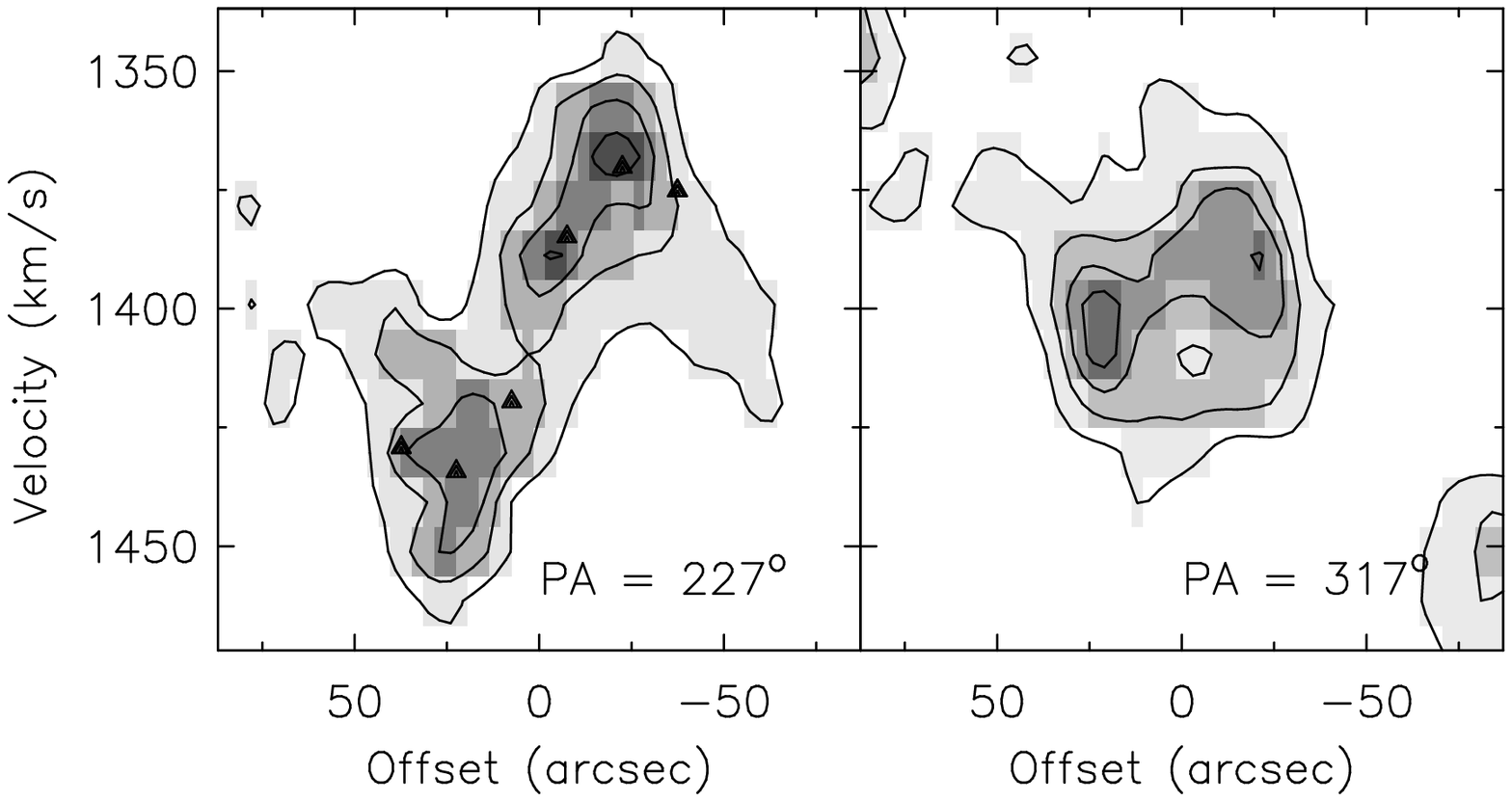} 
\end{array}$
\end{center}
\caption{As in Figure~\ref{fig:ngc0337_vel}, but for NGC~4561. The
position-velocity diagram contours are from 3~${\rm mJy \, beam^{-1}}$
to 17~${\rm mJy \, beam^{-1}}$ in steps of 3~${\rm mJy \,
beam^{-1}}$.}
\label{fig:ngc4561_vel}
\end{figure}

\begin{figure}
\begin{center}$
\begin{array}{c}
\includegraphics[width=\textwidth]{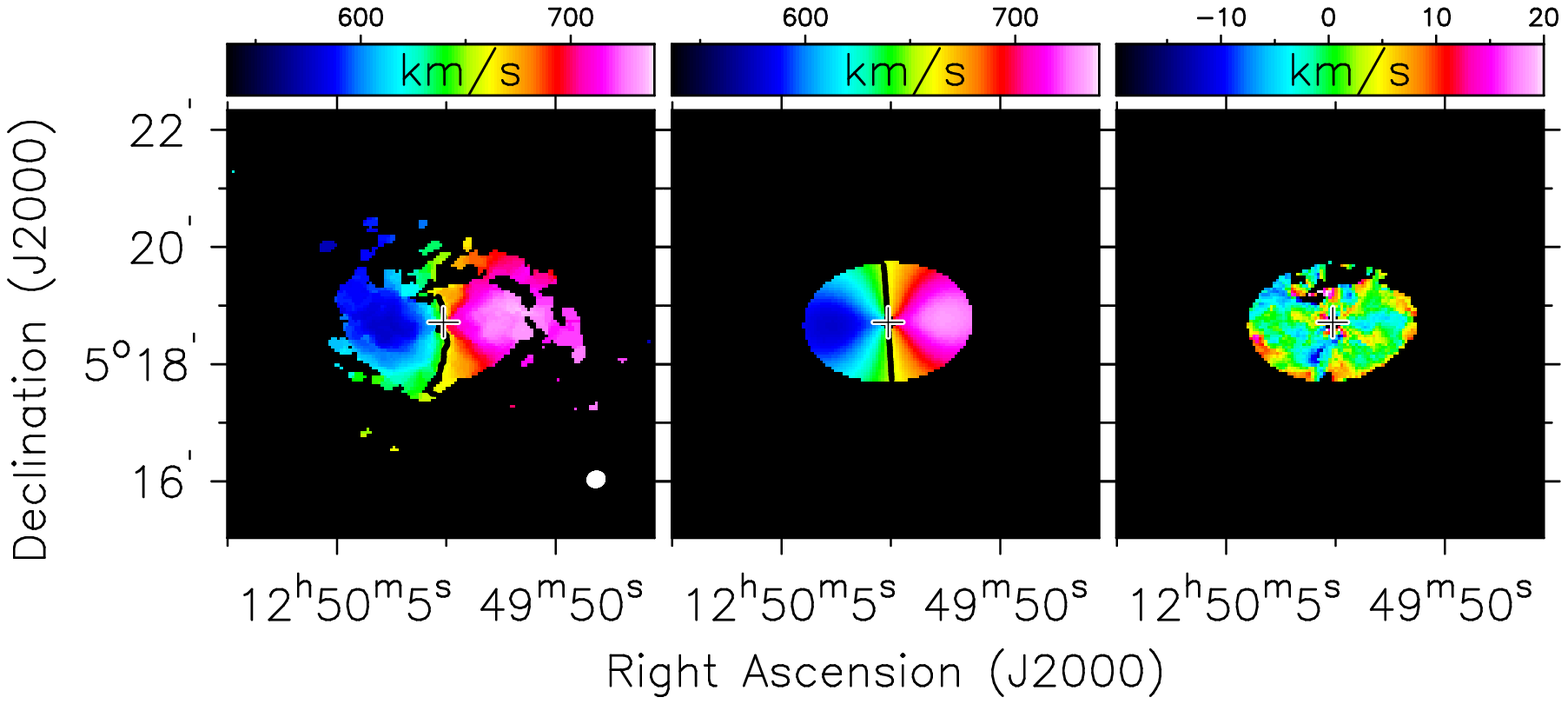} \\
\\
\includegraphics[width=\textwidth]{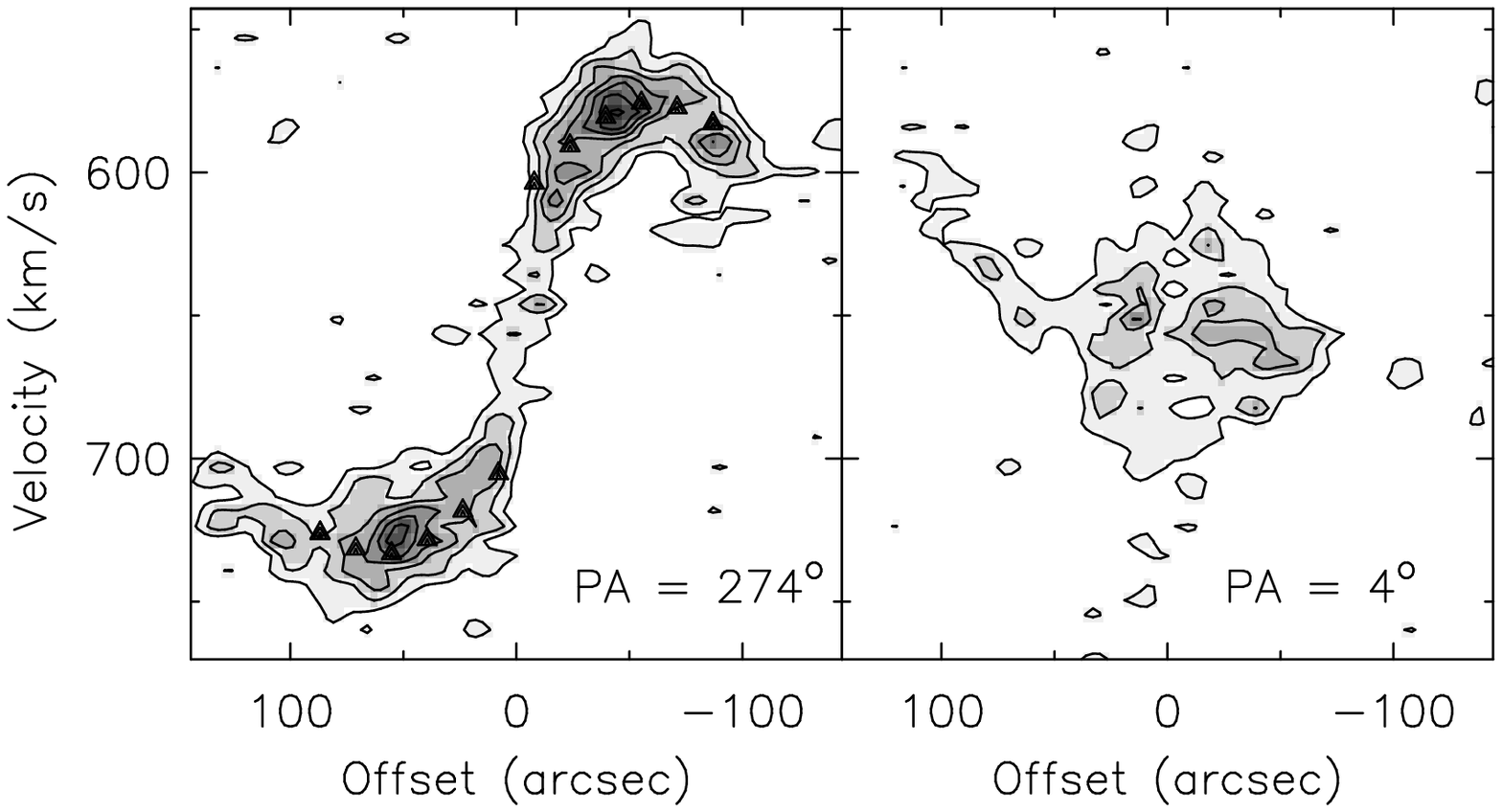} 
\end{array}$
\end{center}
\caption{As in Figure~\ref{fig:ngc0337_vel}, but for NGC~4713. The
position-velocity diagram contours are from 2~${\rm mJy \, beam^{-1}}$
to 16~${\rm mJy \, beam^{-1}}$ in steps of 2~${\rm mJy \,
beam^{-1}}$.}
\label{fig:ngc4713_vel}
\end{figure}

\begin{figure}
\begin{center}$
\begin{array}{c}
\includegraphics[width=\textwidth]{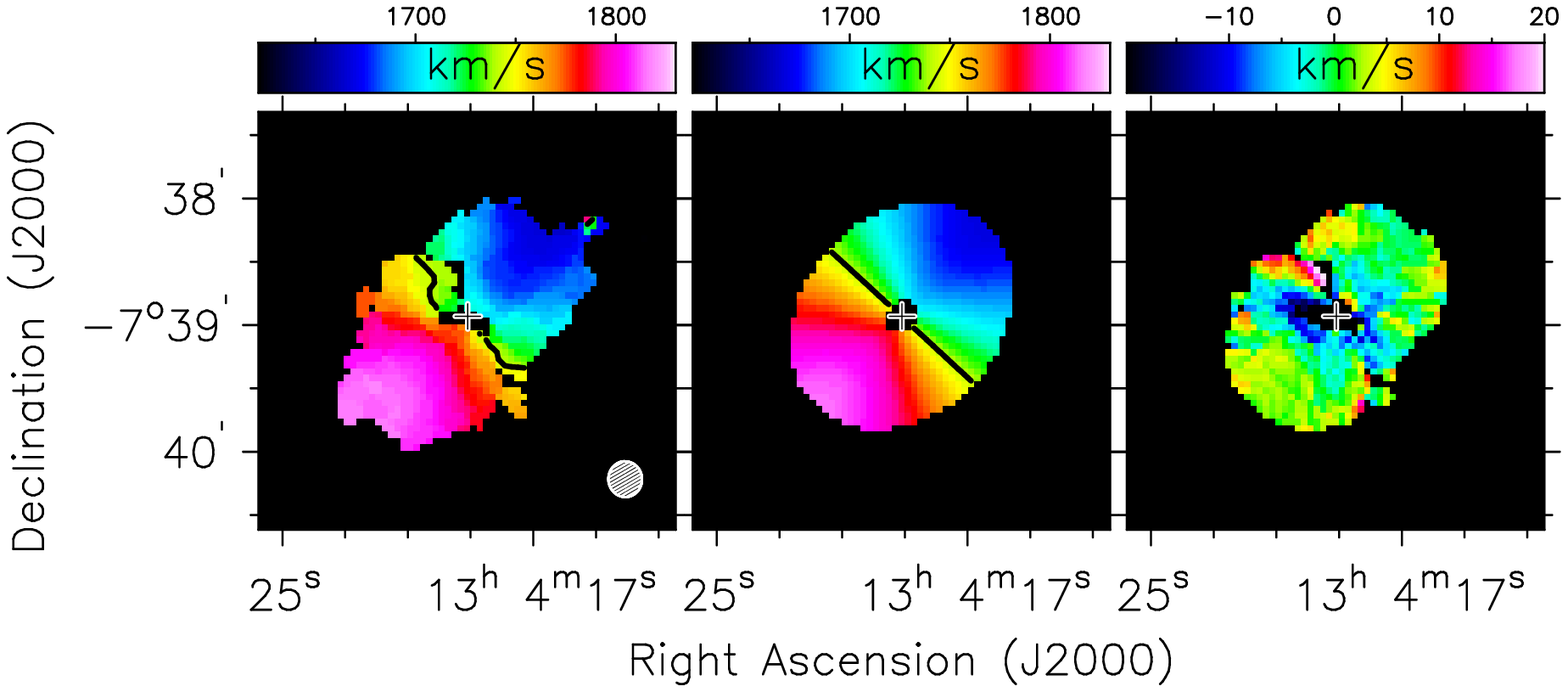} \\
\\
\includegraphics[width=\textwidth]{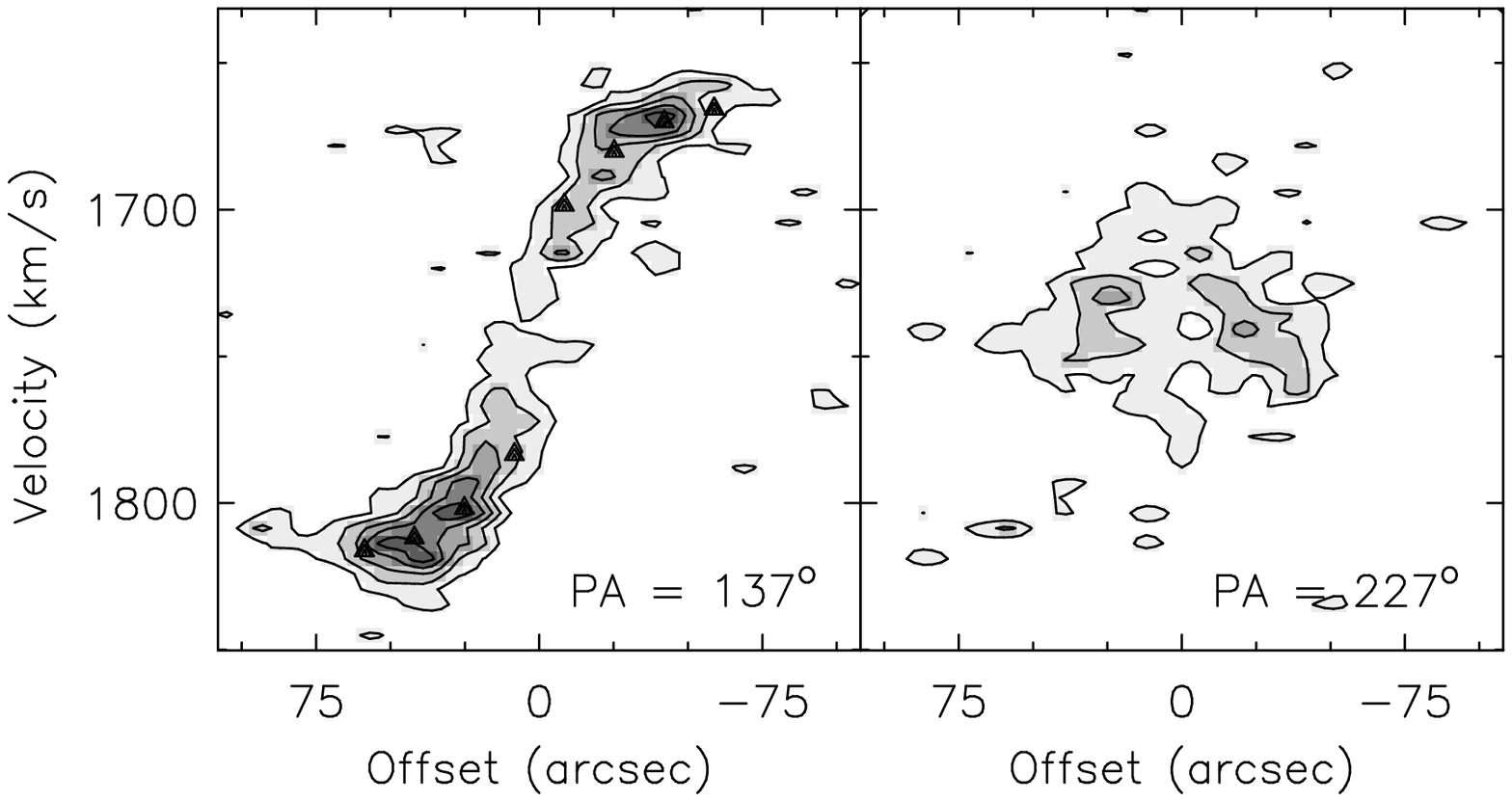} 
\end{array}$
\end{center}
\caption{As in Figure~\ref{fig:ngc0337_vel}, but for NGC~4942. The
position-velocity diagram contours are from 2~${\rm mJy \, beam^{-1}}$
to 14~${\rm mJy \, beam^{-1}}$ in steps of 2~${\rm mJy \,
beam^{-1}}$.}
\label{fig:ngc4942_vel}
\end{figure}

\begin{figure}
\begin{center}$
\begin{array}{c}
\includegraphics[width=\textwidth]{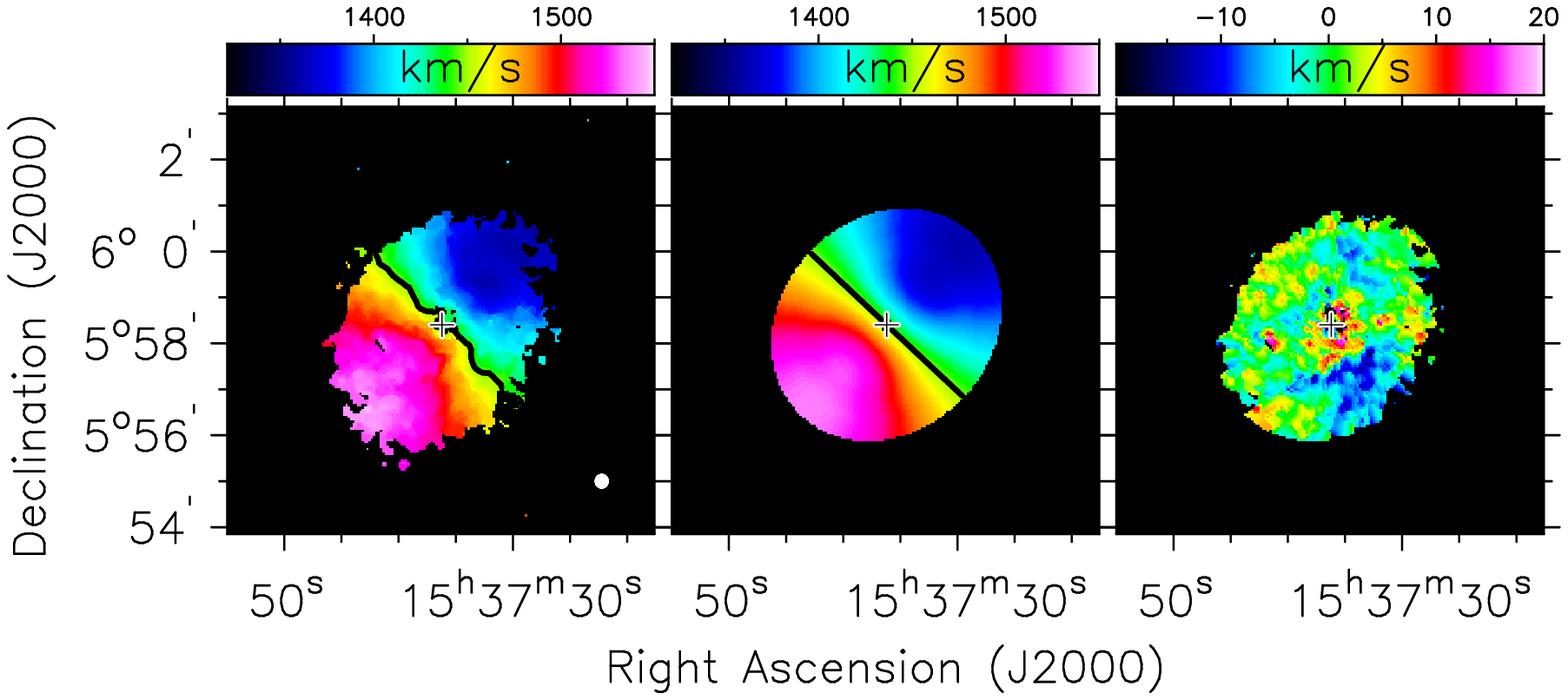} \\
\\
\includegraphics[width=\textwidth]{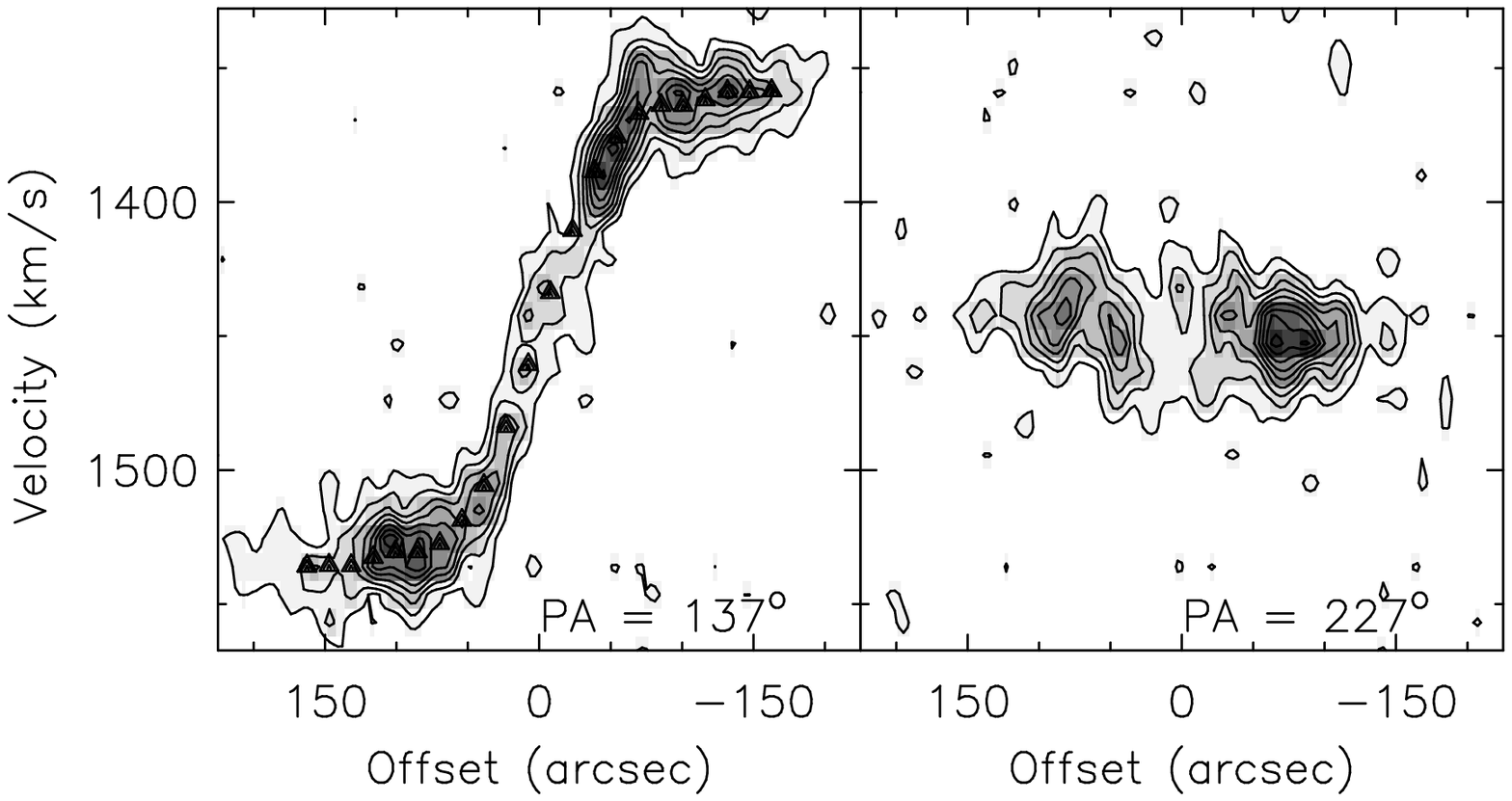} 
\end{array}$
\end{center}
\caption{As in Figure~\ref{fig:ngc0337_vel}, but for NGC~5964.  The
position-velocity diagram contours are from 1~${\rm mJy \, beam^{-1}}$
to 10~${\rm mJy \, beam^{-1}}$ in steps of 1~${\rm mJy \,
beam^{-1}}$.}
\label{fig:ngc5964_vel}
\end{figure}

\begin{figure}
\begin{center}$
\begin{array}{c}
\includegraphics[width=\textwidth]{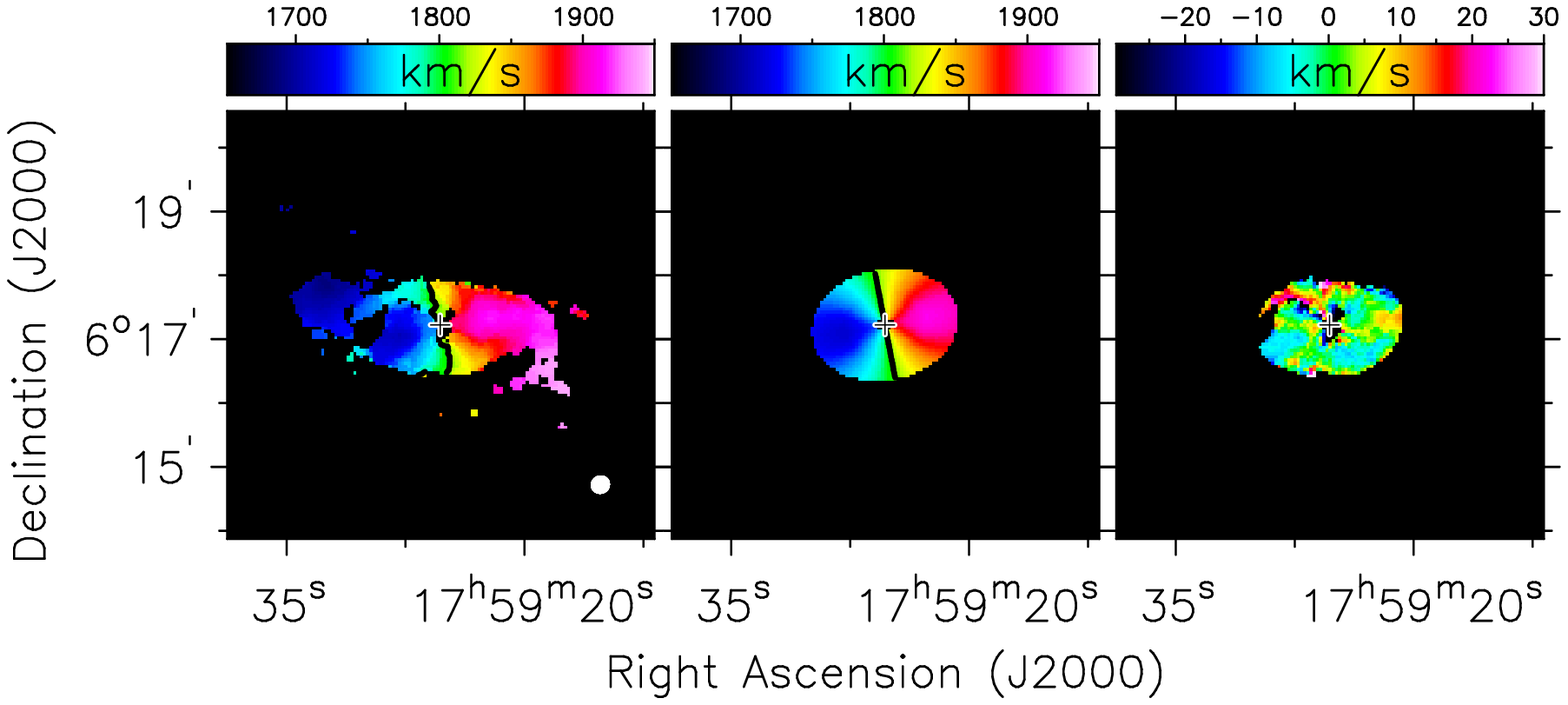} \\
\\
\includegraphics[width=\textwidth]{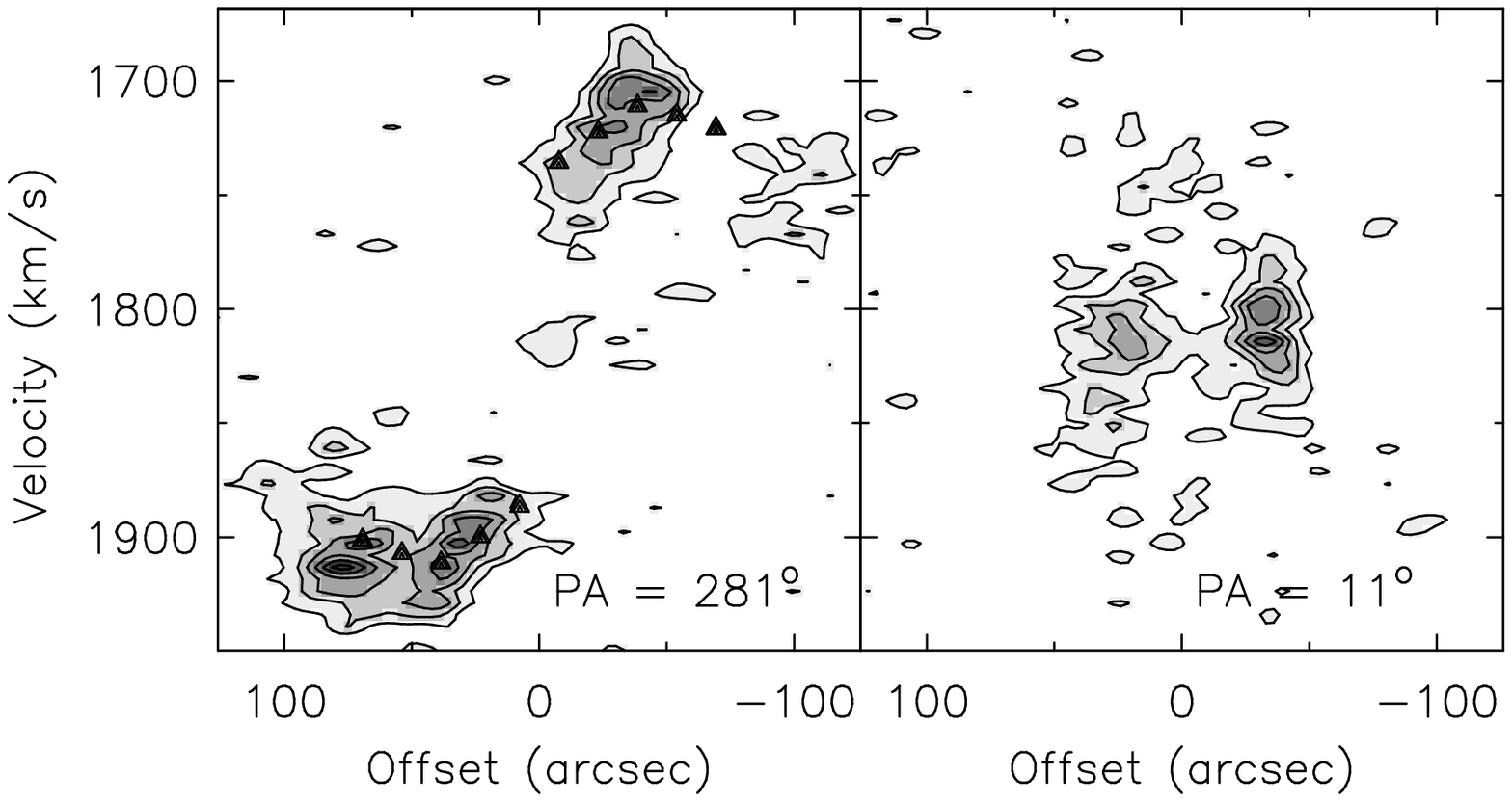} 
\end{array}$
\end{center}
\caption{As in Figure~\ref{fig:ngc0337_vel}, but for NGC~6509. The
position-velocity diagram contours are from 2~${\rm mJy \, beam^{-1}}$
to 15~${\rm mJy \, beam^{-1}}$ in steps of 2~${\rm mJy \,
beam^{-1}}$.}
\label{fig:ngc6509_vel}
\end{figure}

\begin{figure}
\begin{center}$
\begin{array}{c}
\includegraphics[width=\textwidth]{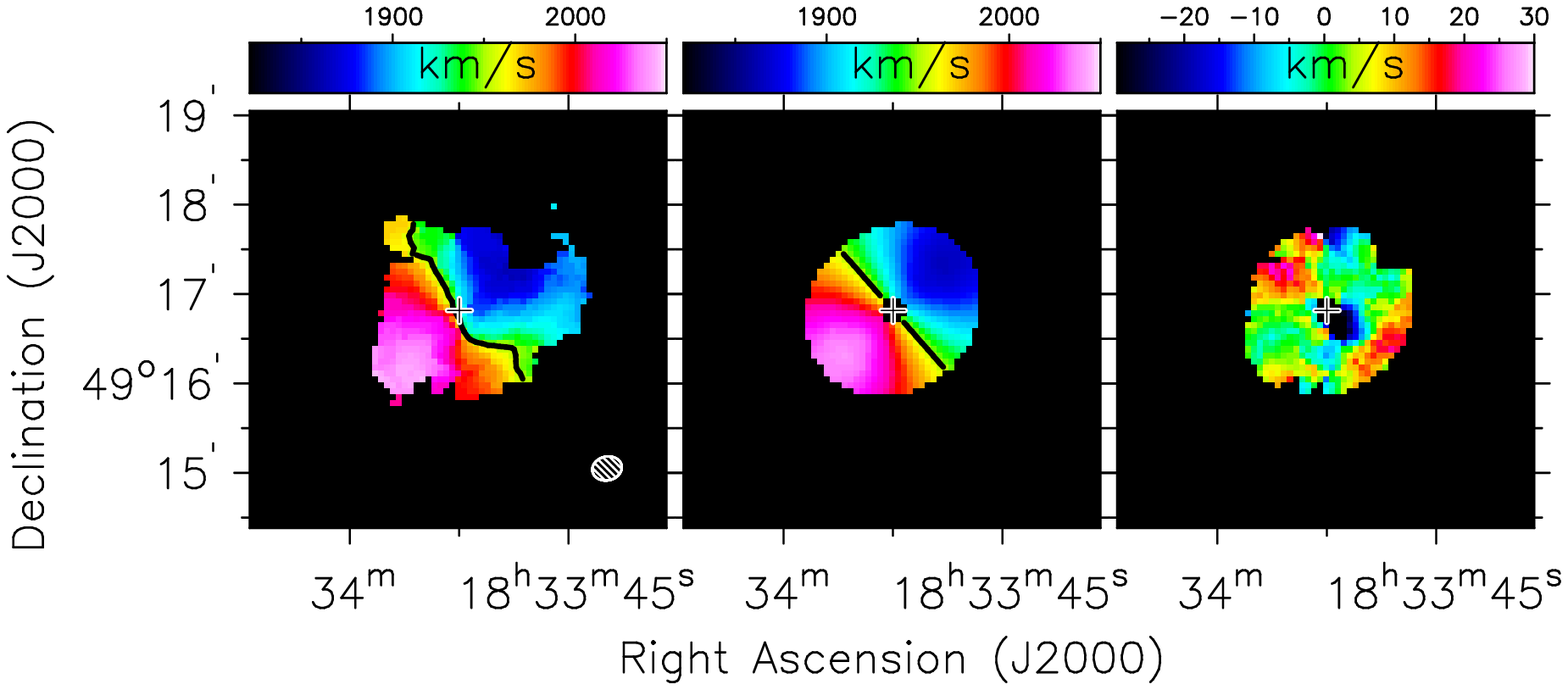} \\
\\
\includegraphics[width=\textwidth]{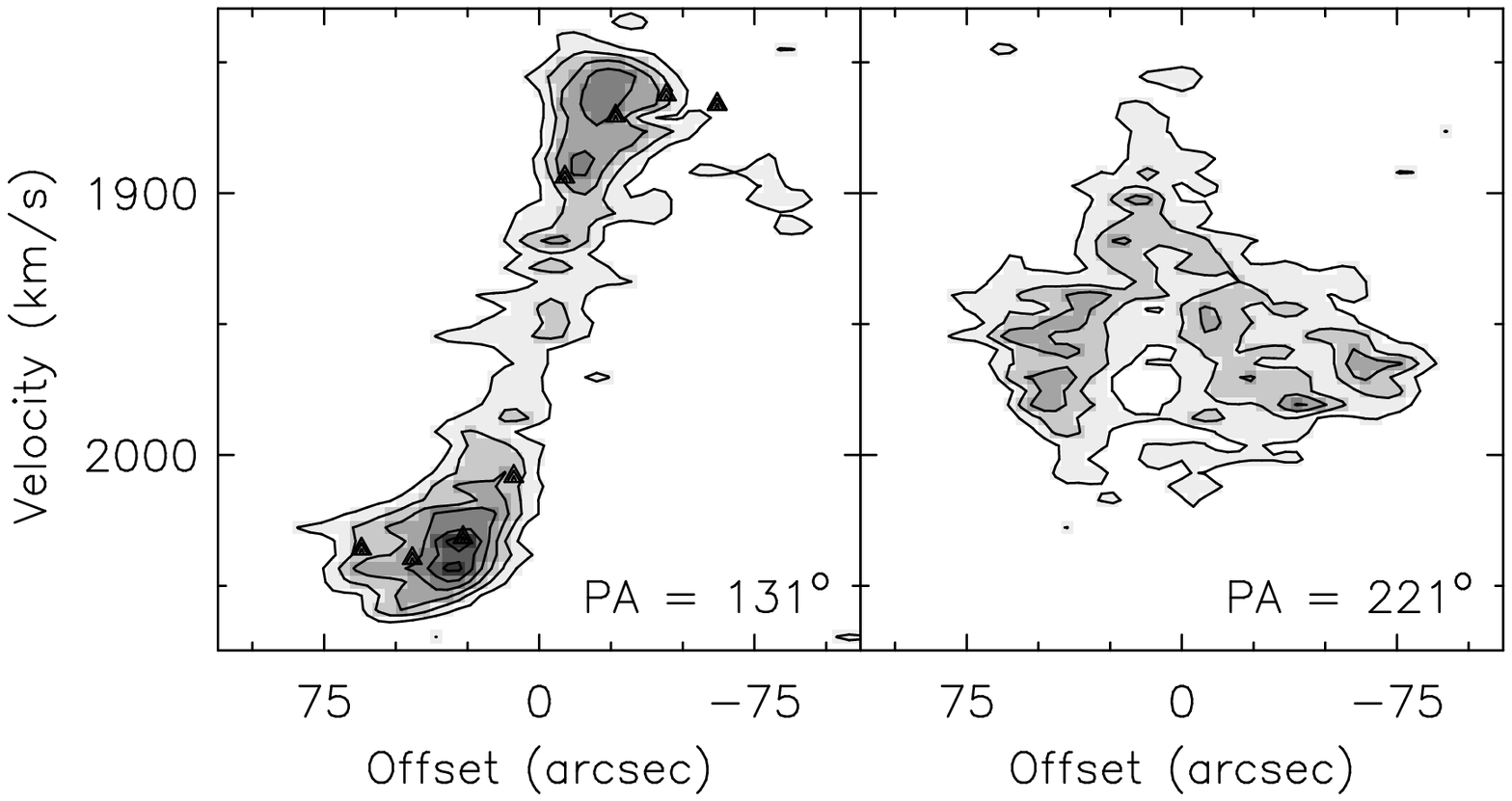} 
\end{array}$
\end{center}
\caption{As in Figure~\ref{fig:ngc0337_vel}, but for IC~1291.  The
position-velocity diagram contours are from 2~${\rm mJy \, beam^{-1}}$
to 15~${\rm mJy \, beam^{-1}}$ in steps of 2~${\rm mJy \,
beam^{-1}}$.}
\label{fig:ic1291_vel}
\end{figure}

Figures~\ref{fig:ngc0337_vel}-\ref{fig:ic1291_vel} show the velocity
fields and position-velocity diagrams for the sample.  In each figure,
the top row, left panel shows the first-moment map, created using the
method described above in this section and covering the same area as
the channel maps for the object.  The middle panel shows the tilted
ring model velocity field and the right panel shows the residual
velocity map -- the first-moment map minus the model velocity field.
The dynamical center (from Columns 5 and 6 in
Table~\ref{tab:final_params}) is shown as a cross in all three panels
and the systemic velocity (from Column 2 of
Table~\ref{tab:final_params}) is shown as a thick black contour in the
left and middle panels.  The bottom row of each figure shows the
position-velocity diagram along the major axis in the left panel and
along the minor axis in the right panel.  These figures display data
extracted from a slice that is a single spatial pixel in width (see
Column 8 of Table~\ref{tab:im_summary} for pixel size in arcseconds).
The contours start at twice the image noise, increase in intervals of
twice the image noise, and end at the maximum surface brightness along
the major axis slice plus that interval.  We give the specific range
in each figure.  For comparison, the projected tilted ring model
values along the major axis are overplotted as solid triangles in the
left panel.

We are particularly interested in deriving accurate circular
velocities from our rotation curve analysis because in future work we
will study star formation and secular evolution above and below the
dust scale height transition velocity of ${\rm v_{circ}} = 120 \, {\rm
km \, s^{-1}}$, which was observed in \citet{dalcanton04}.  We take
the maximum velocity from the deprojected rotation curve as a measure
of the circular velocity.  These values are listed in Column 8 of
Table~\ref{tab:final_params}.  \citet{dalcanton02} primarily measured
the circular velocity as half $W_{50}$ (the line width at 50\% of the
peak flux density) from single-dish \ion{H}{1} data.  For about one
quarter of their sample, they used the maximum velocity in long-slit
H$\alpha$ rotation curves.  No inclination correction was necessary
because they studied edge-on disks.  The maximum velocity from the
rotation curve is not an optimal measure of the circular velocity
because it is sensitive to details of the rotation curve shape
\citep{persic96,kannappan02}.  Alternative methods are used in
Tully-Fisher relation studies, which often evaluate a parametrization
of the rotation curve at a relatively large radius \citep[see,
e.g.,][]{courteau97,kannappan02,pizagno07}.  The form of the
parametrization and the radius at which the fit is evaluated vary.
For simplicity and for consistency with \citet{dalcanton02}, we used
the maximum velocity from the rotation curve as a measure of the
circular velocity.  As a check, we compared our circular velocities to
values estimated as $W_{50}/2/{\rm sin}(i)$, where we measured
$W_{50}$ from our integrated \ion{H}{1} line profiles and $i$ is from
Column~4 of Table~\ref{tab:final_params}, and found good agreement.

We calculated the errors listed in Table~\ref{tab:final_params} in the
following ways.  The error on an IRAC parameter is the standard
deviation of the parameter values for the outer few isophotes of the
ellipse fits.  The error on the systemic velocity, center, PA, or
inclination determined from the velocity field modeling is the
standard deviation of the individual ring values about the weighted
mean, divided by the square-root of the number of rings.  We excluded
the first and last ring in this calculation.  We chose this error
estimate because it is most well-matched with the IRAC errors.  Note
that the ESO~501-G023 inclination is from RC3, but RC3 does not give
an error, so the inclination error is the value from the velocity
field modeling.  The final inclination for NGC~4561 and PA for
PGC~6667 are from the IRAC analysis, but we quote errors from the
velocity field modeling because the outer few isophotes in the IRAC
data have the same values, which gives a standard deviation of zero.
The systemic velocity for UGC~1862 is from a single ring and the error
is just the ROTCUR formal least squares error for that ring.

The circular velocity errors in Column~8 of
Table~\ref{tab:final_params} are propagated inclination errors.
Specifically, we ran ROTCUR using our final inclination plus and minus
the inclination error (from Column~4 of Table~\ref{tab:final_params}),
and the circular velocity errors are the deviation of the maximum
rotation velocity in these runs from the nominal value.  These errors
are a measure of the uncertainty associated with converting the
measured velocity into a deprojected value.  We do not include the
formal least squares error returned by ROTCUR, which is a measure of
the error in the observed rotation velocity.  The formal least squares
error returned by ROTCUR accounts for correlations between parameters
that are allowed to vary as a function of radius.  Because we hold the
systemic velocity, PA, center, and particularly the inclination fixed
for all annuli when calculating the final rotation curve, the output
formal least squares error on the rotation velocities are
underestimates.  These values, albeit underestimated, are typically
significantly smaller than the propagated inclination errors.  There
is also an uncertainty associated with whether the maximum velocity is
a good measure of the circular velocity, which is typically the case
when a galaxy rotation curve shows a clear turnover. Six of our
galaxies do not fall in this category.  To address this concern, we
assigned each galaxy a rotation curve quality assessment, listed in
Column 9 of Table~\ref{tab:final_params}.  Rankings A, B, and C refer
to galaxies with clear, moderate, and no rotation curve turnovers,
respectively.

\section{Morphology and Evidence for Interactions}
\label{sec:morph}

\begin{figure}
\begin{center}$
\begin{array}{cc}
\includegraphics[width=0.31\textwidth]{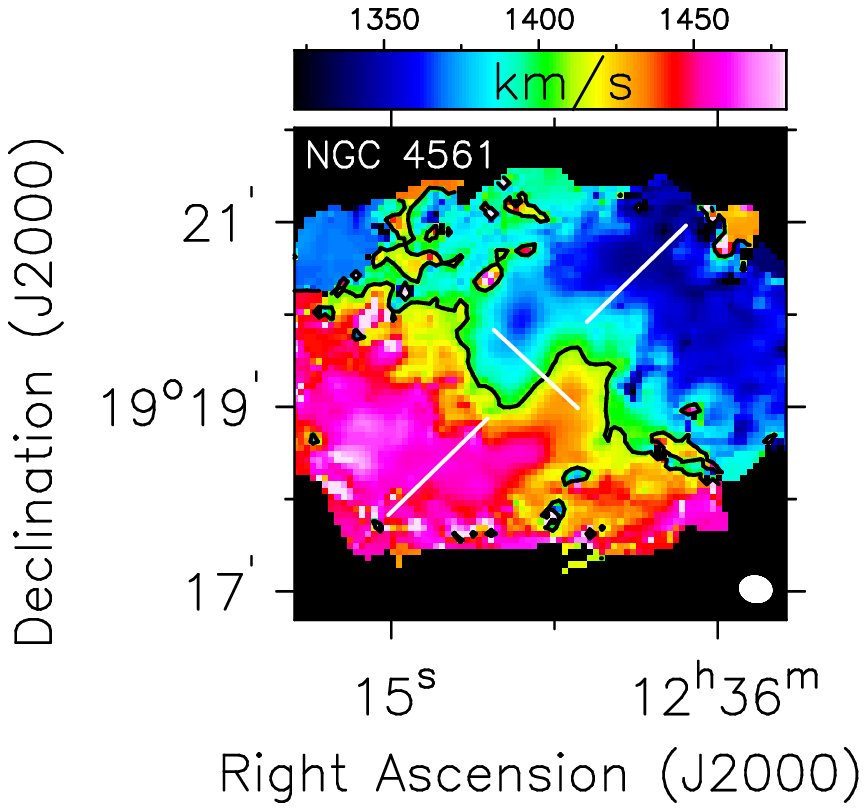} & 
\includegraphics[width=0.51\textwidth]{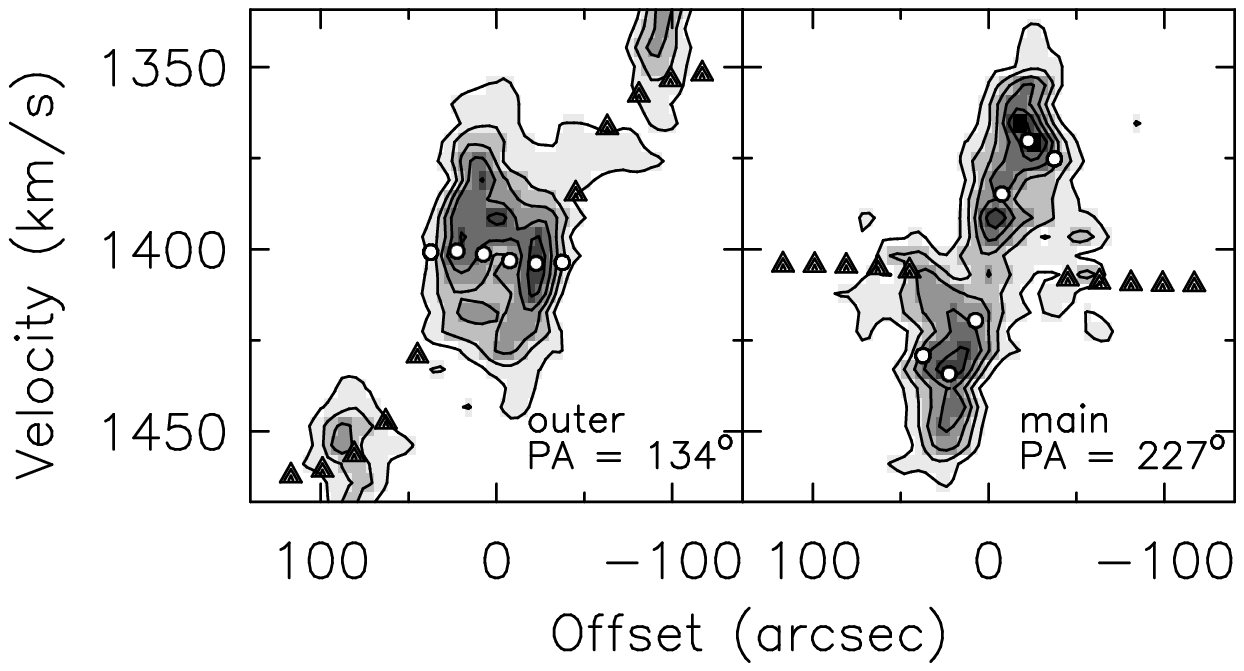} \\ [0.4cm]
\includegraphics[width=0.31\textwidth]{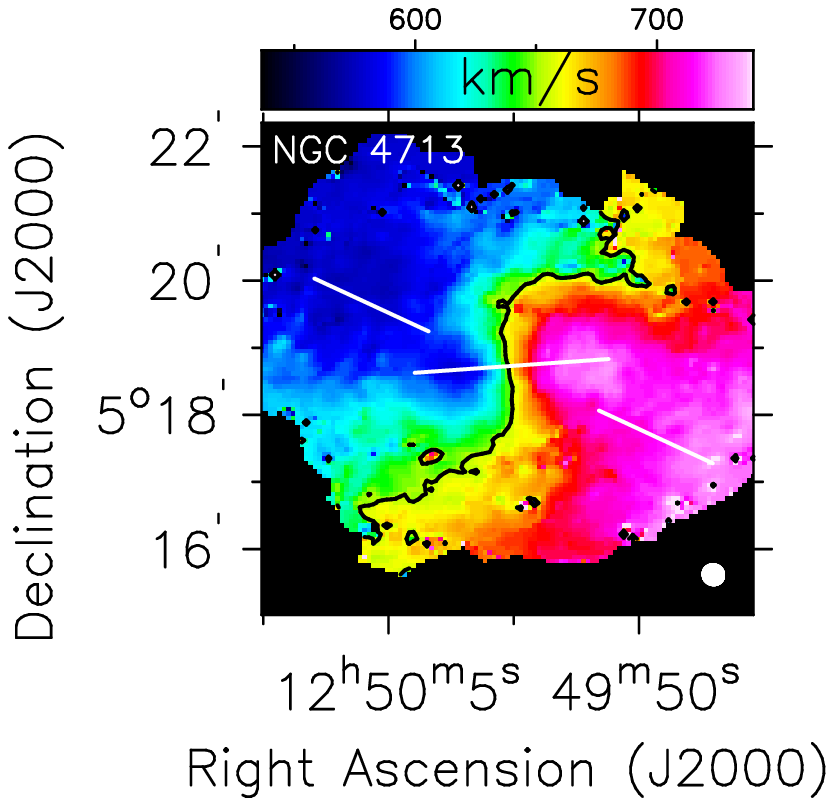} &
\includegraphics[width=0.51\textwidth]{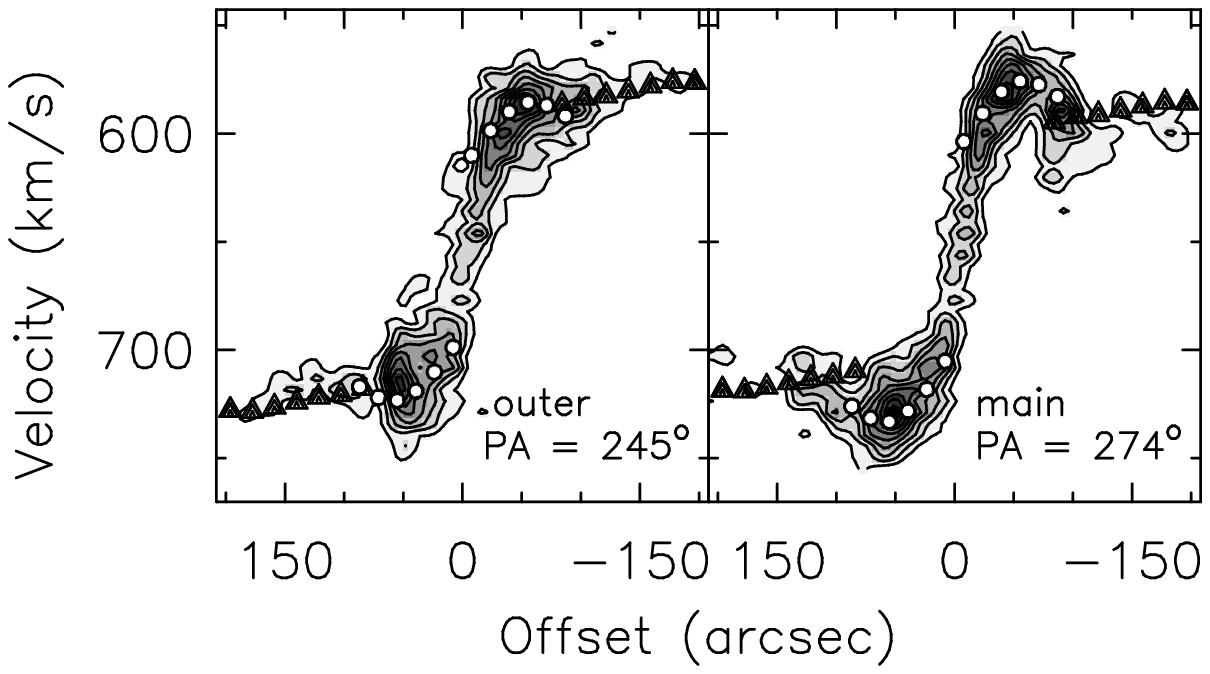} \\ [0.4cm]
\includegraphics[width=0.31\textwidth]{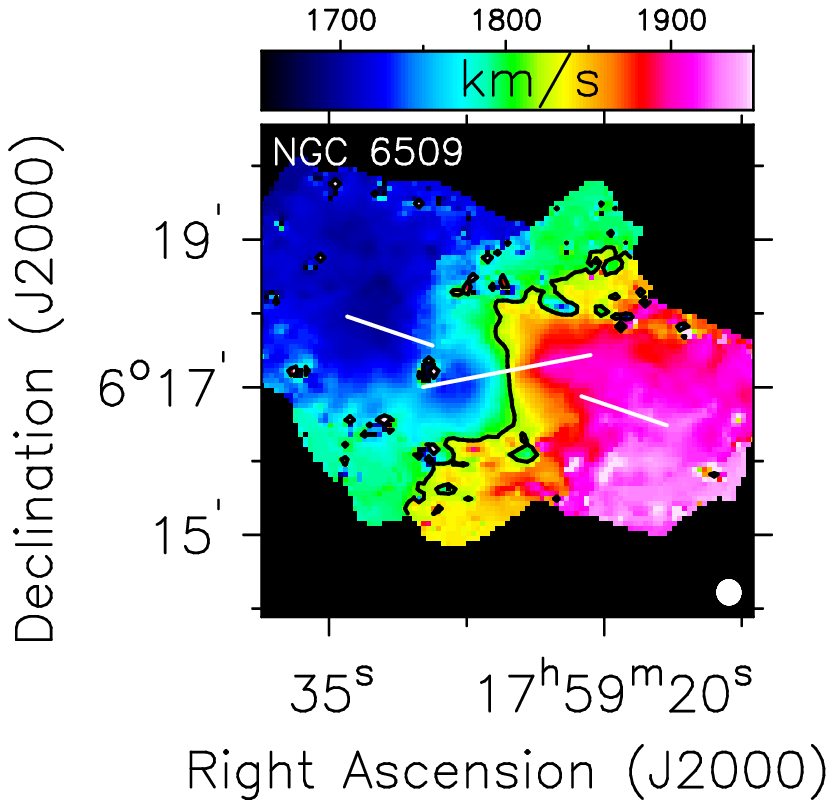} &
\includegraphics[width=0.51\textwidth]{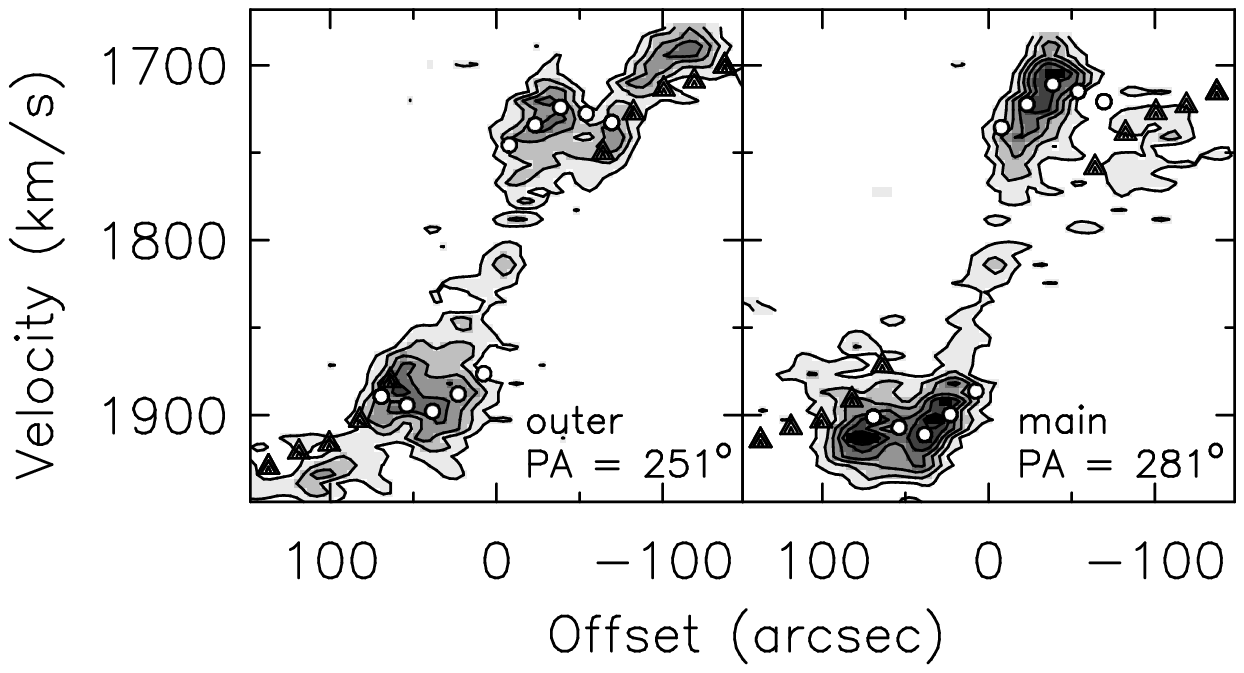} \\
\end{array}$
\end{center}
\caption{First-moment map ({\it left}) and position-velocity diagrams
  ({\it right}) showing the main and outer kinematic components in
  NGC~4561 ({\it top row}), NGC~4713 ({\it middle row}), and NGC~6509
  ({\it bottom row}).  These figures were created from the
  naturally-weighted data cubes that were blanked using the method
  described in Section~~\ref{sec:dp1}.  In the first-moment maps, the
  beam size is shown in the lower right corner, the black contour
  shows the systemic velocity of the outer component, and the white
  lines lie along the major axis of the main and outer components,
  over the range used in the rotation curve fits.  The {\it middle}
  ({\it right}) {\it panel} of each row shows the position-velocity
  diagram along the major axis of the outer (main) component.
  Contours begin at $2\sigma$ and end at the maximum surface
  brightness along the major axis of the outer component plus
  $2\sigma$, in steps of $2\sigma$, where $\sigma$ is the image noise.
  The projected titled ring model fit for the outer (main) component
  is overplotted as {\it black triangles} ({\it white circles}). }
\label{fig:outer_disk}
\end{figure}

Three of our objects have two kinematic components: NGC~4561,
NGC~4713, and NGC~6509.  The main component covers most of or more
than the optical disk and the outer component is characterized by low
surface brightness emission in the outer disk.  We carried out the
velocity field modeling separately for the two
components. Table~\ref{tab:final_params} quotes only the kinematic
parameters determined for the main component, as it corresponds to the
optical disk.  For the outer component fit, we used the
naturally-weighted data cubes, blanked using the method described in
Section~\ref{sec:dp1}, which better probes the low surface brightness
emission.  The left panel of the top, middle, and bottom rows in
Figure~\ref{fig:outer_disk} shows the main and outer components in the
first-moment map of NGC~4561, NGC~4713, and NGC~6509, respectively.
The position angle of the major axis of the outer component differs by
between about $30\degr$ and $90\degr$ from that of the main component.

The middle panel of each row shows the position-velocity diagram along
the major axis of the outer component while the right panel of each
row shows the position-velocity diagram along the major axis of the
main component.  The filled triangles and circles show the projected
tilted ring model values for the outer and main components,
respectively.  Although much of the emission in the first-moment maps
is of low significance, the position-velocity diagrams show that there
is significant emission in the outer-disk components.  These outer
components are likely due to a warp, which may indicate that these
galaxies have undergone a recent interaction.

We detected a companion(s) to NGC~0337, NGC~2805, ESO~501-G023,
NGC~3906, and IC~1291 within the $\sim 30\arcmin$ field of view and
$\sim 500 \, {\rm km \, s^{-1}}$ bandwidth of our final \ion{H}{1}
data cubes.  These objects may deserve further investigation into
whether the galaxies are interacting with the companions.  However, we
see no obvious evidence of bridges or tidal tails in any of the data
to suggest an ongoing interaction.

NGC~6509 shows \ion{H}{1} in absorption on the east side of the galaxy
because it is in the foreground of the radio source 4C~+06.63, which
is the northwest component of a pair of radio lobes, the southeast
component being VLSS~J1759.5+0615.  4C~+06.63 is centered at
approximately RA 17:59:29.6 and DEC 06:17:13 (J2000), about
$65\arcsec$ east of the galaxy center.  It is slightly resolved, with
a FWHM of $20\arcsec$ in a continuum image with a beam size of
$15.97\arcsec \times 14.76\arcsec$.  The full extent of the source is
approximately $1\arcmin$.  Because of this absorption, the \ion{H}{1}
mass in NGC~6509 is a lower limit.  We made no adjustments to the
rotation curve analysis for this object.  Especially note that we
still used both the approaching and receding side of the galaxy,
including about $25\degr$ where absorption is evident in position
velocity diagrams.  Even so, the rotation curve model is acceptable.

\section{Summary}
\label{sec:summary}

We have presented \ion{H}{1} channel maps, line profiles, and
integrated intensity maps for a sample of 20 nearby, moderately
inclined, bulgeless disk galaxies, which were selected to bracket the
circular velocity transition for dust scale heights found in
\citet{dalcanton04}.  We have also calculated \ion{H}{1} fluxes and
\ion{H}{1} masses and have carried out rotation curve fitting to
derive kinematic properties, in particular the circular velocity, for
each galaxy.  We found that three objects in our sample have two
kinematic components, which may indicate a recent interaction.  The
remaining objects appear undisturbed, which makes them well suited for
studying the internal evolution of galaxies with quiescent merger
histories.  We will use the data presented here, as well as data from
the IRAM $30 \, {\rm m}$, MDM, the {\it Spitzer Space Telescope}, and
{\it HST} in studies of star formation, dust properties, and secular
evolution in bulgeless disk galaxies as a function of circular
velocity.

\acknowledgements 

We thank the LITTLE THINGS group, especially Dana Ficut-Vicas and
Elias Brinks for generously providing us with their data reduction
recipe for VLA-EVLA transition data.  We are also grateful to
Sebastian Haan and Fabian Walter for advice on the data reductions, to
Eric Greisen for help with AIPS, and to the VLA for rescheduling
observations of UGC~6930 and NGC~4519.  We thank the referee for
helpful comments and the MPIA for hospitality and support during
several productive visits.  L.C.W. gratefully acknowledges support
from an NSF Graduate Research Fellowship.  P.M. is grateful for
support from the NSF via award AST-0705170.  U.L. acknowledges
financial support from the research project AYA2007-67625-C02-02 from
the Spanish Ministerio de Ciencia y Educaci\'on and from the Junta de
Andaluc\'\i a.

\clearpage

\begin{deluxetable}{lccccccc}
\tablewidth{0pt}
\tabletypesize{\scriptsize}
\tablecaption{General Galaxy Properties}
\tablehead{
\colhead{Source} &
\colhead{RA (J2000.0)} &
\colhead{DEC (J2000.0)} &
\colhead{D} &
\colhead{$D_{25}$} &
\colhead{$m_{B}$} &
\colhead{$M_{B}$} &
\colhead{Type} \\
\colhead{} &
\colhead{(hh:mm:ss.s)} &
\colhead{(dd:mm:ss)} &
\colhead{(Mpc)} &
\colhead{(arcsec)} &
\colhead{(mag)} &
\colhead{(mag)} &
\colhead{} \\
\colhead{(1)} &
\colhead{(2)} &
\colhead{(3)} &
\colhead{(4)} &
\colhead{(5)} &
\colhead{(6)} &
\colhead{(7)} &
\colhead{(8)}
}
\startdata

NGC~0337      &  00:59:50.0   &   -07:34:41	&  20.7 [T88]	&  173  & 11.44  & -20.14 & 7.0 \\
PGC~3853      &  01:05:04.8   &   -06:12:46	&  11.4 [T08]	&  250  & 11.98  & -18.30 & 7.0 \\
PGC~6667      &  01:49:10.3   &   -10:03:45    &  24.6 [T88]   &  173  &  12.92  & -19.03 & 6.7       \\
ESO~544-G030  &  02:14:57.2   &   -20:12:40    &  13.9 [T08]   &  123  &  13.25  & -17.47 & 7.7        \\
UGC~1862      &  02:24:24.8   &   -02:09:41    &  22.3 [T08]   &  99.6 &  13.47  & -18.27 & 7.0        \\
ESO~418-G008  &  03:31:30.8   &   -30:12:46    &  23.6 [T08]   &  70.5 &  13.65  & -18.21 & 8.0        \\
ESO~555-G027  &  06:03:36.6   &   -20:39:17    &  24.3 [T88]   &  138  &  13.18  & -18.75 & 7.0        \\
NGC~2805      &  09:20:20.4   &   +64:06:12    &  28.0 [T88]   &  379  &  11.17  & -21.07 & 7.0        \\
ESO~501-G023  &  10:35:23.6   &   -24:45:21    &  7.01 [T08]   &  208  &  12.86  & -16.37 & 8.0       \\
UGC~6446      &  11:26:40.6   &   +53:44:58    &  18.0 [T08]   &  213  &  13.30  & -17.98 & 7.0       \\
NGC~3794      &  11:40:54.8   &   +56:12:10    &  19.2 [T08]   &  134  &  13.23  & -18.19 & 6.5        \\
NGC~3906      &  11:49:40.2   &   +48:25:30    &  18.3 [...]   &  112  &  13.50  & -17.81 & 7.0        \\
UGC~6930      &  11:57:17.2   &   +49:17:08    &  17.0 [T88]   &  262  &  12.38  & -18.77 & 7.0        \\
NGC~4519      &  12:33:30.5   &   +08:39:16    &  19.6 [T08]   &  190  &  12.15  & -19.31 & 7.0        \\
NGC~4561      &  12:36:08.6   &   +19:19:26    &  12.3 [T88]   &  90.8 &  12.82  & -17.63 & 8.0        \\
NGC~4713      &  12:49:58.1   &   +05:18:39    &  14.9 [T08]   &  162  &  11.85  & -19.02 & 7.0        \\
NGC~4942      &  13:04:19.2   &   -07:39:00    &  28.5 [T88]   &  112  &  13.27  & -19.00 & 7.0        \\
NGC~5964      &  15:37:36.3   &   +05:58:28    &  24.7 [T88]   &  250  &  12.28  & -19.68 & 7.0        \\
NGC~6509      &  17:59:24.9   &   +06:17:12    &  28.2 [T88]   &  95.1 &  12.12  & -20.13 & 7.0        \\
IC~1291	      &  18:33:51.5   &   +49:16:45    &  31.5 [T88]   &  109  &  13.28  & -19.21 & 8.0        \\

\enddata

\tablecomments{Column 1: Object name; Column 2 and 3: Right ascension
  (RA) and declination (DEC) from \citet{RC3}; Column 4: Distance and
  distance reference.  Distances are derived using the Tully-Fisher
  relation, except for NGC~3906.  T08: \citet{tully08}, T88:
  \citet{tully88}, and the NGC~3906 distance is from the \citet{RC3}
  heliocentric velocity, corrected for Virgo infall using
  \citet{mould00} and using $H_{0} = 71 \, {\rm km \, s^{-1} \,
  Mpc^{-1}}$.  Note that the NGC~2805 distance may be overestimated.
  See Section~\ref{sec:lineprof} for details. Column 5: Major
  isophotal diameter at $25 \, {\rm mag \, arcsec^{-2}}$ in the B
  band, from \citet{RC3}.  Column 6: Apparent blue magnitude,
  corrected for Galactic and internal extinction and redshift.  Values
  are from \citet{RC3}, except for NGC~4942 and PGC~6667, which are
  from \citet{doyle05} and are only corrected for Galactic extinction.
  Column 7: Absolute blue magnitude, calculated from the apparent
  magnitude in column 6 and the distance in column 4.  Column 8:
  Morphological type from \citet{RC3}.}

\label{tab:sample}
\end{deluxetable}

\clearpage

\begin{landscape}
\begin{deluxetable}{lcccccccccccccc}
\tablewidth{0pt}
\tabletypesize{\scriptsize}
\tablecaption{Summary of Observations}
\tablehead{
\colhead{Source} &
\colhead{Conf.} &
\colhead{Date} &
\colhead{Start} &
\colhead{End} &
\colhead{On-source} &
\colhead{Mode} &
\colhead{Chan.} &
\colhead{${\rm \Delta \nu}$} &
\colhead{${\rm \Delta v}$} &
\colhead{${\rm v_{hel,1}}$} &
\colhead{${\rm v_{hel,2}}$} &
\colhead{Flux} &
\colhead{Phase} &
\colhead{Flux Density} \\
\colhead{} &
\colhead{} &
\colhead{(yyyy-mm-dd)} &
\colhead{(hh:mm)} &
\colhead{(hh:mm)} &
\colhead{(hh:mm)} &
\colhead{} &
\colhead{(No.)} &
\colhead{(kHz)} &
\colhead{(${\rm km \, s^{-1}}$)} &
\colhead{(${\rm km \, s^{-1}}$)} &
\colhead{(${\rm km \, s^{-1}}$)} &
\colhead{Calibrator} &
\colhead{Calibrator} &
\colhead{(Jy)} \\
\colhead{(1)} &
\colhead{(2)} &
\colhead{(3)} &
\colhead{(4)} &
\colhead{(5)} &
\colhead{(6)} &
\colhead{(7)} &
\colhead{(8)} &
\colhead{(9)} &
\colhead{(10)} &
\colhead{(11)} &
\colhead{(12)} &
\colhead{(13)} &
\colhead{(14)} &
\colhead{(15)} 
}
\startdata
NGC~0337     &	C    &	 2006-11-04  &	  00:23 &  05:26  &   01:54 &  4    &	 64   &   24.41 &  5.21  &  1520.4	 &   1770.7	 &    3C48  	 &     0059+001   &	$2.318  \pm 0.007 $  \\
PGC~3853     &	C    &	 2006-11-04  &	  00:39 &  05:29  &   02:00 &  4    &	 64   &   24.41 &  5.20  &  968.4	 &   1217.8	 &    3C48  	 &     0059+001   &	$2.316  \pm 0.007 $  \\
PGC~6667     &	C    &	 2006-11-04  &	  05:30 &  07:44  &   01:57 &  4    &	 64   &   24.41 &  5.23  &  1857.4	 &   2108.2	 &    3C48  	 &     0157-107   &	$2.000  \pm 0.010 $  \\
ESO~544-G030 &	CnB  &	 2006-10-08  &	  07:08 &  10:33  &   02:59 &  4    &	 64   &   24.41 &  5.21  &  1487.4	 &   1737.6	 &    3C48  	 &     0240-231   &	$6.13	\pm 0.02  $  \\
	     &	CnB  &	 2006-10-12  &	  06:57 &  10:26  &   02:59 &  4    &	 64   &   24.41 &  5.21  &  1487.4	 &   1737.6	 &    3C48  	 &     0240-231   &	$6.025  \pm 0.011 $  \\
UGC~1862     &	C    &	 2006-11-04  &	  07:46 &  10:03  &   01:58 &  4    &	 64   &   24.41 &  5.20  &  1257.4	 &   1507.2	 &    3C48  	 &     0231+133   &	$1.511  \pm 0.002 $  \\
ESO~418-G008 &	CnB  &	 2006-10-10  &	  07:31 &  10:56  &   02:57 &  4    &	 64   &   24.41 &  5.20  &  1067.4	 &   1316.9	 &    3C48  	 &     0416-209   &	$2.576  \pm 0.010 $  \\
	     &	CnB  &	 2006-10-15  &	  08:17 &  11:06  &   02:35 &  4    &	 64   &   24.41 &  5.20  &  1067.4	 &   1316.9	 &    ...	&     0416-209   &	$2.58	  $  \\
ESO~555-G027 &	CnB  &	 2006-10-12  &	  10:26 &  14:00  &   03:04 &  4    &	 64   &   24.41 &  5.23  &  1855.4	 &   2106.2	 &    3C48  	 &     0609-157   &	$2.848  \pm 0.010 $  \\
NGC~2805     &	C    &	 2008-05-14  &	  01:51 &  04:12  &   01:59 &  4    &	 64   &   24.41 &  5.21  &  1617.7	 &   1867.7	 &    3C286  	 &     0841+708   &	$3.278  \pm 0.012 $  \\
ESO~501-G023 &	CnB  &   2006-10-12  &    14:02 &  17:17  &   03:00 &  4    &	 64   &   24.41 &  5.19  &  914.4	 &   1163.7	 &    3C48  	 &     1057-245   &	$1.050  \pm 0.003 $  \\
UGC~6446     &	B    &	 2002-06-22  &	  19:10 &  07:06  &   09:40 &  2AD  &	 128  &   12.21 &  2.59  &  643.7	 &   ...	 &    3C286  	 &     1035+564   &	$1.816  \pm 0.004 $  \\
	     &	C    &	 2002-11-18  &	  11:39 &  17:51  &   05:05 &  2AD  &	 128  &   12.21 &  2.59  &  643.7	 &   ... 	 &    3C286  	 &     1035+564   &	$1.788  \pm 0.003 $  \\
NGC~3794     &	C    &	 2008-05-17  &	  01:40 &  04:00  &   01:58 &  4    &	 64   &   24.41 &  5.20  &  1254.7	 &   1504.1	 &    3C286  	 &     1035+564   &	$1.838  \pm 0.010 $  \\
NGC~3906     &	C    &	 2008-05-16  &	  01:44 &  04:04  &   01:59 &  4    &	 64   &   24.41 &  5.18  &  832.7	 &   1081.4	 &    3C286  	 &     1219+484   &	$0.602  \pm 0.003 $  \\
UGC~6930     &	C    &	 2009-07-30  &	  23:37 &  02:05  &   01:57 &  4    &	 64   &   24.41 &  5.18  &  647.5	 &   896.2	 &    3C286  	 &     1219+484   &	$0.656  \pm 0.004 $  \\
NGC~4519     &	C    &	 2009-08-05  &	  00:15 &  02:46  &   02:00 &  4    &	 64   &   24.41 &  5.19  &  1097.1	 &   1346.2	 &    3C286  	 &     1254+116   &	$0.765  \pm 0.004 $  \\
NGC~4561     &	C    &	 2008-05-14  &	  04:13 &  06:37  &   01:59 &  4    &	 64   &   24.41 &  5.20  &  1279.6	 &   1529.1	 &    3C286  	 &     1254+116   &	$0.771  \pm 0.007 $  \\
NGC~4713     &	C    &	 2008-05-16  &	  04:04 &  06:28  &   01:59 &  4    &	 64   &   24.41 &  5.17  &  524.6	 &   772.8	 &    3C286  	 &     1254+116   &	$0.761  \pm 0.006 $  \\
NGC~4942     &	C    &	 2008-05-17  &	  04:00 &  06:24  &   01:58 &  4    &	 64   &   24.41 &  5.21  &  1623.6	 &   1873.7	 &    3C286  	 &     1246-075   &	$0.495  \pm 0.003 $  \\
NGC~5964     &	C    &	 2001-08-20  &	  20:45 &  04:31  &   06:21 &  4    &	 64   &   24.41 &  5.21  &  1312.1	 &   1582.9	 &    3C286  	 &     1557-000   &	$0.6275 \pm 0.0013$  \\
NGC~6509     &	C    &	 2008-05-13  &	  08:54 &  11:14  &   01:58 &  4    &	 64   &   24.41 &  5.21  &  1686.5	 &   1936.6	 &    3C286,3C48  &     1751+096   &	$1.373  \pm 0.007 $  \\
IC~1291	     &	C    &	 2008-05-13  &	  11:14 &  13:38  &   01:58 &  4    &	 64   &   24.41 &  5.22  &  1858.2	 &   2108.6	 &    3C286,3C48  &     1845+401   &	$0.941  \pm 0.003 $  \\
\enddata

\tablecomments{Column 1: Object name; Column 2: VLA Configuration;
  Column 3: Date observed in International Atomic Time (IAT); Column
  4: Start time of observation (IAT); Column 5: End time of
  observation (IAT); Column 6: Time spent on source; Column 7:
  Correlator mode; Column 8: Number of channels per IF; Column 9:
  Channel width in kHz; Column 10: Channel width in ${\rm km \,
  s^{-1}}$; Column 11: Central heliocentric velocity of IF1; Column
  12: Central heliocentric velocity of IF2; Column 13: Flux
  calibrator: 3C48 and 3C286 are also known as 0137+331 1331+305 in
  J2000, respectively; Column 14: Phase calibrator (J2000); Column 15:
  Flux density of phase calibrator, from GETJY on IF1.  Note that the
  observations of ESO~418-G008 on 2006 October 15 were made without a
  primary calibrator.  We set the flux density of the phase calibrator
  to the value from 2006 October 10.  \\ }

\label{tab:obs_summary}
\end{deluxetable}
\clearpage
\end{landscape}

\clearpage

\begin{deluxetable}{lcccccccc}
\tablewidth{0pt}
\tabletypesize{\scriptsize}
\tablecaption{Summary of Data Cube Properties}
\tablehead{
\colhead{Source} &
\colhead{Weighting} &
\colhead{$B_{\rm maj}$} &
\colhead{$B_{\rm min}$} &
\colhead{PA} &
\colhead{Noise} &
\colhead{Image Size} &
\colhead{Pixel Size} &
\colhead{Channel Width} \\
\colhead{} &
\colhead{} &
\colhead{(arcsec)} &
\colhead{(arcsec)} &
\colhead{($\degr$)} &
\colhead{(${\rm mJy \, beam^{-1}}$)} &
\colhead{(pix)} &
\colhead{(arcsec)} &
\colhead{(${\rm km \, s^{-1}}$)} \\
\colhead{(1)} &
\colhead{(2)} &
\colhead{(3)} &
\colhead{(4)} &
\colhead{(5)} &
\colhead{(6)} &
\colhead{(7)} &
\colhead{(8)} &
\colhead{(9)}
}
\startdata
NGC~0337      &  NA   &   25.77 &  15.06 &  -18.30 & 1.7 & 512  &   4.0  &   5.2    \\
	      &  RO   &   21.23 &  13.58 &  -15.37 & 1.1 & 512  &   3.0  &   10.4   \\
PGC~3853      &  NA   &   25.33 &  15.47 &  -16.86 & 1.4 & 512  &   4.0  &   5.2    \\
	      &  RO   &   20.68 &  13.89 &  -16.00 & 0.99 & 512  &   3.0  &   10.4   \\
PGC~6667      &  NA   &   25.73 &  14.78 &  3.81   & 1.2 & 512  &   4.0  &   5.2    \\
	      &  RO   &   20.57 &  12.75 &  3.19   & 0.82 & 512  &   3.0  &   10.5   \\
ESO~544-G030  &  NA   &   27.78 &  12.01 &  10.49  & 1.0 & 512  &   4.0  &   5.2    \\
	      &  RO   &   13.25 &  7.64  &  66.48  & 0.75 & 512  &   2.0  &   10.4   \\
UGC~1862      &  NA   &   23.97 &  15.32 &  22.14  & 1.4 & 512  &   4.0  &   5.2    \\
	      &  RO   &   19.30 &  13.88 &  15.43  & 0.88 & 512  &   3.0  &   10.4   \\
ESO~418-G008  &  NA   &   31.17 &  13.00 &  -8.18  & 1.3 & 512  &   4.0  &   5.2    \\
	      &  RO   &   13.12 &  9.99  &  51.01  & 0.90 & 512  &   2.0  &   10.4   \\
ESO~555-G027  &  NA   &   28.07 &  12.72 &  11.28  & 1.3 & 512  &   4.0  &   5.2    \\
	      &  RO   &   13.21 &  7.86  &  68.75  & 0.88 & 512  &   2.0  &   10.5   \\
NGC~2805      &  NA   &   19.84 &  16.79 &  -57.86 & 1.1 & 512  &   4.0  &   5.2    \\
	      &  RO   &   17.02 &  13.76 &  -54.49 & 0.85 & 512  &   3.0  &   10.4   \\
ESO~501-G023  &  NA   &   30.30 &  12.51 &  -14.66 & 1.4 & 512  &   4.0  &  5.2	\\
	      &  RO   &   12.32 &  8.65  &  -76.74 & 1.0 & 512  &   2.0   & 10.4	\\
UGC~6446      &  NA   &   11.68 &  11.56 &  -75.23 & 0.5 & 512  &   3.0  &   2.6    \\
NGC~3794      &  NA   &   19.28 &  16.86 &  -57.66 & 1.0 & 512  &   4.0  &   5.2    \\
	      &  RO   &   15.72 &  14.23 &  -46.75 & 1.1 & 512  &   3.0  &   5.2    \\
NGC~3906      &  NA   &   20.69 &  16.16 &  -64.87 & 0.80 & 1024 &   4.0  &   10.4   \\
	      &  RO   &   16.24 &  13.80 &  -54.94 & 0.85 & 1024 &   3.0  &   10.4   \\
UGC~6930      &  NA   &   19.36 &  14.81 &  -62.87 & 1.1 & 512  &   4.0  &   5.2    \\
	      &  RO   &   16.18 &  12.50 &  -59.72 & 1.2 & 512  &   3.0  &   5.2    \\
NGC~4519      &  NA   &   51.91 &  18.77 &  -34.50 & 1.3 & 512  &   4.0  &   5.2    \\
	      &  RO   &   17.69 &  14.43 &  53.53  & 1.4 & 512  &   3.0  &   5.2    \\
NGC~4561      &  NA   &   19.97 &  16.23 &  79.10  & 1.5 & 512  &   4.0  &   5.2    \\
	      &  RO   &   16.23 &  13.83 &  -86.37 & 1.3 & 512  &   3.0  &   10.4   \\
NGC~4713      &  NA   &   18.99 &  18.40 &  64.18  & 1.1 & 512  &   4.0  &   5.2    \\
	      &  RO   &   16.63 &  15.06 &  -73.24 & 1.2 & 512  &   3.0  &   5.2    \\
NGC~4942      &  NA   &   21.52 &  18.91 &  9.28   & 1.1 & 512  &   4.0  &   5.2    \\
	      &  RO   &   17.42 &  16.29 &  3.45   & 1.2 & 512  &   3.0  &   5.2    \\
NGC~5964      &  NA   &   20.15 &  17.29 &  -7.11  & 0.64 & 512  &   4.0  &   5.2    \\
	      &  RO   &   16.09 &  14.94 &  -5.03  & 0.52 & 512  &   3.0  &   10.4   \\
NGC~6509      &  NA   &   18.61 &  17.91 &  4.03   & 1.0 & 512  &   4.0  &   5.2    \\
	      &  RO   &   15.93 &  14.95 &  -67.69 & 1.2 & 512  &   3.0  &   5.2    \\
IC~1291	      &  NA   &   19.62 &  16.02 &  -78.49 & 1.2 & 2048 &   4.0  &   5.2    \\
	      &  RO   &   16.76 &  13.46 &  -69.28 & 1.2 & 2048 &   3.0  &   5.2    \\
\enddata    

\tablecomments{Column 1: Object name; Column 2: Image weighting: NA
uses {\it robust} = 5 in IMAGR; RO uses {\it robust} = 0.5 in IMAGR;
Column 3: Beam major axis FWHM; Column 4: Beam minor axis FWHM; Column
5: Position angle of beam major axis; Column 6: Image noise; Column 7:
Spatial dimensions of data cube; Column 8: Pixel size; Column 9:
Channel width.  See Section~\ref{sec:dp1} for more details.  \\ }

\label{tab:im_summary}
\end{deluxetable}
\clearpage

\begin{deluxetable}{lccccccc}
\tablewidth{0pt}
\tabletypesize{\scriptsize}
\tablecaption{\ion{H}{1} Flux, Mass, and Line Width Measurements}
\tablehead{
\colhead{Source} &
\colhead{$S_{\rm HI}$} &
\colhead{$M_{\rm HI}$} &
\colhead{$S_{\rm HI}^{\rm HIPASS}$} &
\colhead{$S_{\rm HI}^{\rm Springob}$} &
\colhead{$W_{20}$} &
\colhead{$W_{20}^{\rm HIPASS}$} &
\colhead{$W_{20}^{\rm RC3}$} \\
\colhead{} &
\colhead{$({\rm Jy \, km \, s^{-1}}$)} &
\colhead{($10^{8} \, M_{\odot}$)} &
\colhead{(${\rm Jy \, km \, s^{-1}}$)} &
\colhead{$({\rm Jy \, km \, s^{-1}}$)} &
\colhead{(${\rm km \, s^{-1}}$)} &
\colhead{(${\rm km \, s^{-1}}$)} &
\colhead{(${\rm km \, s^{-1}}$)} \\
\colhead{(1)} &
\colhead{(2)} &
\colhead{(3)} &
\colhead{(4)} &
\colhead{(5)} &
\colhead{(6)} &
\colhead{(7)} &
\colhead{(8)} 
}
\startdata

NGC~0337      &  75.6	& 76.4	&  50.6  &  50.84 &  261  &   277.3 &  $261   \pm  6 $     \\
PGC~3853      &  125.3  & 38.4	&  61.9  &  ...   &  192  &   194.5 &  $182   \pm  7 $     \\
PGC~6667      &  33.3	& 47.6	&  21.4  &  ...   &  198  &   219.8 &  $213   \pm  16$    \\
ESO~544-G030  &  11.2	& 5.09	&  8.4   &  ...   &  146  &   148.4 &  $104   \pm  16$     \\
UGC~1862      &  4.7	& 5.47	&  3.4   &  3.51  &  125  &   121.6 &  ...		   \\
ESO~418-G008  &  8.2	& 10.7	&  9.6   &  ...   &  140  &   185.4 &  ...		   \\
ESO~555-G027  &  34.0	& 47.3	&  24.0  &  ...   &  162  &   180.8 &  $168   \pm  16$     \\
NGC~2805      &  120.3  & 223	&  ...   &  ...   &  120  &   ...   &  $118   \pm  4 $     \\
ESO~501-G023  &  24.9	& 2.89	&  29.3  &  ...   &  83   &   88.3  &  $79    \pm  16$    \\   
UGC~6446      &  43.8	& 33.5	&  ...   &  34.55 &  150  &   ...   &  $143   \pm  6 $    \\
NGC~3794      &  19.4	& 16.8	&  ...   &  20.32 &  182  &   ...   &  $187   \pm  6 $     \\
NGC~3906      &  4.8	& 3.76	&  ...   &  11.38 &  49   &   ...   &  $48    \pm  6 $     \\
UGC~6930      &  43.5	& 29.7	&  ...   &  ...   &  140  &   ...   &  $133   \pm  8 $     \\
NGC~4519      &  50.9	& 46.2	&  ...   &  ...   &  218  &   ...   &  $209   \pm  6 $     \\
NGC~4561      &  27.5	& 9.83	&  ...   &  26.16 &  171  &   ...   &  $140   \pm  4 $     \\
NGC~4713      &  36.0	& 18.8	&  ...   &  55.70 &  176  &   ...   &  $192   \pm  5 $     \\
NGC~4942      &  10.1	& 19.4	&  11.4  &  ...   &  177  &   203.4 &  $159   \pm  16$     \\
NGC~5964      &  54.5	& 78.4	&  ...   &  40.84 &  208  &   ...   &  $192   \pm  4 $     \\
NGC~6509      &  30.4 & $>57.0$	&  ...   &  31.02 &  266  &   ...   &  $283   \pm  9 $     \\
IC~1291	      &  14.5	& 34.0	&  ...   &  23.16 &  209  &   ...   &  $226   \pm  7 $     \\
	
\enddata

\tablecomments{Column 1: Object name; Column 2: Integrated HI line
  flux.  We assign an error of 12\% to our fluxes to account for flux
  calibration and aliasing.  Column 3: Total HI mass.  Note that the
  NGC~6509 mass is a lower limit due to absorption.  See
  Section~\ref{sec:morph} for details; Column 4: Integrated flux from
  HIPASS, which uses the single-dish 64-m Parkes telescope
  \citep{meyer04}; Column 5: Integrated flux from a variety of large
  single-dish telescopes, compiled in \citet{springob05}; Column 6:
  Width of the HI line at 20\% of the peak flux density, corrected for
  the spectral resolution assuming $W_{20} = \sqrt{W_{20, {\rm
  obs}}^{2} - W_{20,{\rm res}}^{2}}$, $W_{20, {\rm res}} = W_{50,{\rm
  res}} \sqrt{{\rm ln \, 5/ln \, 2}}$, and $W_{50,{\rm res}} = \Delta
  {\rm v}$, where $W_{50,{\rm res}}$ is the spectral resolution FWHM,
  $\Delta {\rm v}$ is the channel width of the naturally-weighted data
  cube from column~9 of Table~\ref{tab:im_summary}, $W_{20,{\rm res}}$
  is the spectral resolution at 20\% of the peak flux density, and
  $W_{20, {\rm obs}}$ is the observed width at 20\% of the peak flux
  density.  These assumptions are appropriate for data that has been
  Hanning smoothed and for which the spectral response function is
  Gaussian.  $W_{20}$ is not corrected for turbulent broadening or
  inclination.  The uncertainty in $W_{20}$ is the channel width,
  which is $5\, {\rm km \, s^{-1}}$ for all objects except UGC~6446
  and NGC~3906, where the uncertainty is $3 \, {\rm km \, s^{-1}}$ and
  $10\, {\rm km \, s^{-1}}$, respectively. Column 7: $W_{20}$ from
  \citet{meyer04}; Column 8: $W_{20}$ values compiled in
  \citet{RC3}. \\ }

\label{tab:flux}
\end{deluxetable}

\clearpage

\begin{deluxetable}{lccccccccccc}
\tablewidth{0pt}
\tabletypesize{\scriptsize}
\tablecaption{Comparison of Kinematic and Photometric Parameters}
\tablehead{
\colhead{} &
\multicolumn{3}{c}{Center} &
\colhead{} &
\multicolumn{3}{c}{Position Angle} &
\colhead{} &
\multicolumn{3}{c}{Inclination} \\
\cline{2-4} \cline{6-8} \cline{10-12} \\
\colhead{Source} &
\colhead{RA (J2000.0)} &
\colhead{DEC (J2000.0)} &
\colhead{Offset} &
\colhead{} &
\colhead{ROTCUR} &
\colhead{IRAC} &
\colhead{RC3} &
\colhead{} &
\colhead{ROTCUR} &
\colhead{IRAC} &
\colhead{RC3} \\
\colhead{} &
\colhead{(hh:mm:ss.s)} &
\colhead{(dd:mm:ss)} &
\colhead{(arcsec)} &
\colhead{} &
\colhead{($\degr$)} &
\colhead{($\degr$)} &
\colhead{($\degr$)} &
\colhead{} &
\colhead{($\degr$)} &
\colhead{($\degr$)} &
\colhead{($\degr$)} \\
\colhead{(1)} &
\colhead{(2)} &
\colhead{(3)} &
\colhead{(4)} &
\colhead{} &
\colhead{(5)} &
\colhead{(6)} &
\colhead{(7)} &
\colhead{} &
\colhead{(8)} &
\colhead{(9)} &
\colhead{(10)} \\
}
\startdata
NGC~0337     &	 00:59:50.0 &	-07:34:42   &	 3.2	     && 119.0  & 118.4  &  130 & & 49.9  &  43.7 &  51 \\
PGC~3853     &	 01:05:05.1 &	-06:12:58   &	 12.4	     && 106.2  & 105.3  &  ... & & 40.7  &  41.4 &  32 \\
PGC~6667     &	 01:49:10.6 &	-10:03:48   &	 3.7	     && 125.2  & 122.9  &  ... & & 48.0  &  34.0 &  37 \\
ESO~544-G030 &   02:14:57.6 &	-20:12:44   &	 11.2	     && 106.1  & 107.6  &  103 & & 12.0  &  48.5 &  51 \\
UGC~1862     &	 02:24:24.8 &	-02:09:47   &	 3.8	     && 21.4   & 21.7	&  10  & & 39.0  &  43.4 &  39 \\
ESO~418-G008 &	 03:31:30.8 &	-30:12:46   &	 0.8	     && 317.9  & 321.6  &  329 & & 49.9  &  55.6 &  49 \\
ESO~555-G027 &   06:03:35.9 &	-20:39:10   &	 27.6	     && 221.5  & 245.3  &  230 & & 31.0  &  20.9 &  36 \\
NGC~2805     &	 ...	    &	...	    &	 ...	     && 300.0  & ...	&  305 & & 37.5  &  ...  &  41 \\
ESO~501-G023 &   ...	    &	...	    &    ...	     && 223.6  & ...	&  194 & & 49.0  &  ...  &  37 \\
UGC~6446     &	 ...	    &	...	    &    ...	     && 189.4  & ...	&  190 & & 52.5  &  ...  &  50 \\
NGC~3794     &	 11:40:54.4 &	+56:12:07   &    0.8	     && 123.1  & 116.2  &  120 & & 54.2  &  54.8 &  50 \\
NGC~3906     &	 11:49:39.8 &	+48:25:27   &    11.1        && 183.1  & 185.0  &  ... & & 40.0  &  16.5 &  27 \\
UGC~6930     &	 11:57:16.8 &	+49:16:58   &    3.5	     && 39.5   & 24.2	&  ... & & 31.3  &  25.4 &  51 \\
NGC~4519     &	 12:33:30.4 &	+08:39:18   &    9.6         && 354.9  & 326.3  &  325 & & 42.5  &  42.4 &  39 \\
NGC~4561     &	 12:36:08.3 &	+19:19:24   &    7.1	     && 221.3  & 227.3  &  210 & & 33.0  &  34.3 &  32 \\
NGC~4713     &	 12:49:57.5 &	+05:18:39   &    4.7         && 274.1  & 274.0  &  280 & & 41.4  &  45.2 &  51 \\
NGC~4942     &	 13:04:19.1 &	-07:38:56   &    1.1	     && 137.3  & 146.9  &  145 & & 45.1  &  37.3 &  45 \\
NGC~5964     &	 15:37:36.2 &	+05:58:24   &    3.4         && 136.7  & 160.5  &  145 & & 37.7  &  31.9 &  39 \\
NGC~6509     &	 17:59:25.3 &	+06:17:13   &    0.5         && 280.8  & 286.2  &  285 & & 47.1  &  41.0 &  42 \\
IC~1291	     &	 ...	    &	...	    &    ...	     && 131.4  & ...	&  210 & & 28.0  &  ...  &  34 \\
\enddata

\tablecomments{Column 1: Object name.  Columns 2 and 3: RA and DEC of
  the photometric center, derived from ellipse fits on the Channel 1
  IRAC data.  Empty columns here and in subsequent IRAC columns
  correspond to objects where the ellipse fits did not
  converge. Column 4: Offset of the kinematic center determined in the
  rotation curve analysis relative to the photometric center.  Column
  5: PA derived from the rotation curve analysis (degrees N to E to
  receding side). Column 6: PA derived from ellipse fits on the
  Channel 1 IRAC data. Column 7: RC3 PA, converted to the definition
  used in column 5 when necessary. Column 8: Inclination derived from
  the rotation curve analysis.  Column 9: Inclination derived from
  ellipse fits on the Channel 1 IRAC data.  Column 10: RC3
  inclination. \\ }

\label{tab:rot_comp}
\end{deluxetable}

\clearpage

\begin{deluxetable}{lccccccccc}
\tablewidth{0pt}
\tabletypesize{\scriptsize}
\tablecaption{Final Kinematic Parameters}
\tablehead{
\colhead{Source} &
\colhead{$V_{\rm sys}$} &
\colhead{PA} &
\colhead{{\it i}} &
\colhead{Center RA} &
\colhead{Center DEC} &
\colhead{Center error} &
\colhead{${\rm v_{circ}}$} &
\colhead{RC Quality} &
\colhead{Note} \\
\colhead{} &
\colhead{(${\rm km\, s^{-1}}$)} &
\colhead{($\degr$)} &
\colhead{($\degr$)} &
\colhead{(hh:mm:ss.s)} &
\colhead{(dd:mm:ss)} &
\colhead{(arcsec)} &
\colhead{(${\rm km\, s^{-1}}$)} &
\colhead{Assessment} &
\colhead{} \\
\colhead{(1)} &
\colhead{(2)} &
\colhead{(3)} &
\colhead{(4)} &
\colhead{(5)} &
\colhead{(6)} &
\colhead{(7)} &
\colhead{(8)} &
\colhead{(9)} &
\colhead{(10)}
}
\startdata
NGC~0337     &	 $1646   \pm 2   $  &  $118    \pm 5    $ &  $44    \pm  2    $ &   00:59:50.0   &	  -07:34:42  &    3    & $145    ^{+5    }_{-4    } $ &  A  & i,c,p   \\
PGC~3853     &	 $1094.7 \pm 0.4 $  &  $105.3  \pm 0.2  $ &  $41.4  \pm  1.1  $ &   01:05:04.8   &	  -06:12:47  &    1.3  & $128.1  ^{+1.6  }_{-2    } $ &  A   & i,p     \\
PGC~6667     &	 $1989.2 \pm 0.6 $  &  $122.9  \pm 1.8  $ &  $34.0  \pm  1.1  $ &   01:49:10.5   &	  -10:03:45  &    1.1  & $155    ^{+4    }_{-4    } $ &  A   & i,p     \\
ESO~544-G030 &   $1608.4 \pm 1.0$   &  $107.6  \pm 1.1  $ &  $48.5  \pm  1.2  $ &   02:14:56.9   &	  -20:12:40  &    3    & $100.9  ^{+1.4  }_{-2    } $ &  C   & i,p     \\
UGC~1862     &	 $1382.9 \pm 0.4 $  &  $21.7   \pm 1.7  $ &  $43    \pm  4    $ &   2:24:24.8	 &	  -2:09:47   &    1.8  & $55     ^{+6    }_{-5    } $ &  C  & i,c,p   \\
ESO~418-G008 &	 $1195.4 \pm 0.3 $  &  $317.9  \pm 1.1  $ &  $55.6  \pm  1.4  $ &   03:31:30.8   &	  -30:12:46  &    1.2  & $74.1   ^{+0.6  }_{-0.6  } $ &  B   & i,c     \\
ESO~555-G027 &   $1978.7 \pm 0.4 $  &  $221.5  \pm 0.3  $ &  $21    \pm  4    $ &   06:03:35.9   &	  -20:39:10  &    1.9  & $190    ^{+40   }_{-30   } $ &  A   & i,c     \\
NGC~2805     &	 $1732.6 \pm 0.6 $  &  $300    \pm 3    $ &  $38    \pm  4    $ &   09:20:19.6   &	  +64:06:02  &    7    & $81     ^{+7    }_{-8    } $ &  A   & ...     \\
ESO~501-G023 &   $1046.8 \pm 0.7$   &  $224    \pm 2    $ &  $37    \pm  12   $ &   10:35:23.0   &	  -24:45:19  &    9    & $46     ^{+18   }_{-8    } $ &  A   & i*      \\
UGC~6446     &	 $645.5  \pm 0.6 $  &  $189.4  \pm 0.5  $ &  $52.5  \pm  1.9  $ &   11:26:40.6   &	  +53:44:52  &    1.0  & $79.7   ^{+1.3  }_{-0.8  } $ &  A   & ...     \\
NGC~3794     &	 $1384.9 \pm 0.7 $  &  $123.1  \pm 1.1  $ &  $54.8  \pm  1.3  $ &   11:40:54.4   &	  +56:12:07  &    0.11 & $103.3  ^{+1.1  }_{-1.0  } $ &  A   & i,c     \\
NGC~3906     &	 $959.44 \pm 0.7 $  &  $180    \pm 20   $ &  $16    \pm  5    $ &   11:49:40.4   &	  +48:25:36  &    7    & $65     ^{+30   }_{-16   } $ &  C   & i,p     \\
UGC~6930     &	 $776.7  \pm 0.7 $  &  $39.5   \pm 0.5  $ &  $25    \pm  4    $ &   11:57:16.8   &	  +49:16:58  &    4    & $121    ^{+20   }_{-15   } $ &  A   & i,c     \\
NGC~4519     &	 $1218.1 \pm 1.0 $  &  $355    \pm 2    $ &  $42    \pm  3    $ &   12:33:30.4   &	  +08:39:18  &    1.3  & $112    ^{+8    }_{-7    } $ &  A   & i,c     \\
NGC~4561     &	 $1402.2 \pm 0.9 $  &  $227    \pm 8    $ &  $34    \pm  4    $ &   12:36:08.3   &	  +19:19:24  &    3    & $57     ^{+6    }_{-5    } $ &  B   & i,c,p   \\
NGC~4713     &	 $654.5  \pm 0.5 $  &  $274.0  \pm 0.6  $ &  $45.2  \pm  1.2  $ &   12:49:57.7   &	  +05:18:43  &    1.0  & $110.9  ^{+2    }_{-1.8  } $ &  A   & i,p     \\
NGC~4942     &	 $1741   \pm 2   $  &  $137.3  \pm 0.8  $ &  $37    \pm  2    $ &   13:04:19.1   &	  -07:38:56  &    3    & $124    ^{+6    }_{-5    } $ &  A   & i,c     \\
NGC~5964     &	 $1447.1 \pm 1.2 $  &  $136.7  \pm 1.2  $ &  $32    \pm  3    $ &   15:37:36.2   &	  +05:58:24  &    5    & $168    ^{+18   }_{-14   } $ &  A   & i,c     \\
NGC~6509     &	 $1811.0 \pm 0.4 $  &  $280.8  \pm 1.1  $ &  $41    \pm  4    $ &   17:59:25.3   &	  +06:17:13  &    4    & $153    ^{+12   }_{-9    } $ &  A   & i,c     \\
IC~1291	     &	 $1951.0 \pm 1.1 $  &  $131    \pm 2    $ &  $28    \pm  3    $ &   18:33:52.5   &	  +49:16:49  &    2    & $189    ^{+17   }_{-14   } $ &  B   & ...     \\
\enddata 

\tablecomments{Column 1: Object name.  Column 2: Systemic velocity
derived from the rotation curve analysis, corrected to the
heliocentric reference frame. Column 3: Position angle of the major
axis (degrees N to E to receding side).  For columns 3-6, the values
are either from the rotation curve analysis or from ellipse fits on
the Channel 1 IRAC data.  See column 10 and
Section~\ref{sec:rot_curves} for more information.  Column 4:
Inclination.  Column 5 and 6: RA and DEC (J2000.0) of the galaxy
center.  Column 7: Error on the galaxy center.  Column 8: Circular
velocity, defined as the maximum velocity in the rotation curve fits.
Column 9: Rotation curve quality assessment.  A refers to galaxies
with clear turnovers in the rotation curve, B refers to galaxies with a
hint of a turnover, and C refers galaxies with no turnover.  Column
10: The note lists which final parameters are derived from the IRAC
analysis (i=inclination, p=PA, c=center).  The remaining parameters
are from the rotation curve analysis.  * The inclination for
ESO~501-G023 is from RC3.}
\label{tab:final_params}
\end{deluxetable}

\clearpage


\clearpage

\begin{figure}
\begin{center}
\plotone{NGC4942_chmap_final.eps}
\caption{As in Figure~\ref{fig:337_chmap}, but for NGC~4942.  The
  grayscale spans -0.02 to 16.02 ${\rm mJy \, beam^{-1}}$.  Every
  second channel is shown and the channel width is 5.2 \kms.}
\label{fig:4942_chmap}
\end{center}
\end{figure}

\begin{figure}
\includegraphics[width=\textwidth]{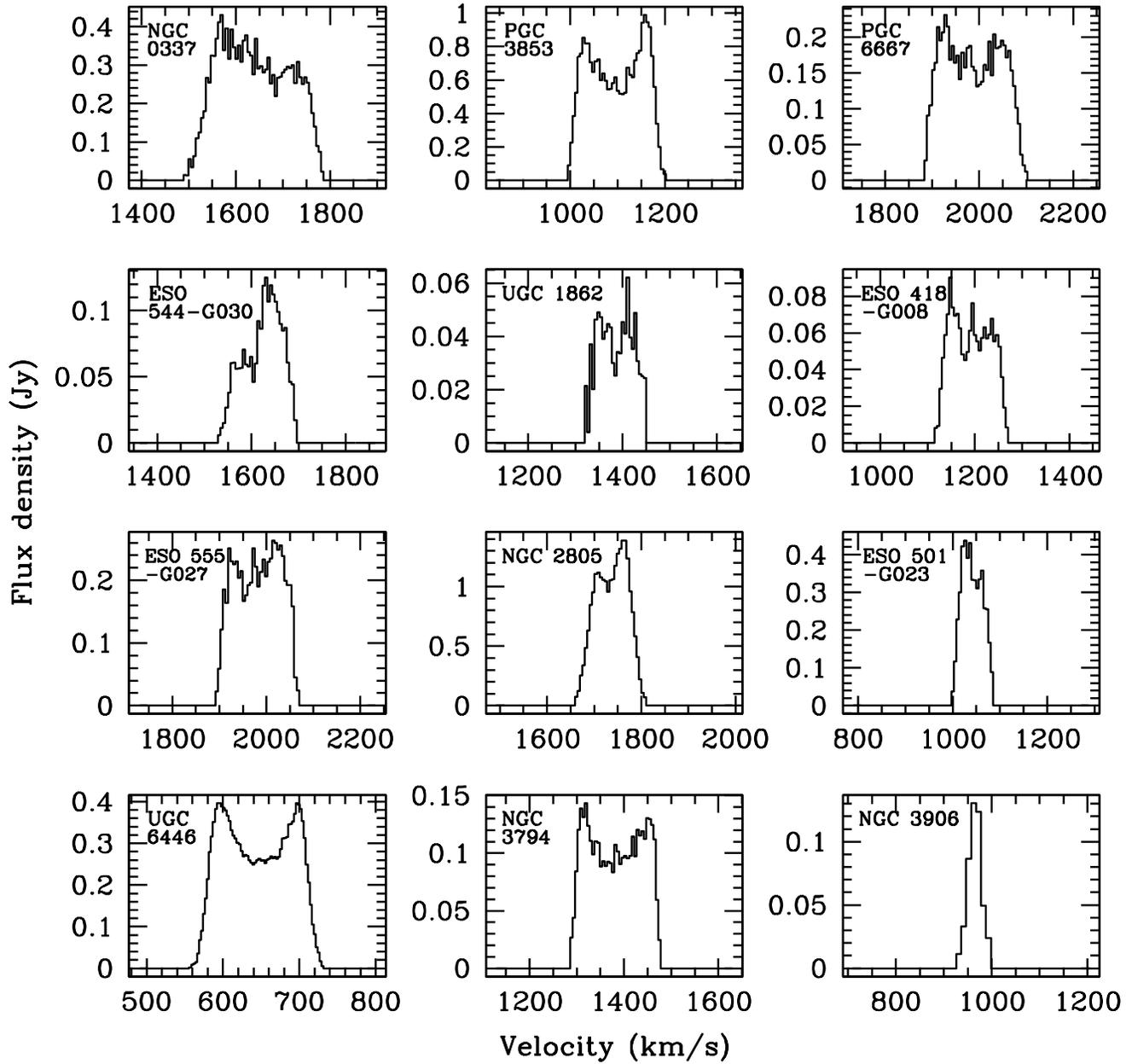}
\caption{Integrated \ion{H}{1} line profiles, created by summing the
  emission in each channel of the naturally-weighted data cubes.  The
  channels outside the line emission show no noise pattern because of
  our blanking procedure.}
\label{fig:int_line_prof}
\end{figure}

\begin{figure}
\figurenum{21}
\includegraphics[width=\textwidth]{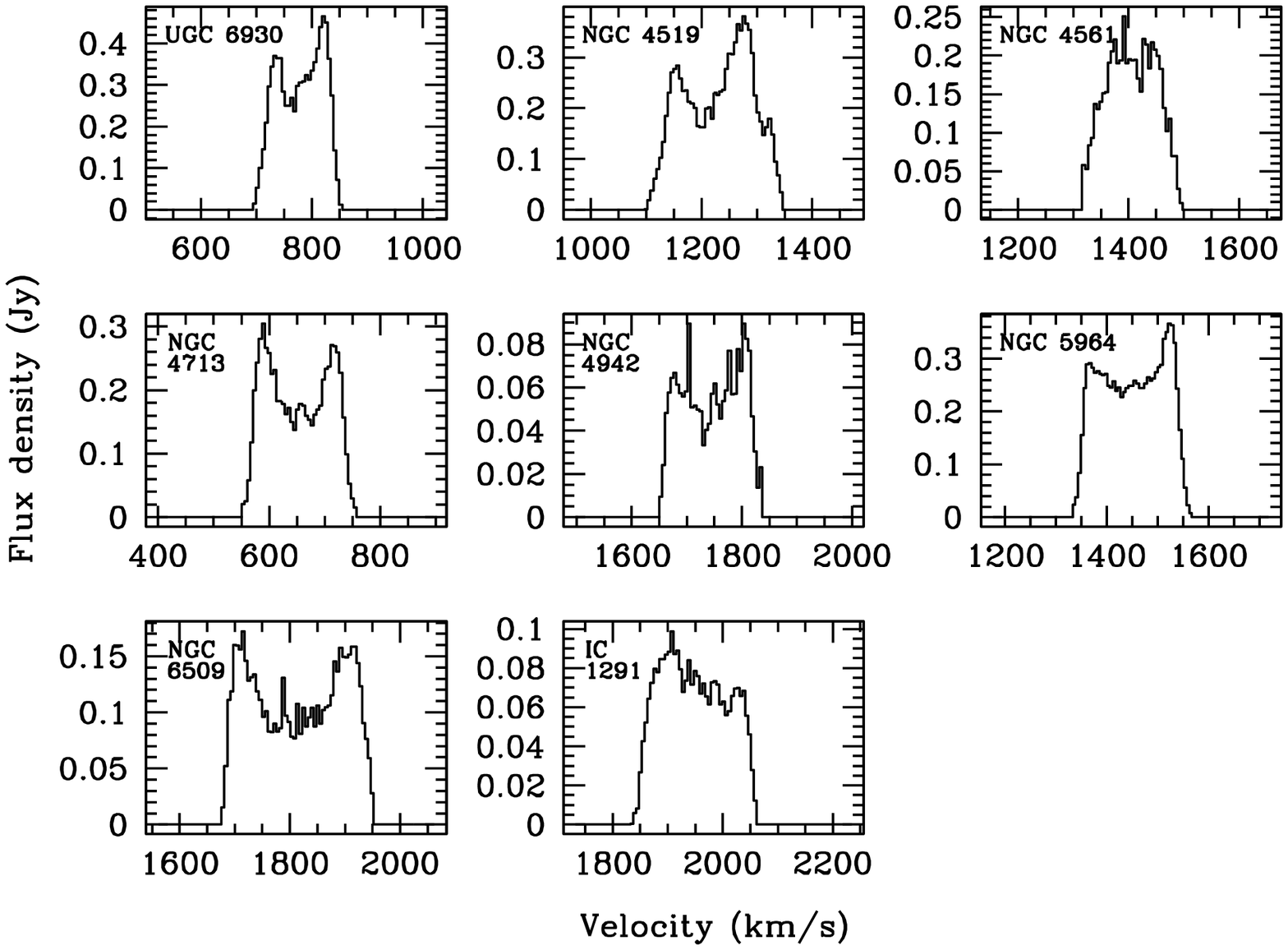}
\caption{(Continued)}
\end{figure}

\clearpage

\begin{figure}
\begin{center}$
\begin{array}{ccc}
\includegraphics[height=0.22\textheight]{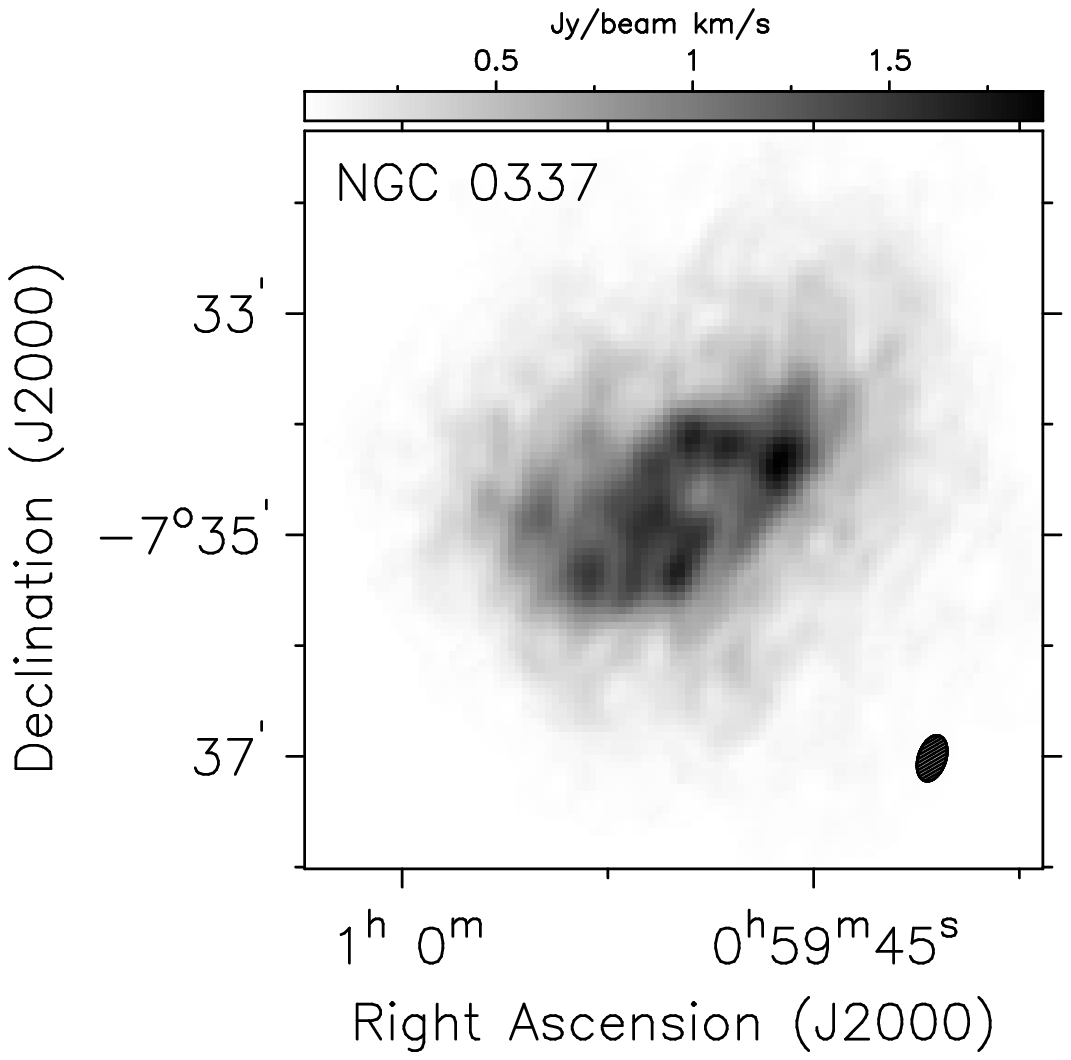} &
\includegraphics[height=0.22\textheight]{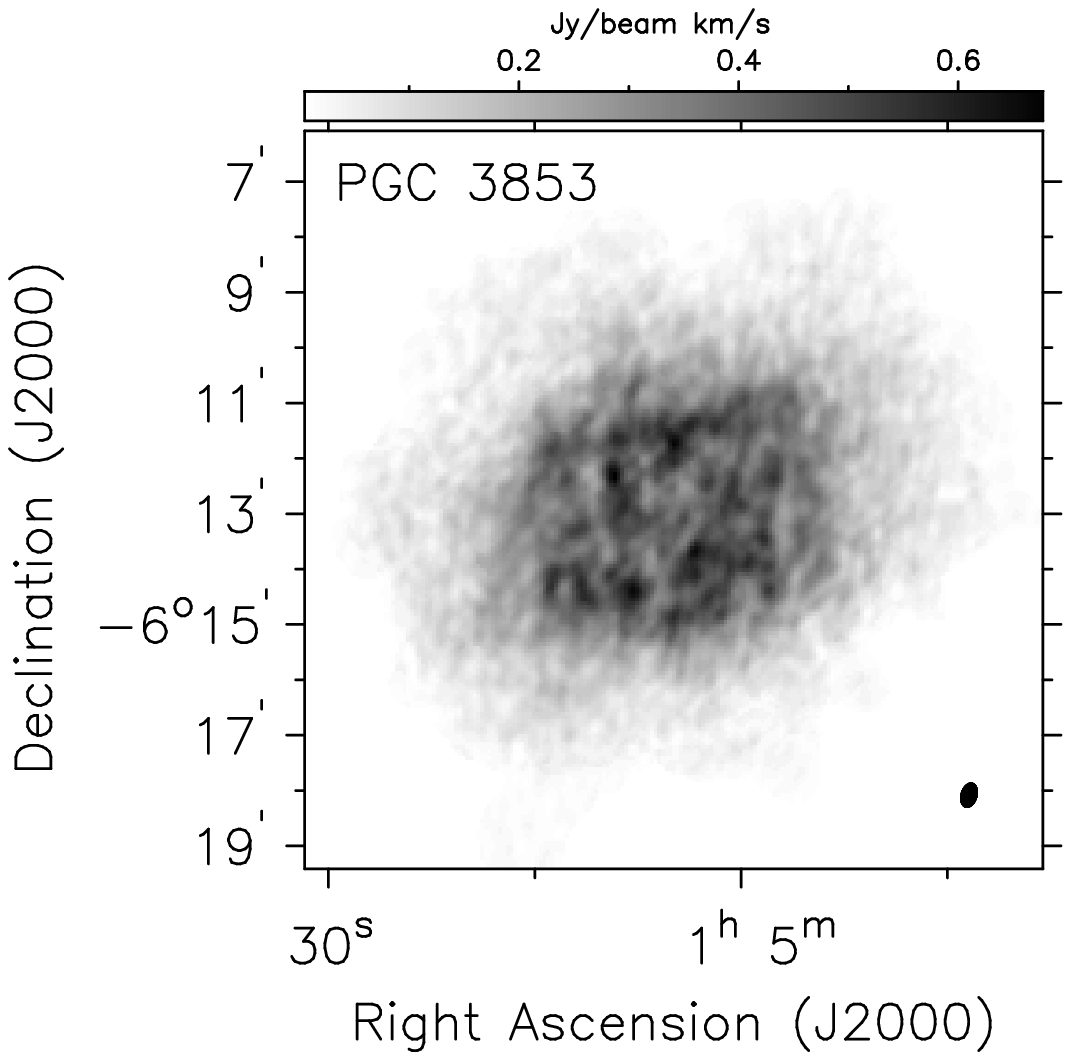} &
\includegraphics[height=0.22\textheight]{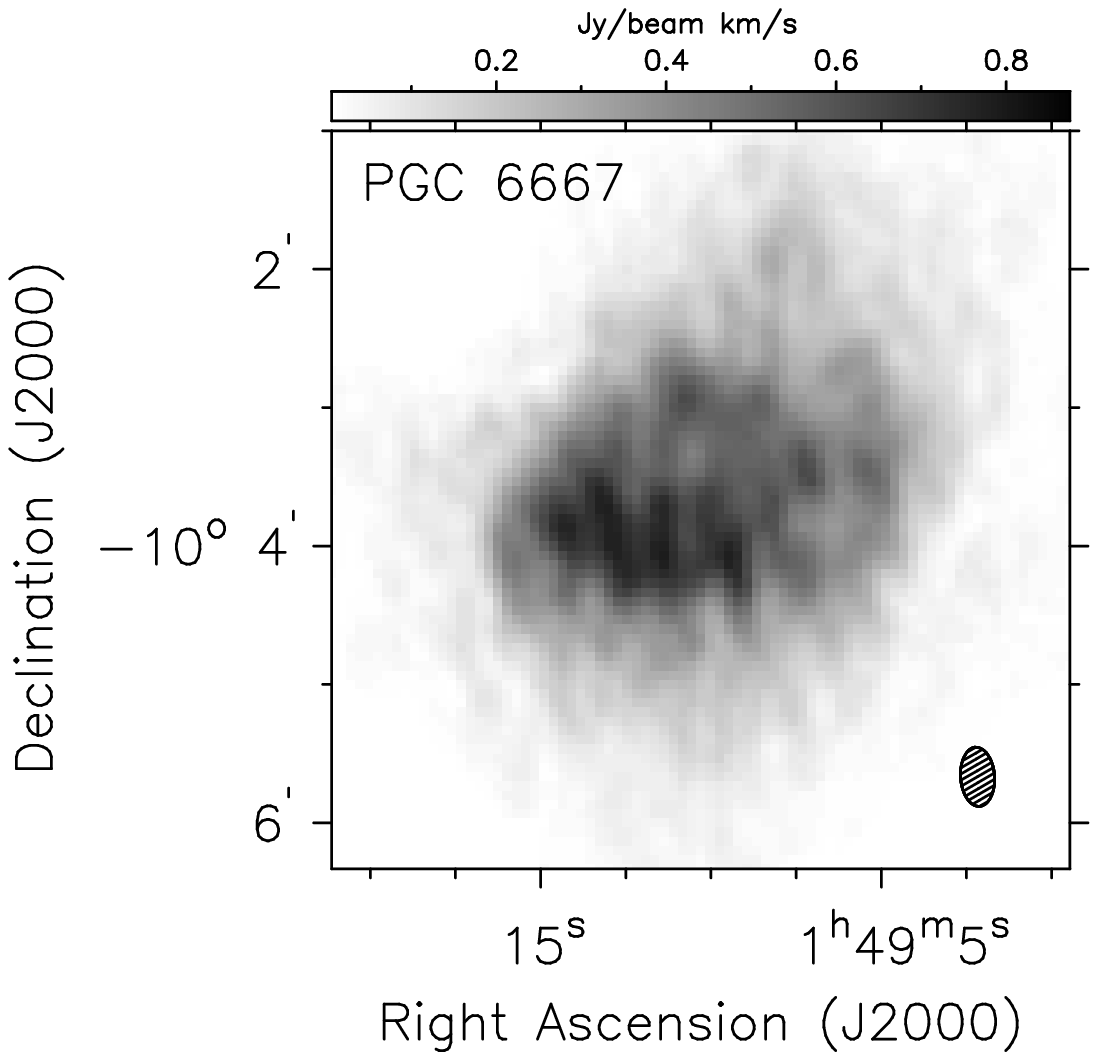} \\
\\
\includegraphics[height=0.22\textheight]{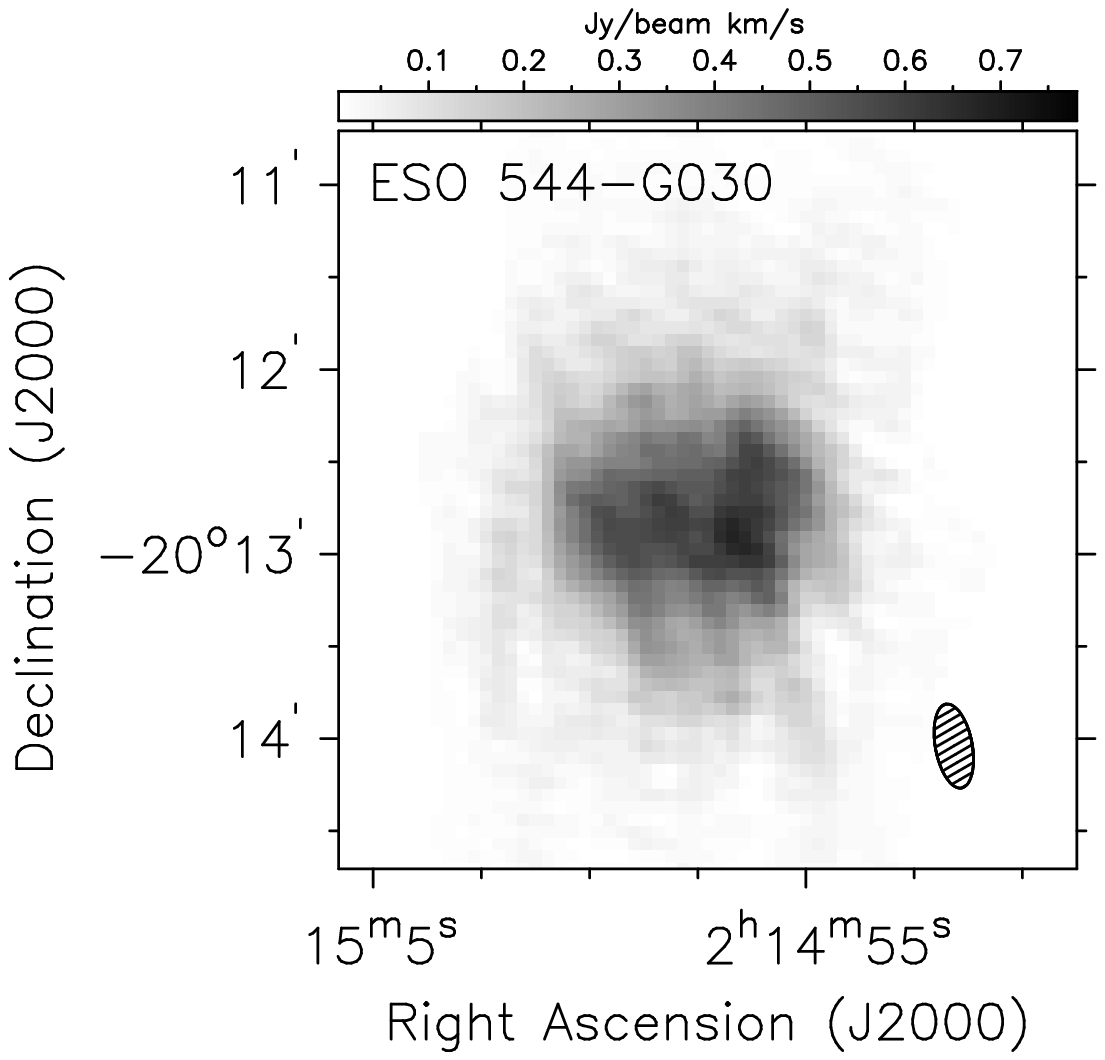} &
\includegraphics[height=0.22\textheight]{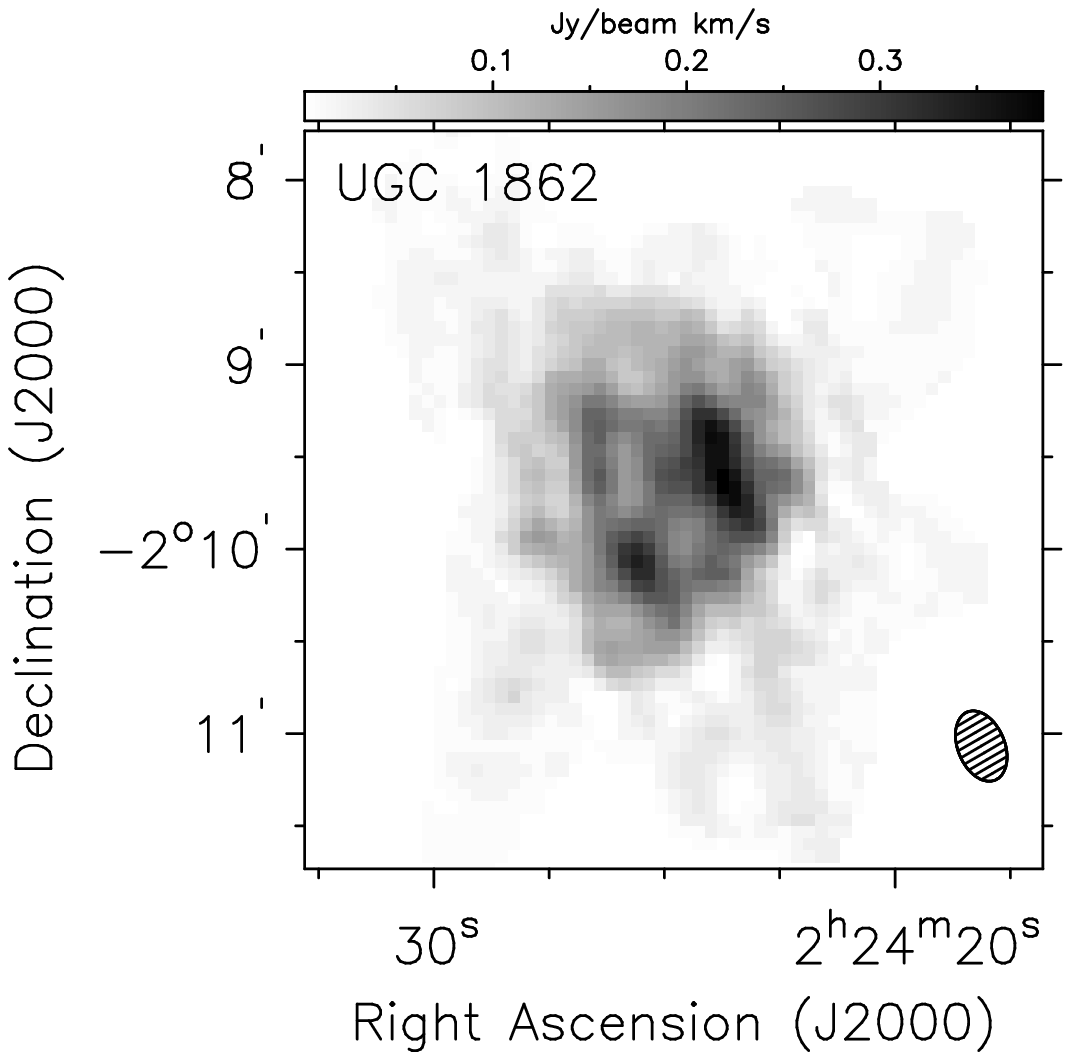} &
\includegraphics[height=0.22\textheight]{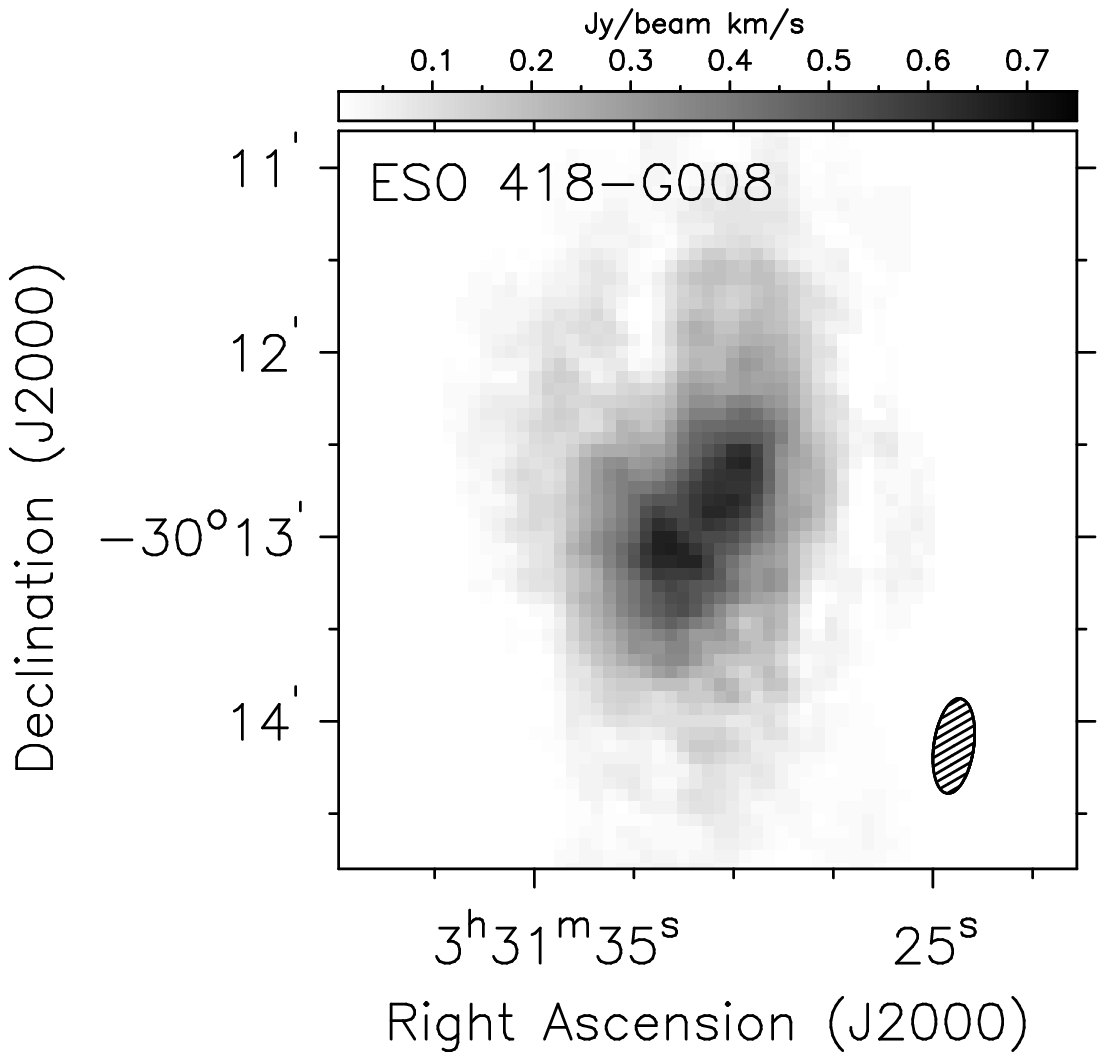} \\
\\
\includegraphics[height=0.22\textheight]{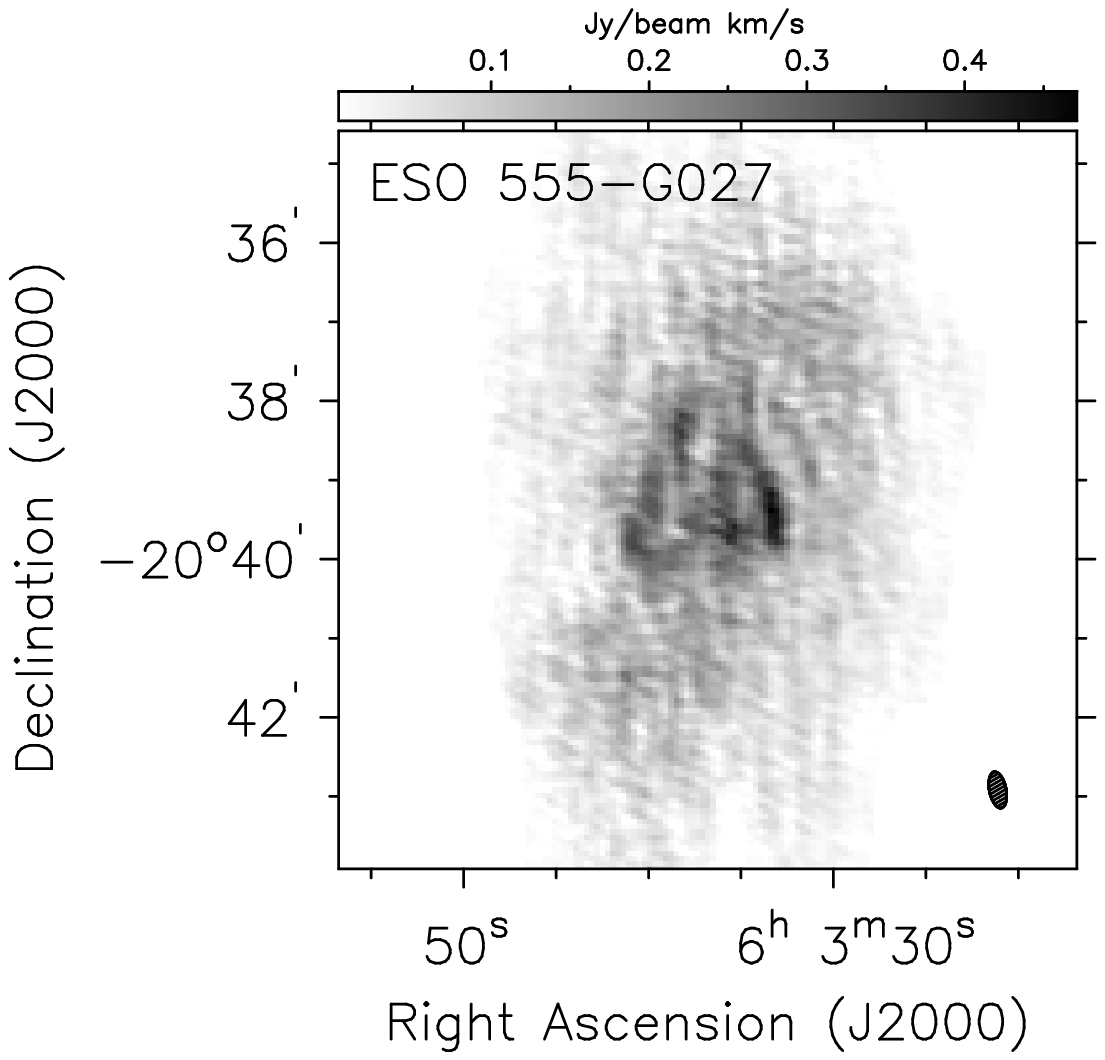} &
\includegraphics[height=0.22\textheight]{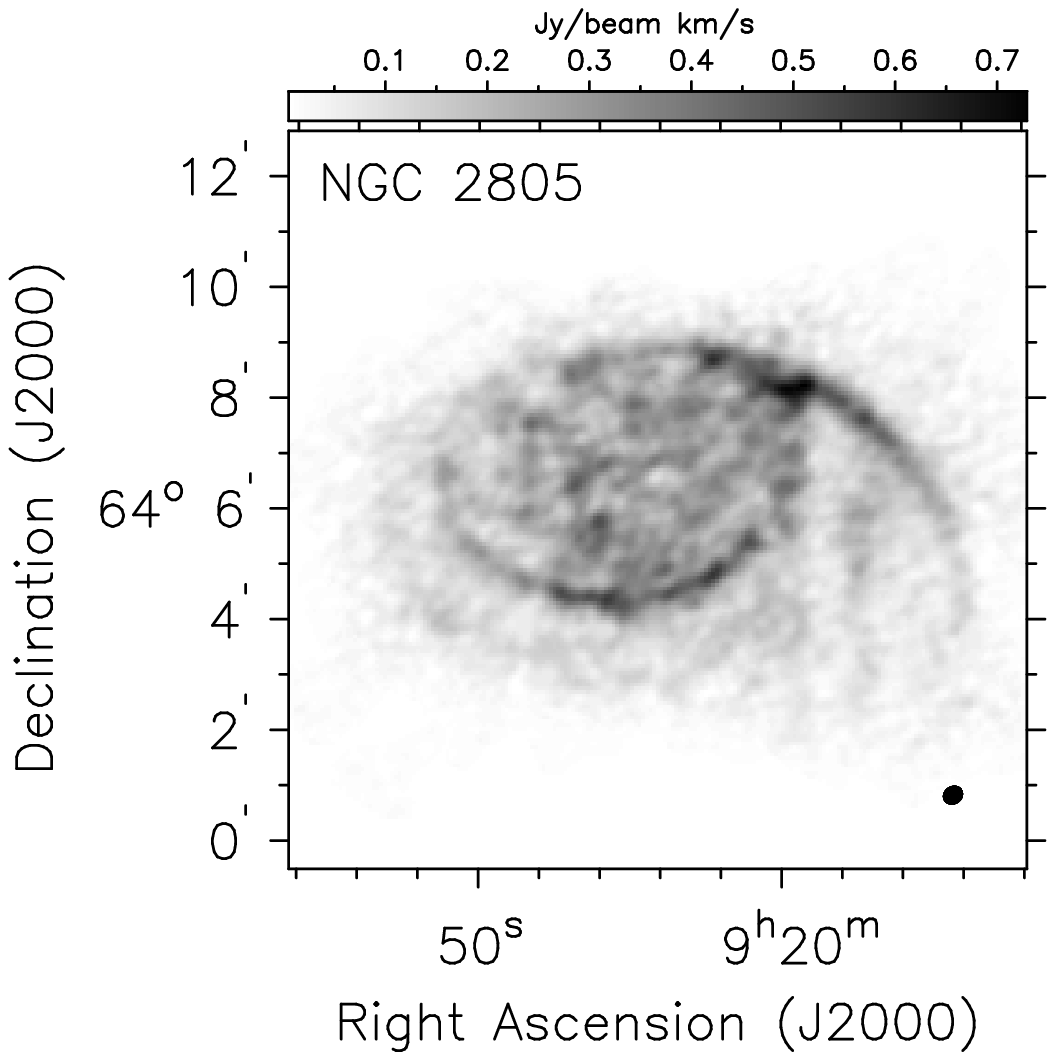} &
\includegraphics[height=0.22\textheight]{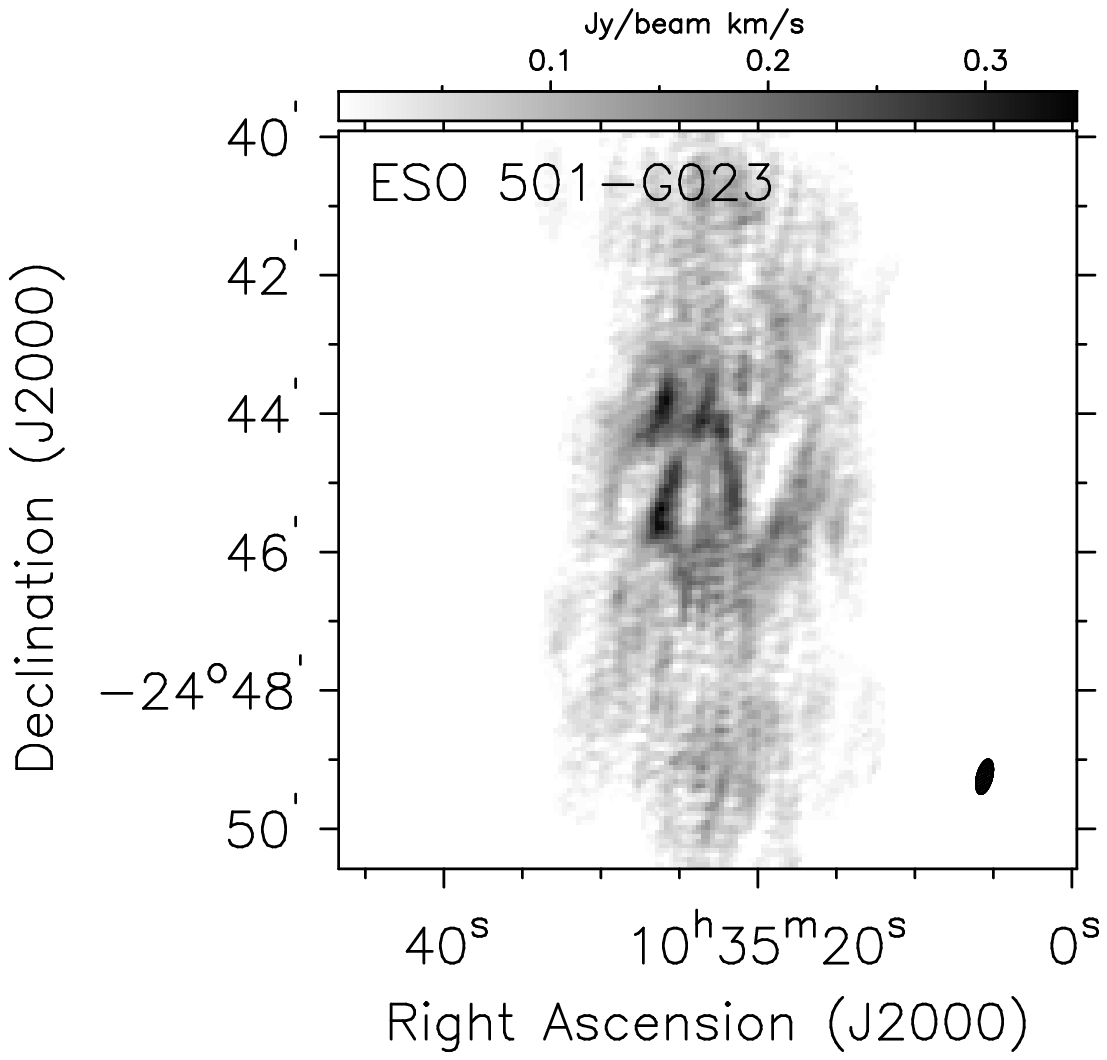} \\
\\
\includegraphics[height=0.22\textheight]{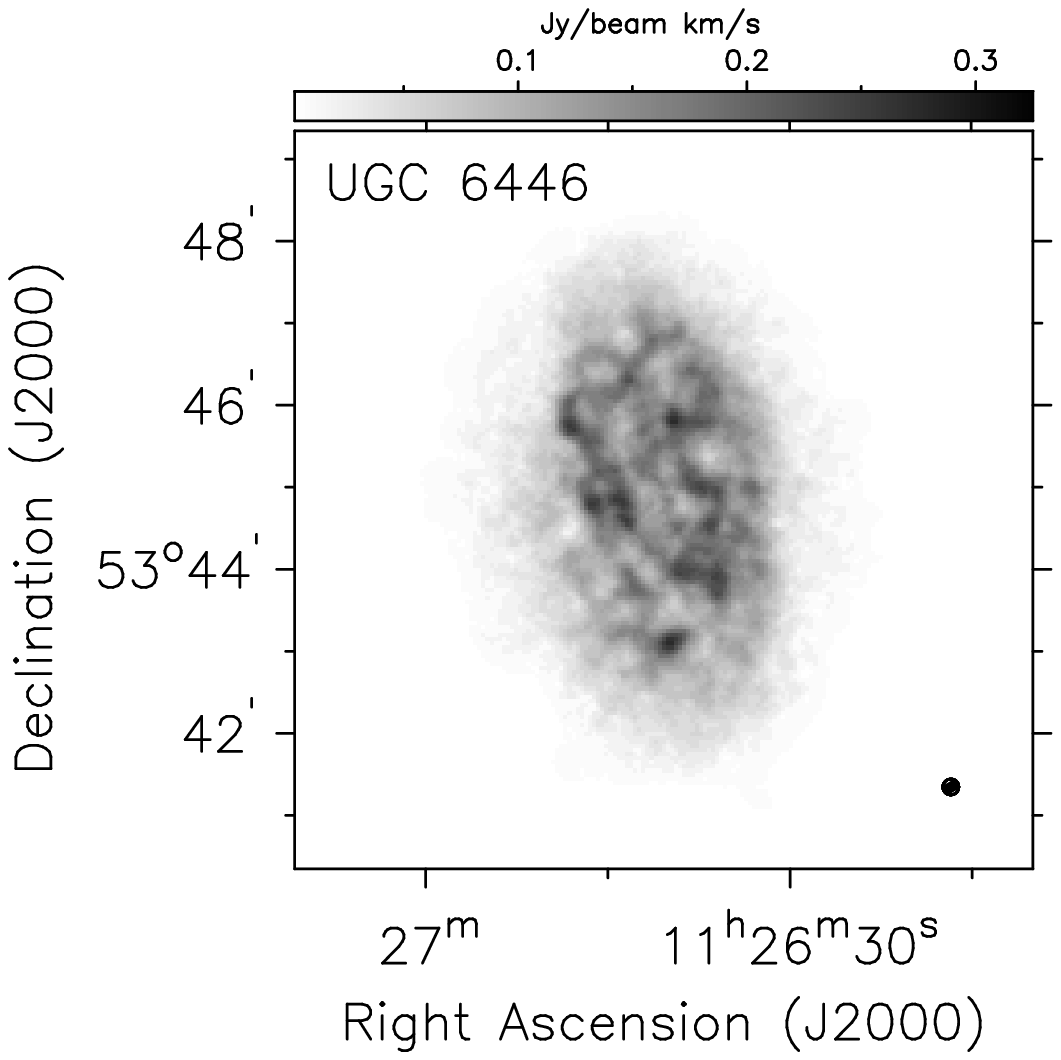} &
\includegraphics[height=0.22\textheight]{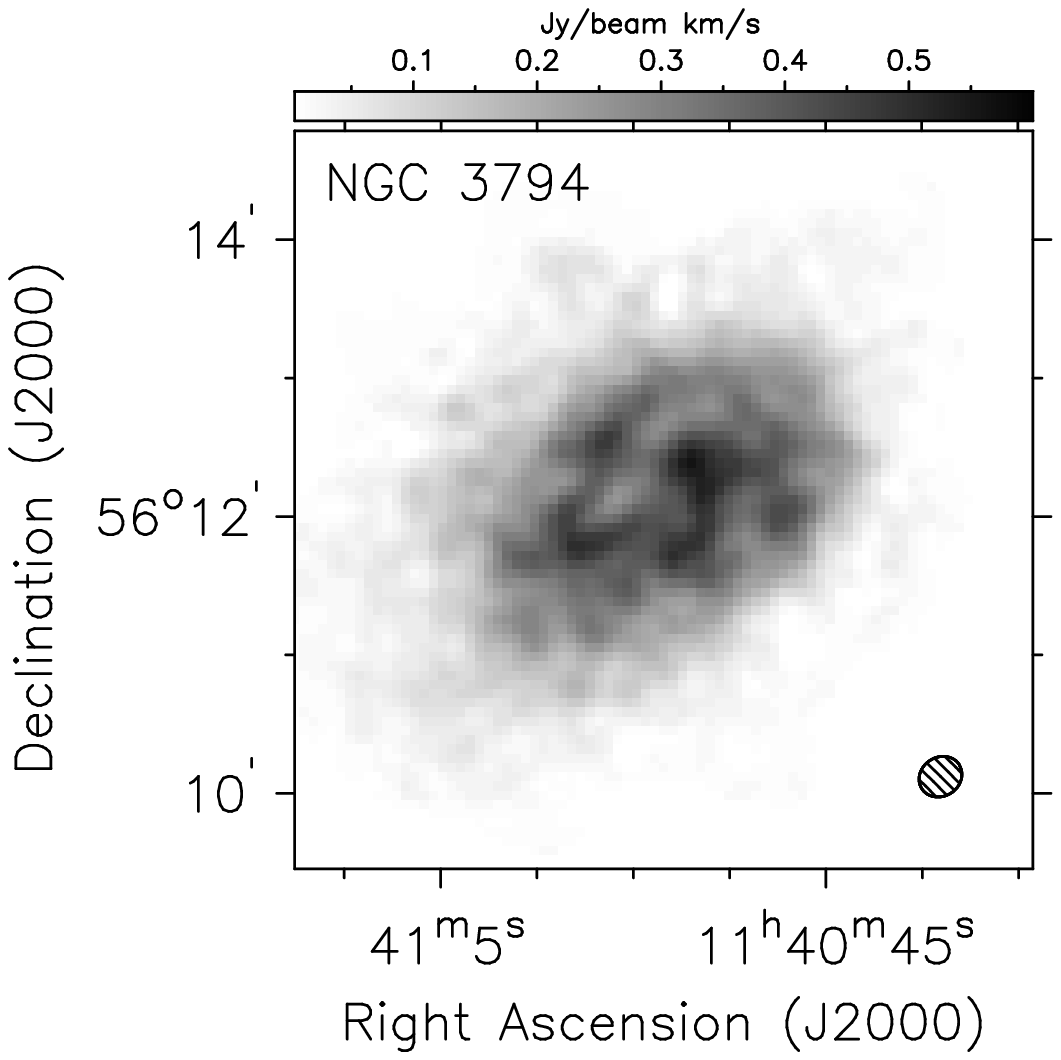} &
\includegraphics[height=0.22\textheight]{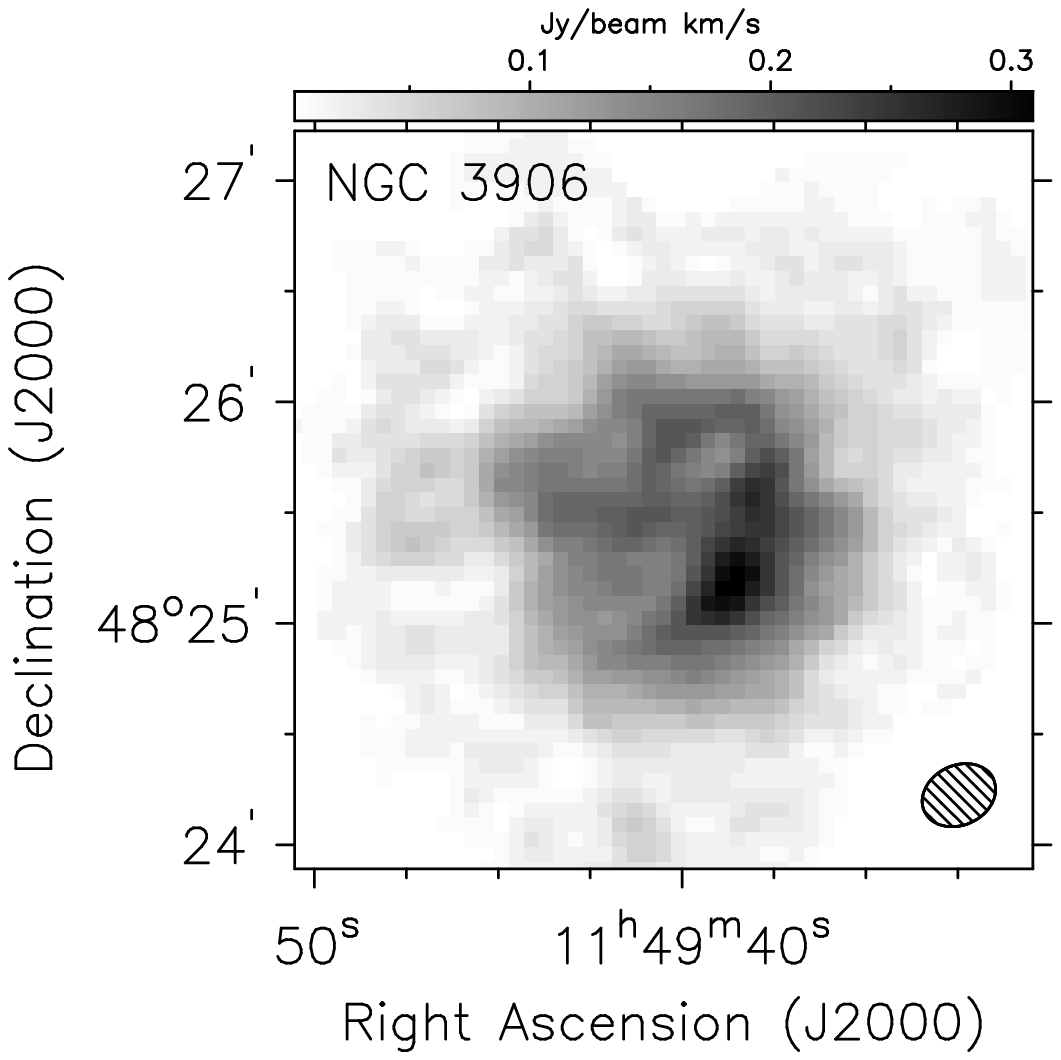} \\
\end{array}$
\end{center}
\caption{Integrated \ion{H}{1} intensity map for each galaxy, derived
  from the naturally-weighted, blanked, and primary beam corrected
  data cube.  The maps cover the same area shown in the channel maps.
  The beam size is shown in the lower right corner of each panel.}
\label{fig:m0}
\end{figure}

\begin{figure}
\figurenum{25}
\begin{center}$
\begin{array}{ccc}
\includegraphics[height=0.22\textheight]{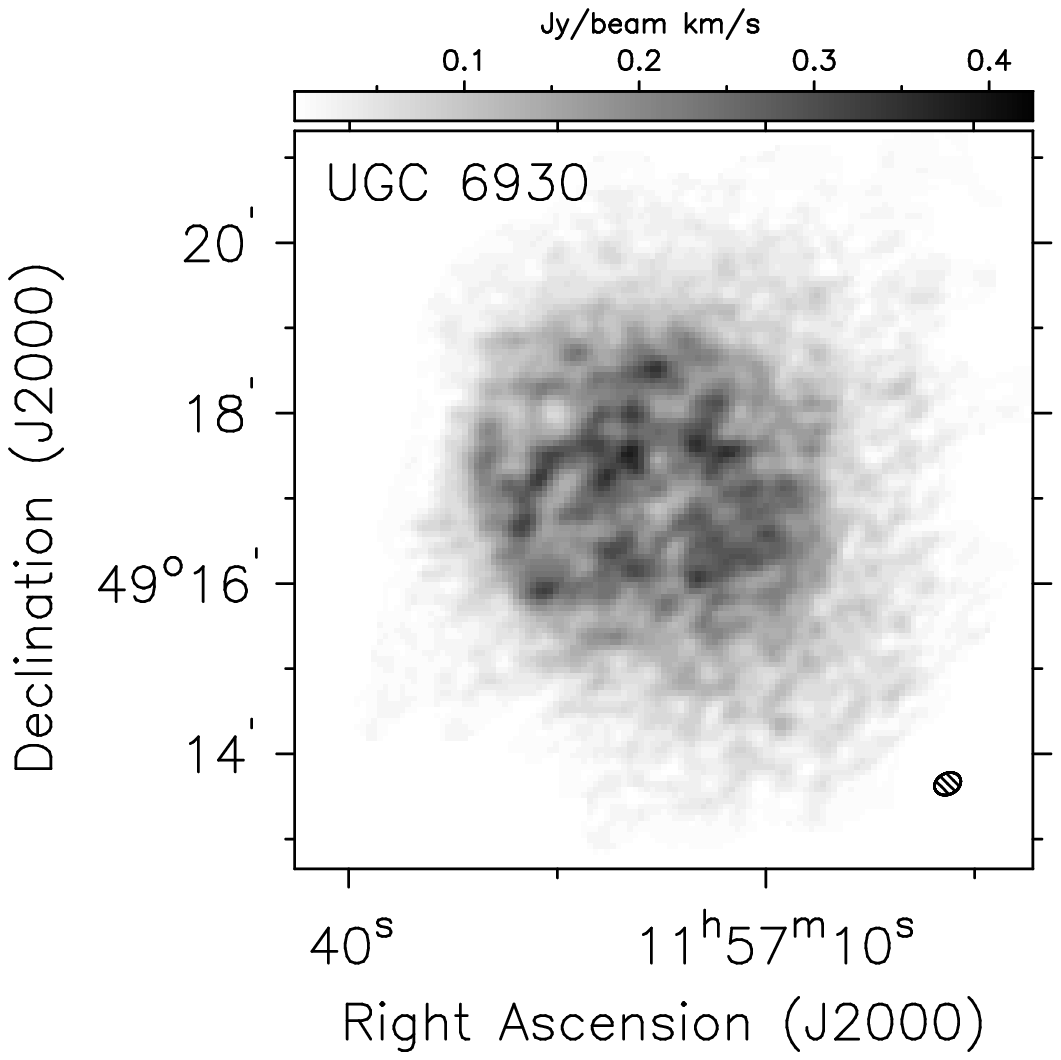} &
\includegraphics[height=0.22\textheight]{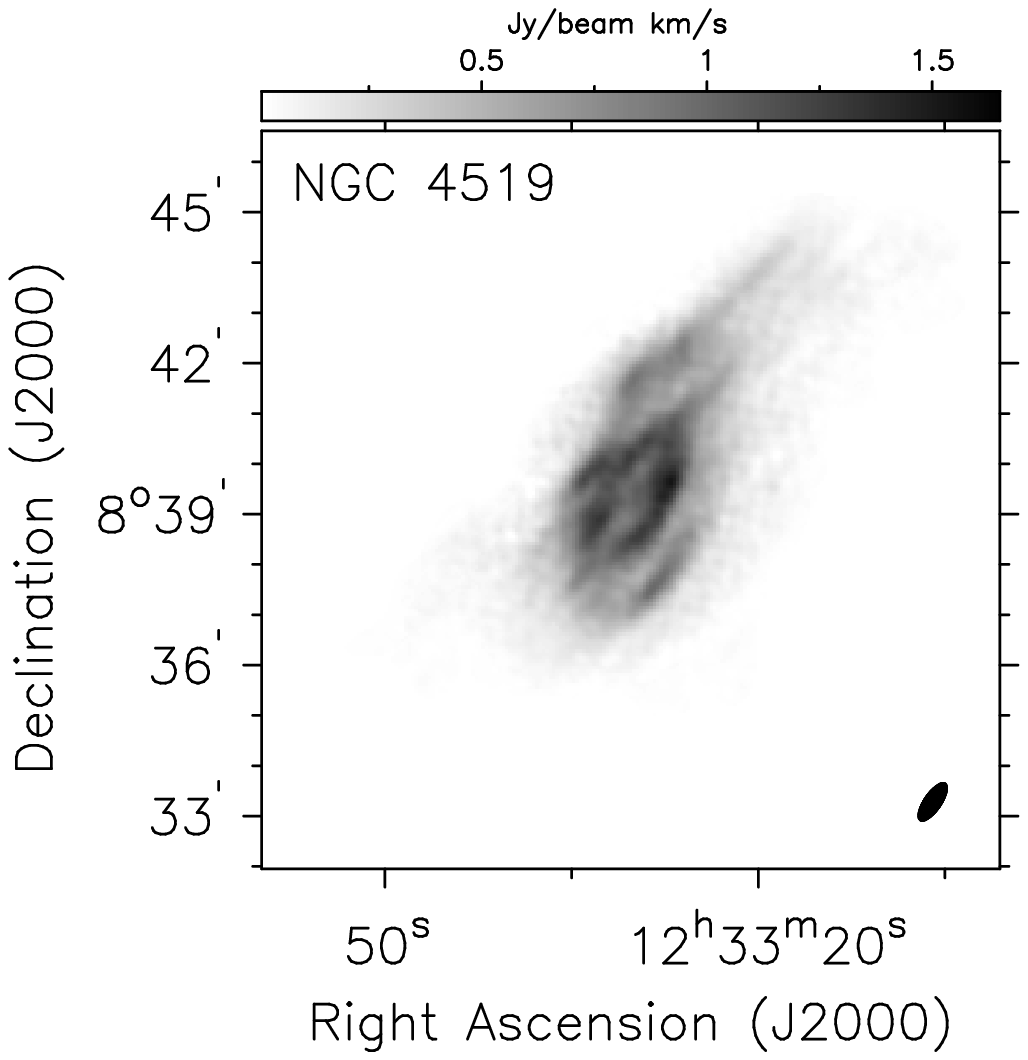} &
\includegraphics[height=0.22\textheight]{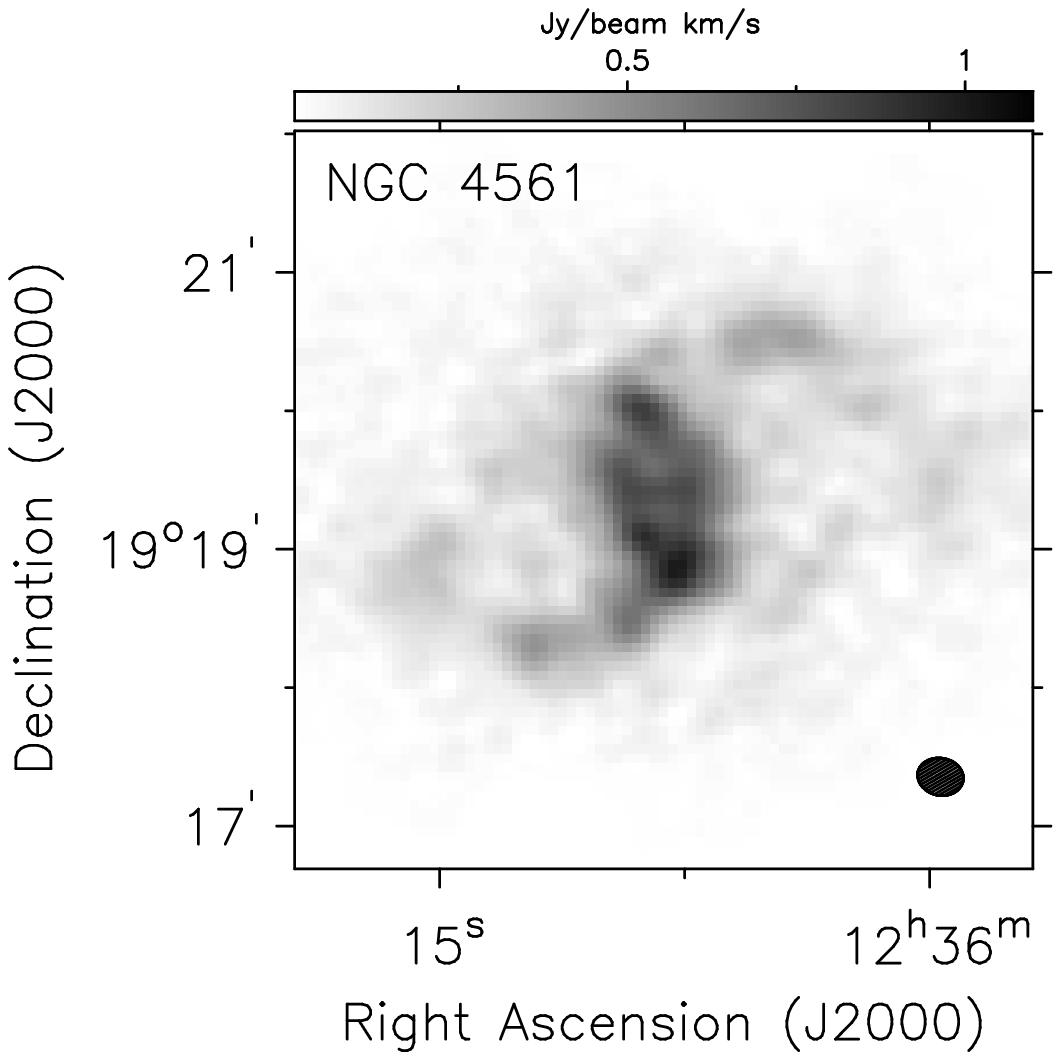} \\
\\
\includegraphics[height=0.22\textheight]{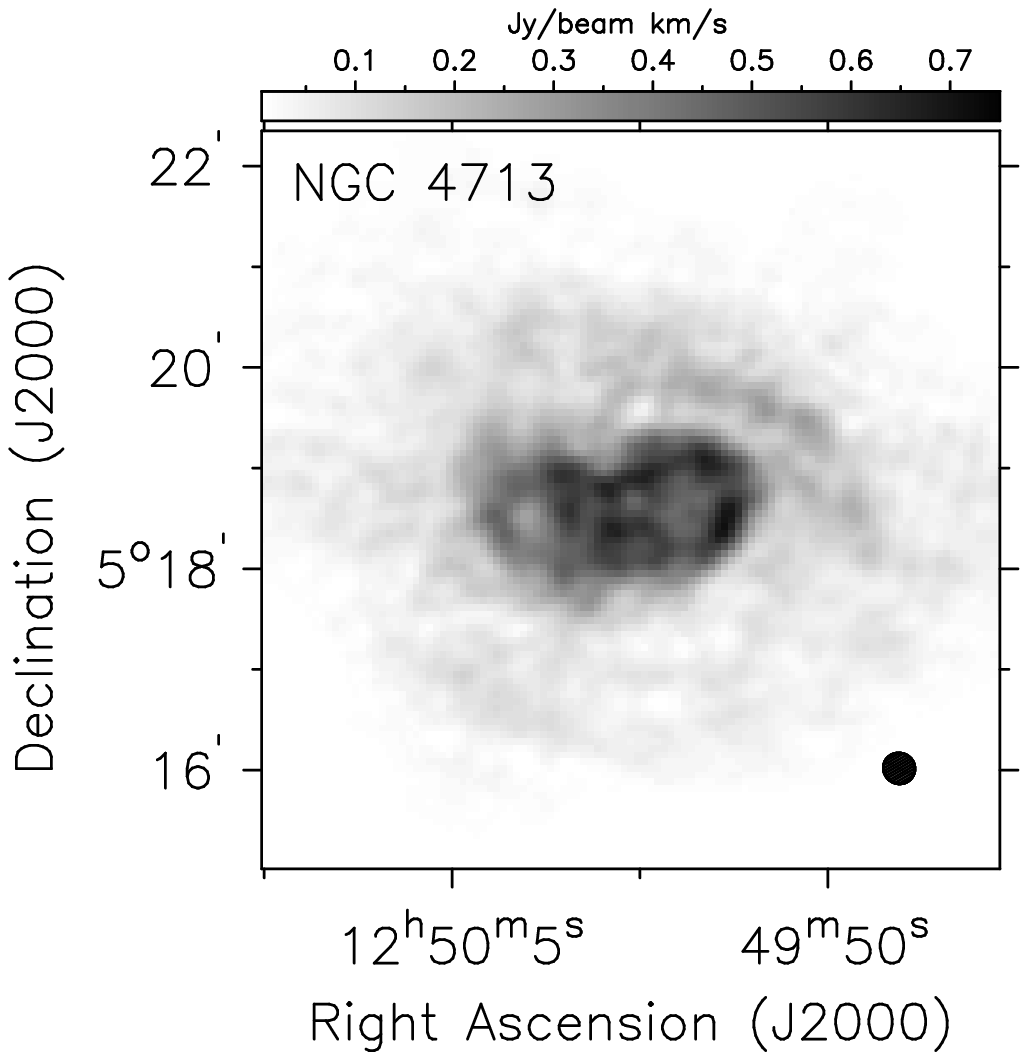} &
\includegraphics[height=0.22\textheight]{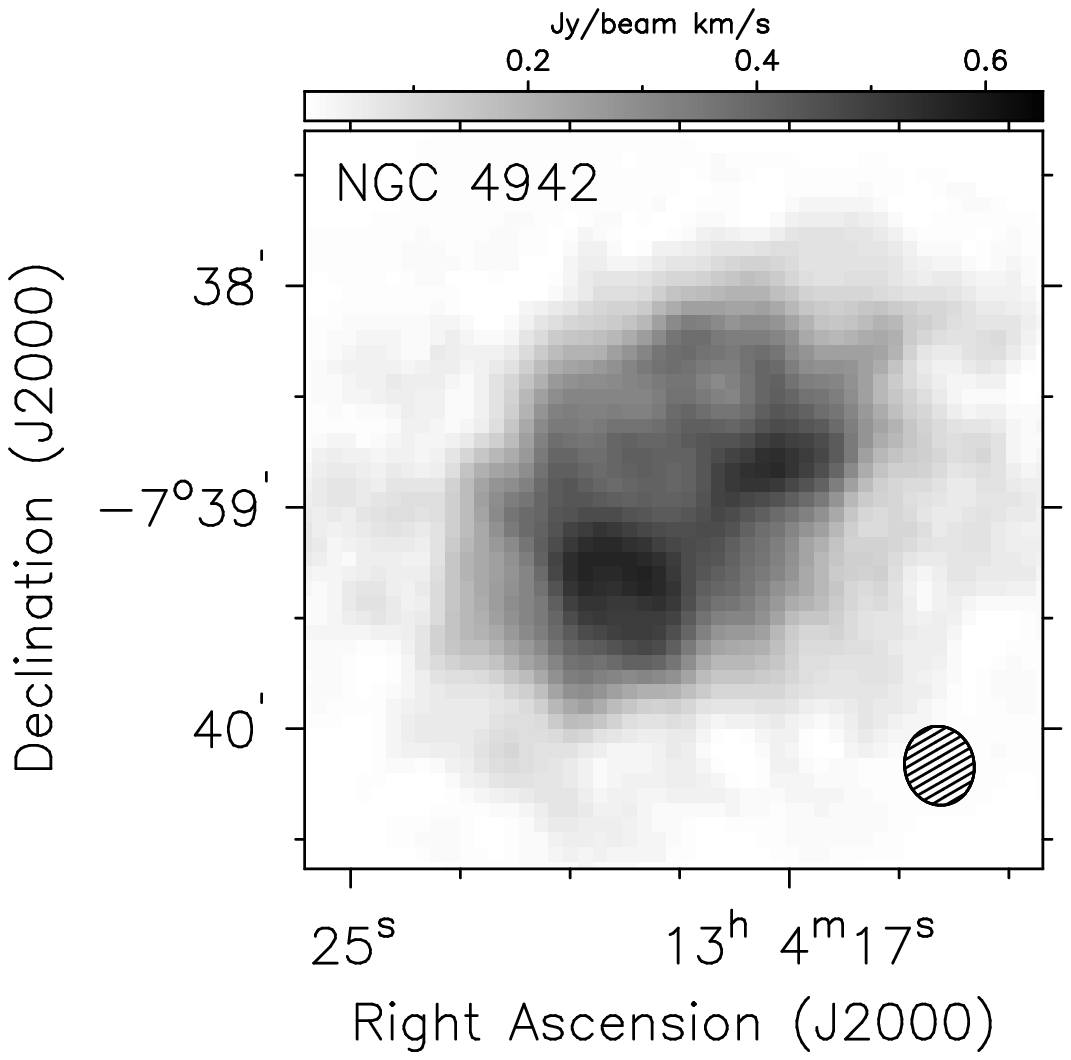} &
\includegraphics[height=0.22\textheight]{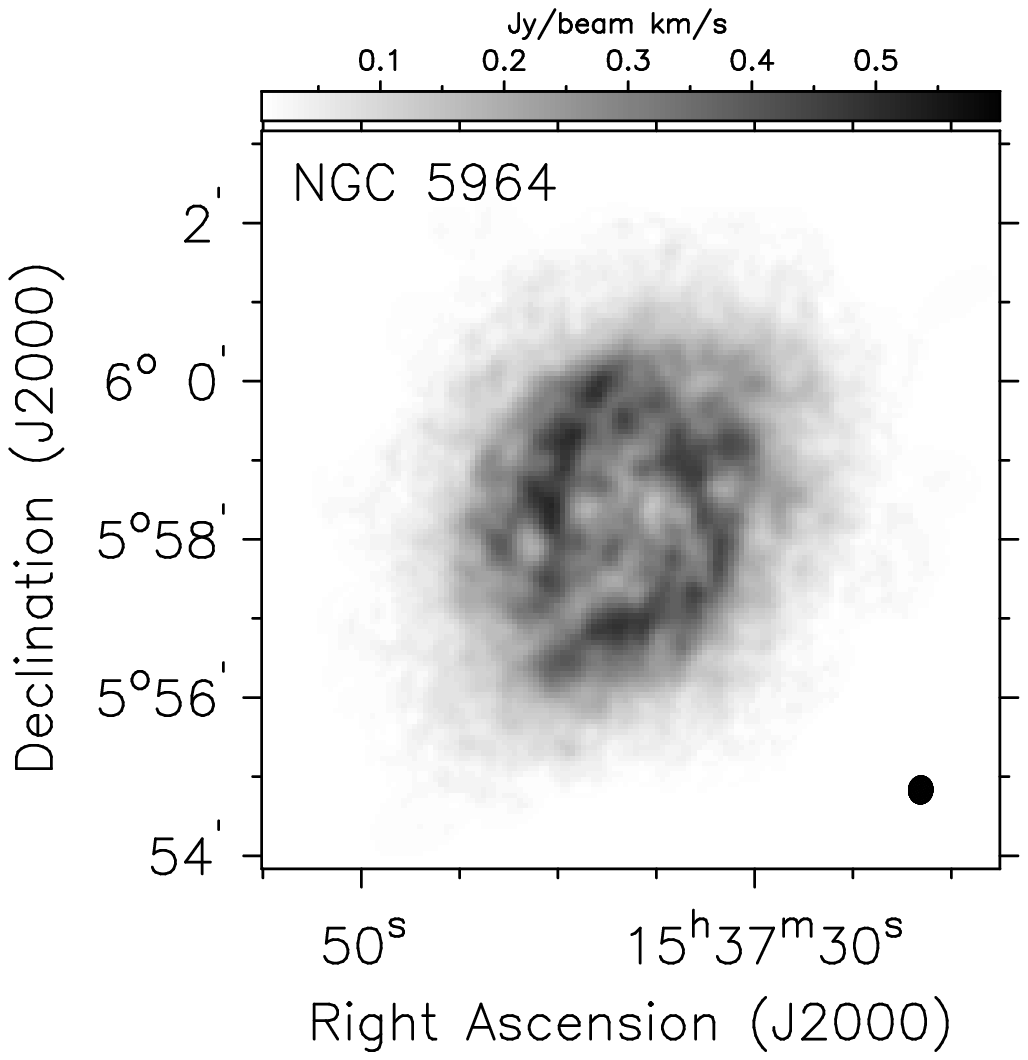} \\
\\
\includegraphics[height=0.22\textheight]{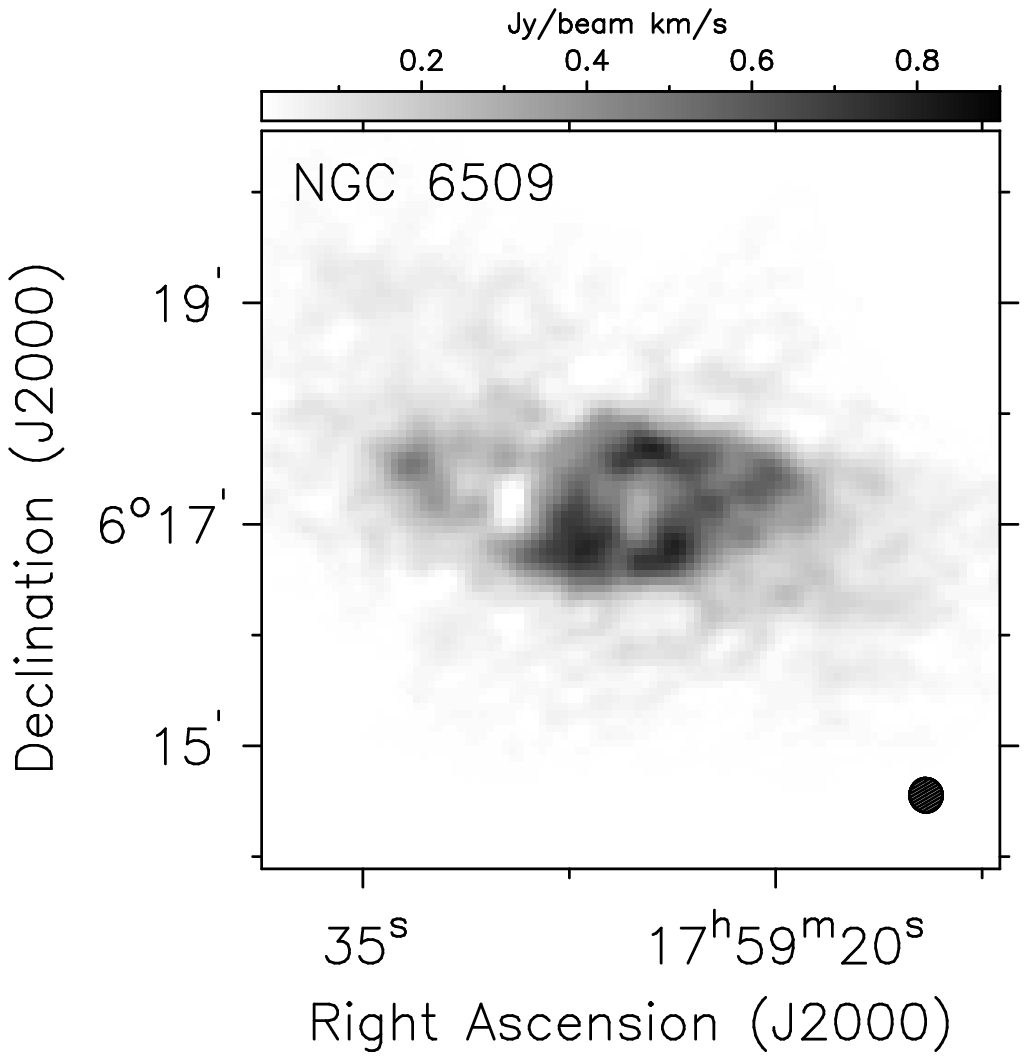} &
\includegraphics[height=0.22\textheight]{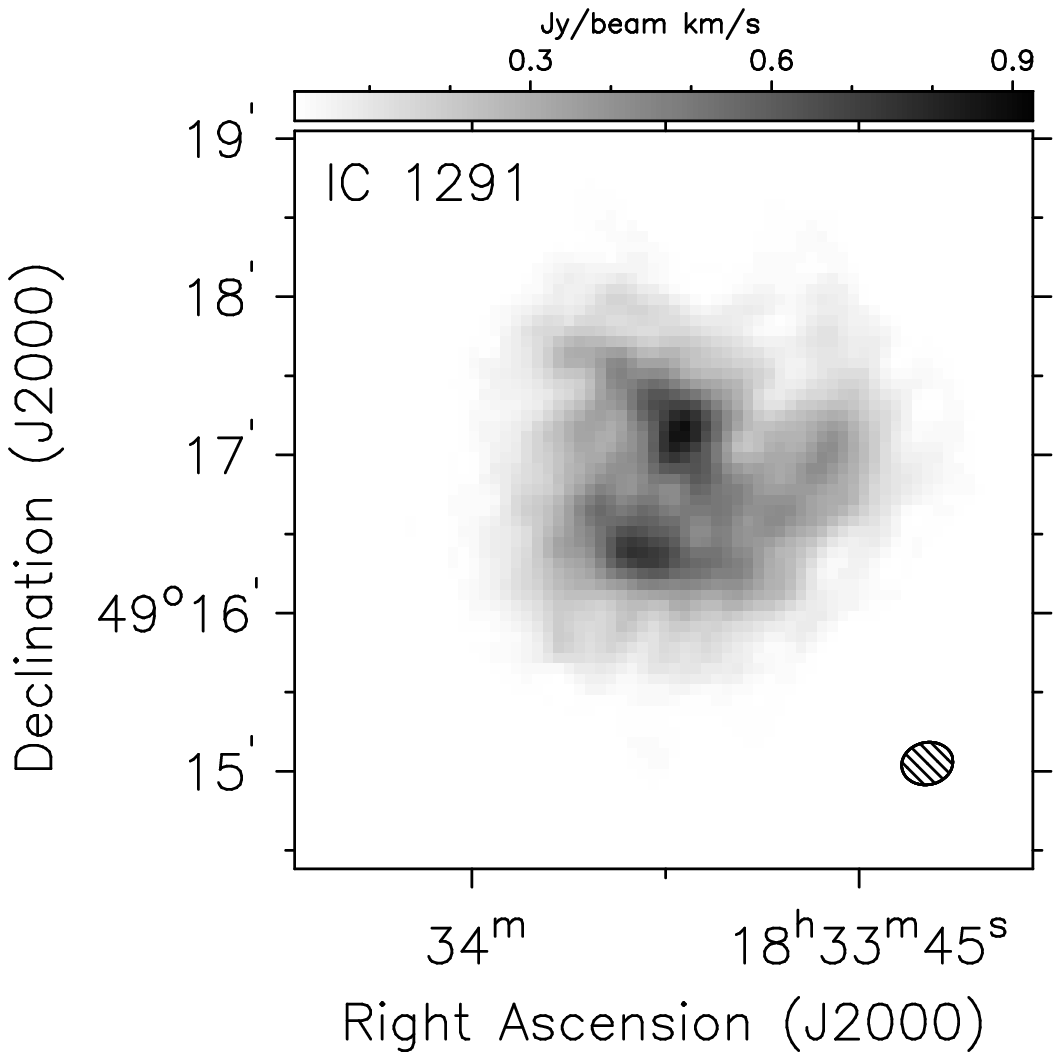} &
 \\
\end{array}$
\end{center}
\caption{(Continued) Note the ``hole'' in NGC~6509.  \ion{H}{1} is in
absorption here because it is in the foreground of a strong radio
lobe.  See Section~\ref{sec:morph} for more details.}
\end{figure}


\clearpage

\end{document}